\documentclass[]{gewu}

\usepackage{amsmath,amsfonts,bm}

\def\eqref#1{equation~\ref{#1}}

\def\1{\bm{1}}

\DeclareMathAlphabet{\mathsfit}{\encodingdefault}{\sfdefault}{m}{sl}
\SetMathAlphabet{\mathsfit}{bold}{\encodingdefault}{\sfdefault}{bx}{n}

\usepackage{graphicx}
\usepackage{wrapfig}
\usepackage{hyperref}
\usepackage{url}
\usepackage{graphicx}
\usepackage{capt-of}
\usepackage{booktabs}
\usepackage{capt-of}
\usepackage{booktabs}
\usepackage{tabularx}
\usepackage{multirow}
\usepackage{array}
\usepackage{booktabs}
\usepackage{tabularx}
\usepackage{multirow}
\usepackage{titletoc}
\usepackage{algorithm}
\usepackage{algpseudocode}
\usepackage[table]{xcolor}
\usepackage{needspace}
\usepackage{amsmath}
\usepackage{multirow}
\usepackage[table]{xcolor}
\usepackage{booktabs}
\usepackage{booktabs}
\usepackage{multirow}
\usepackage{caption}
\usepackage{adjustbox}
\usepackage{booktabs}
\usepackage{multirow}
\usepackage{graphicx}
\usepackage{pifont}
\usepackage{graphicx}
\usepackage{pifont}
\usepackage[table]{xcolor}
\newsavebox{\SteerLeftCaption}
\newsavebox{\SteerRightCaption}
\newlength{\SteerCaptionHeight}
\newlength{\SteerBackboneWidth}
\newlength{\SteerStrategyWidth}
\newlength{\SteerComplexWidth}
\newlength{\SteerSelectiveWidth}
\newlength{\SteerMeasureWidth}

\newcommand{\SteerMaxWidth}[2]{%
	\settowidth{\SteerMeasureWidth}{#2}%
	\ifdim\SteerMeasureWidth>#1\relax
	\setlength{#1}{\SteerMeasureWidth}%
	\fi
}
\definecolor{STEERPurple}{HTML}{6F4FA3}

\definecolor{A1Bg}{HTML}{E9E0F2} 
\definecolor{A2Bg}{HTML}{DDEDE9} 
\definecolor{A3Bg}{HTML}{F5E8CF} 
\definecolor{A4Bg}{HTML}{DFEAF6} 

\usepackage{pifont}
\newcommand{\steertick}{%
	\textcolor{black}{%
		\raisebox{-0.15ex}{\scalebox{1.5}{\ding{51}}}%
	}%
}
\newcommand{\steerdash}{%
\textcolor{black!40}{--}}

\definecolor{OtherModel}{gray}{0.48}
\definecolor{LightPurple}{HTML}{F1ECF8}

\newcommand{\sg}{\operatorname{sg}}
\definecolor{etptpurple}{HTML}{F1ECF8}

\title{
	\bfseries Steer
	Unified Audio Editing 
	\\
	Toward Reliable Source Selection
}

\author[1,2,*]{Youquan Fu}
\author[1]{Dongnuan Cai}
\author[1]{Yu Pan}
\author[2]{Henghui Du}
\author[2]{Yujia Zhang}
\author[3]{Ping Luo}
\author[4]{Quande Liu}
\author[1,\dagger]{Yake Wei}
\author[2,\dagger]{Xi Chen}
\author[3,\dagger]{Hongyuan Zhang}
\author[1]{Di Hu}

\affiliation[1]{Renmin University of China}
\affiliation[2]{Tencent}
\affiliation[3]{The University of Hong Kong}
\affiliation[4]{\mbox{The Chinese University of Hong Kong}}

\contribution[*]{Work done during an internship at Tencent.}
\contribution[\dagger]{Corresponding author.}

\checkdata[Project Page]{\url{https://gewu-lab.github.io/SteerSound/}}
\checkdata[Code]{\url{https://github.com/GeWu-Lab/SteerSound}}
\checkdata[Checkpoint]{\url{https://huggingface.co/GeWu-Lab/SteerSound}}
\checkdata[Dataset]{\url{https://huggingface.co/datasets/GeWu-Lab/SteerSound}}

\abstract{
Unified audio editors support diverse operations, yet often struggle to modify the intended source while preserving non-target content. We show that source selection in audio editing is shaped by early denoising states, while incorrect early selections can hardly be corrected during subsequent denoising. Motivated by these findings, we propose \textit{Source-selection Training via Early-trajectory Emphasis and Recovery (STEER)}, a two-stage post-training algorithm for reliable selective audio editing. STEER-Emphasis prioritizes training on task-critical early states to reduce source-selection errors, while STEER-Recovery learns corrective denoising from self-generated rollout states to recover from early deviations. To systematically assess source-selection reliability, we introduce the Selective Audio Editing Benchmark, enabling large-scale automatic evaluation of target manipulation and non-target preservation. Applying STEER to our pretrained multitask base editor gives SteerSound, a unified audio editor with leading selective editing performance among open-source audio editors. Experiments on our base editor, UNISON, and SAM Audio demonstrate that STEER consistently improves both selective editing capability and overall editing performance. Early-trajectory emphasis further brings substantial improvements in transcript-conditioned target speaker removal and separation, highlighting the broader applicability of this training principle.
}

\begin{document}
\maketitle

\section{Introduction}
\label{introduction}
\begin{figure*}[t]
	\vspace{-4mm}
	\centering
	\includegraphics[width=\textwidth]{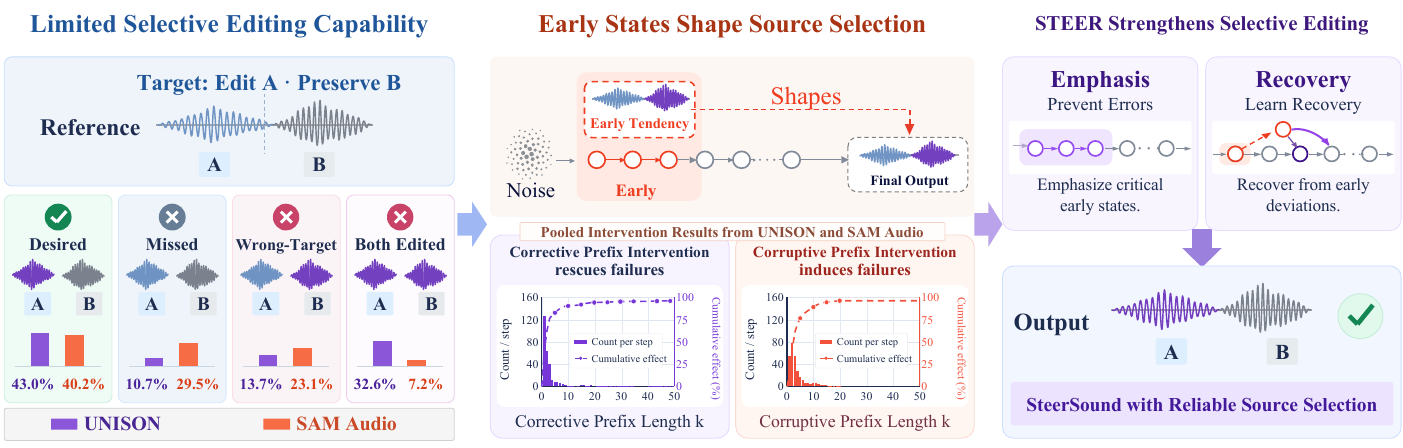}
	\vspace{-5mm}
	\caption{Strengthening selective audio editing with STEER. \textbf{Left}: Existing editors may leave the target unchanged or modify non-target content. \textbf{Middle}: Early states shape final source selection, while later denoising tends to follow the established editing direction rather than revise it. \textbf{Right}: STEER addresses these failures through early-trajectory emphasis (Emphasis) and recovery training (Recovery), yielding SteerSound with more reliable selective editing.}
	\label{figure1}
	\vspace{-3mm}
\end{figure*}

Recent advances in diffusion and Flow Matching have enabled unified audio editors to follow diverse instructions and perform multiple editing operations within a single model~\cite{wang2023audit,tao2025mmedit,tian2026audio,li2026unison}. However, reliable unified editing requires more than broad task coverage. Editing tasks like removal, replacement, and pitch or volume modification require the model to \textit{identify the intended source}, apply the requested operation, and preserve the remaining content. We refer to these as selective editing tasks. As illustrated in Figure~\ref{figure1} (left), even strong existing models, like UNISON~\cite{li2026unison} and SAM Audio~\cite{shi2025sam}, can leave the intended source unchanged, modify a non-target source, or alter both. These selective editing failures highlight the need to strengthen not only audio generation, but also the model's ability to select the correct editing source.

\Needspace{4\baselineskip}
To understand these failures, we investigate how source selection develops along the generation trajectory. By analyzing the editing tendency at different denoising steps, we find that incorrect source-selection tendencies form early in the generation trajectory. This suggests that establishing the correct source selection early may be critical to the final editing outcome. As shown in Figure~\ref{figure1} (middle), we conduct bidirectional early-trajectory interventions: replacing the early trajectories of failed edits with correct ones, and those of successful edits with deliberately incorrect ones. Across UNISON and SAM Audio, correct early trajectories can rescue failed edits, whereas incorrect ones can induce source-selection failures, showing that early source selection strongly shapes which source is ultimately edited. 

Together, these experiments reveal a common pattern across selective editing tasks: the source ultimately edited is largely determined by the early generation trajectory. Once an incorrect source is selected early in denoising, later denoising tends to follow the resulting trajectory rather than revise the earlier decision. These findings motivate us to address these source-selection failures from two complementary directions: preventing source-selection errors from forming early and improving recovery when they occur. We therefore propose \textit{\textcolor{STEERPurple}Source-selection \textcolor{STEERPurple}Training via \textcolor{STEERPurple}Early-trajectory \textcolor{STEERPurple}Emphasis and \textcolor{STEERPurple}Recovery (\textcolor{STEERPurple}{STEER})}, a two-stage post-training algorithm that strengthens source selection in pretrained audio editors. In the first stage, STEER-Emphasis places greater training emphasis on critical early-trajectory states to reduce the formation of incorrect source selections. In the second stage, STEER-Recovery learns corrective denoising from self-generated rollout states to recover from the early deviations that remain, as illustrated in Figure~\ref{figure1} (right).

To enable more comprehensive and systematic evaluation, we first introduce the Selective Audio Editing Benchmark for large-scale assessment of source-selection reliability. Moreover, existing unified audio editors remain limited in selective task coverage, making broad analysis across diverse editing operations difficult. We therefore develop our base editor, an extensively pretrained unified editor covering six generation and editing tasks, including four selective editing tasks, providing a stronger and more comprehensive foundation for selective audio editing and cross-task analysis. Applying STEER to our base editor gives SteerSound, which substantially improves source-selection reliability while achieving advanced editing performance among open-source unified audio editors. Beyond SteerSound, STEER can be well applied to UNISON and SAM Audio, enhancing their selective editing performance, while early-trajectory emphasis further applies to transcript-conditioned target speaker removal and separation, supporting the broader applicability of our approach.

Our contributions are summarized as follows.
\textbf{Firstly}, through intermediate-state analysis and bidirectional prefix interventions, we show that early denoising states largely determine whether a unified audio editor edits the correct source, and hence its selective editing performance.
\textbf{Secondly}, we propose STEER, a two-stage training algorithm: STEER-Emphasis prioritizes critical early-trajectory states, while STEER-Recovery learns corrective denoising from self-generated rollout states. STEER improves selective and overall editing performance on our base editor, UNISON, and SAM Audio, and the early-trajectory emphasis further extends to related selective-control tasks.
\textbf{Finally}, we introduce the Selective Audio Editing Benchmark, enabling large-scale automatic evaluation of editing success and three types of selective failures across six tasks. We also develop a six-task unified base editor and apply STEER to obtain SteerSound, which achieves leading overall and selective editing performance among open-source audio editors.
\section{Related Work}
\label{sec:related}

\paragraph{Unified and Instruction-Guided Audio Editing.}
Text-guided audio editing has been explored through training-based and training-free diffusion approaches, including AUDIT~\cite{wang2023audit}, ZETA~\cite{manor2024zero}, PPAE~\cite{xu2024prompt}, and AudioEditor~\cite{jia2025audioeditor}, while recent systems such as MMEdit~\cite{tao2025mmedit}, Audio-Omni~\cite{tian2026audio}, and UNISON~\cite{li2026unison} further broaden task coverage and integrate audio-language modeling with audio editing. However, source-selection reliability in unified audio editors remains underexplored. We develop SteerSound, a unified six-task audio editor, and strengthen this selective editing capability.

\paragraph{Denoising Dynamics for Source Selection.}
Prior work has shown that denoising timesteps can play non-uniform roles, through stage-specialized denoisers~\cite{balaji2022ediff}, timestep reweighting or sampling~\cite{hang2023efficient,esser2024scaling}, and task-dependent timestep emphasis in diffusion-based visual perception~\cite{pang2025aligning}. Related studies also use model-generated states to reduce train--inference mismatch or trajectory errors, such as Self Forcing~\cite{huang2026self} for autoregressive video diffusion and SOAR~\cite{qin2026soar} for diffusion post-training. However, these works do not study how an audio editor selects which source to modify, nor how such source-selection decisions evolve along the denoising trajectory. We show that source selection is established along a task-dependent early trajectory, motivating greater training emphasis on critical early states and recovery from early deviations.
\begin{figure*}[t]
	\vspace{-4mm}
	\centering
	\includegraphics[width=\textwidth]{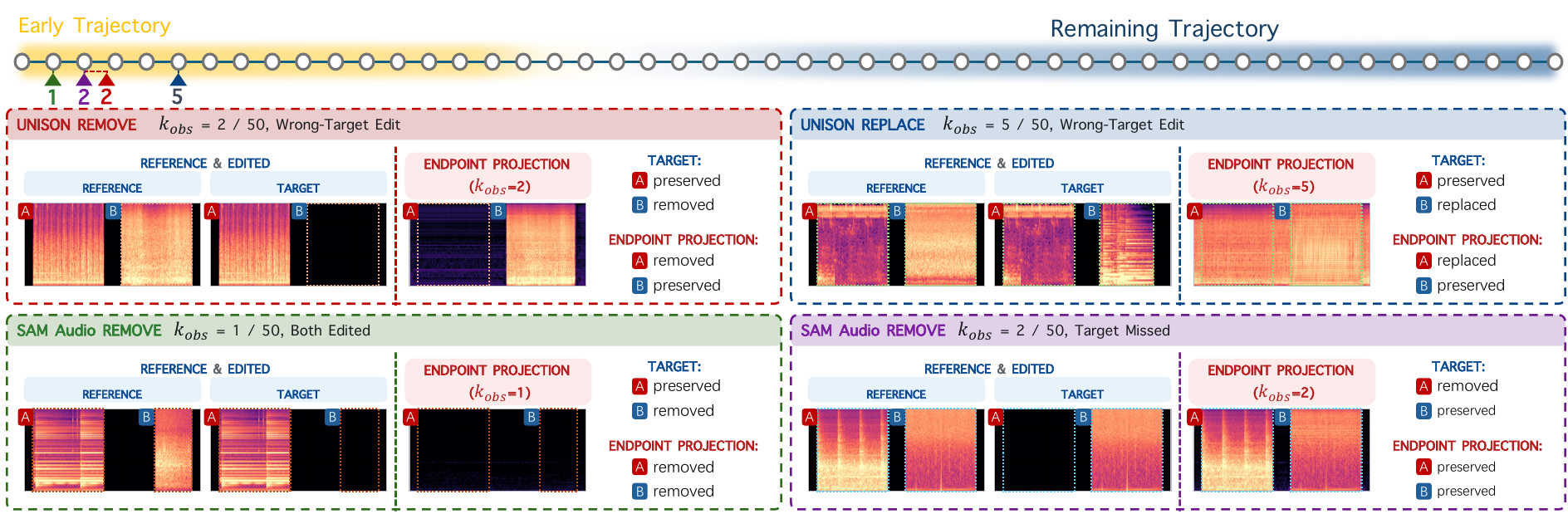}
	\vspace{-5mm}
	\caption{Erroneous source selection is formed and visible within the first few denoising steps. Each panel shows a failed edit: the reference, the target, and the zero-noise estimate decoded at step $k_{\mathrm{obs}}$, the earliest step at which the erroneous tendency becomes perceptible. A and B denote the two sources in the reference audio.}
	\label{figure2}
	\vspace{-3mm}
\end{figure*}

\section{Source Selection Along the Trajectory}
\label{sec:analysis}
To strengthen source selection in unified audio editing, we examine how source-selection decisions evolve during generation. We first investigate when source-selection tendencies emerge and then test the role of early-trajectory states in final source selection on two representative editors: UNISON~\cite{li2026unison}, a representative unified audio editor, and SAM Audio~\cite{shi2025sam}, a representative source separation model that we evaluate for source removal.

\paragraph{Flow Matching and trajectory notation.}
Given a clean target latent $\mathbf{z}_0$ and Gaussian noise
$\boldsymbol{\epsilon}\sim\mathcal{N}(\mathbf{0},\mathbf{I})$, Flow Matching~\cite{lipman2022flow} constructs
$\mathbf{z}_\sigma=(1-\sigma)\mathbf{z}_0+\sigma\boldsymbol{\epsilon}$
and optimizes
\begin{equation}
	\mathcal{L}_{\mathrm{FM}}
	=
	\left\|f_\theta(\mathbf{z}_\sigma,\sigma,c)
	-(\boldsymbol{\epsilon}-\mathbf{z}_0)\right\|_2^2,
	\label{eq:fm}
\end{equation}
where $c$ denotes the editing condition. We denote the denoising schedule by
$1=\sigma_0>\sigma_1>\cdots>\sigma_K=0$ and the state after $k$ steps by
$\mathbf{z}^{(k)}$, with
$\mathbf{z}^{(k)}
=\mathbf{z}^{(k-1)}
+(\sigma_k-\sigma_{k-1})\mathbf{v}_{k-1}$.
Unless otherwise specified, we use $K=50$ denoising steps.

\subsection{Emergence of Source-Selection Tendencies}
\label{sec:observation}
To examine when source-selection tendencies emerge, we inspect the editing direction represented at intermediate stages of the model's denoising trajectory. Since intermediate states remain noisy, we estimate their editing direction by extrapolating each state to the zero-noise endpoint using its instantaneous velocity prediction:
\begin{equation}
	\hat{\mathbf{z}}_0^{(k)}
	=
	\mathbf{z}^{(k)}-\sigma_k\mathbf{v}_k,
\end{equation}
where $\mathbf{v}_k=f_\theta(\mathbf{z}^{(k)},\sigma_k,c)$. Decoding $\hat{\mathbf{z}}_0^{(k)}$ reveals the source-selection tendency at step $k$, allowing us to inspect which source the model tends to edit and which it tends to preserve. This allows us to identify when an erroneous source-selection tendency begins to form during generation. We apply this probe to failed examples from selective editing tasks across the two models. We observe that erroneous source-selection tendencies are already formed and perceptually discernible during early denoising, as illustrated by the representative cases in Figure~\ref{figure2}. In these examples, UNISON edits the wrong source, while SAM Audio either removes both sources or misses the target. Notably, the zero-noise estimates reveal these errors within the first 1--5 of 50 denoising steps, even though the underlying states remain noisy.

\begin{figure*}[t]
	\vspace{-4mm}

	\includegraphics[width=\textwidth]{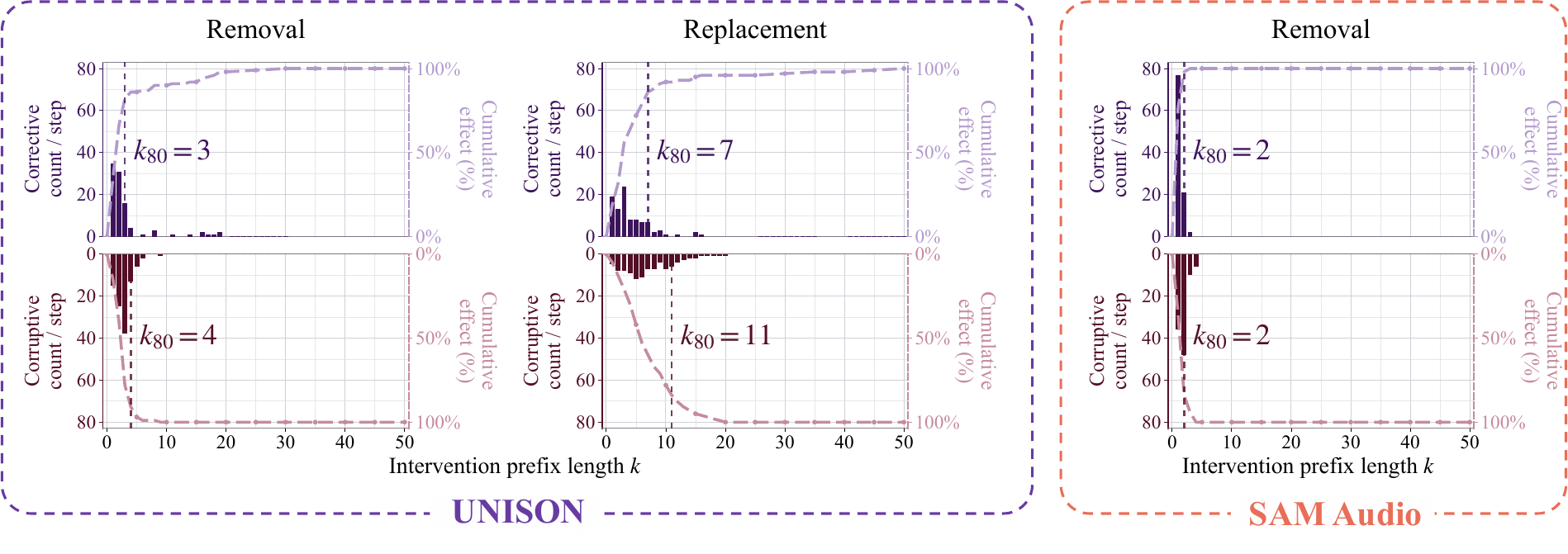}
	\vspace{-8mm}
	\caption{Bidirectional early-trajectory interventions on source selection. Corrective (top) and corruptive (bottom) prefix interventions are evaluated on 100 validation examples per task. Bars show per-step corrective or corruptive counts, while curves show the corresponding cumulative effects. For each task, $k_{80}$ denotes the shortest intervention prefix length at which the cumulative effect reaches 80\%. Removal consistently exhibits a shorter $k_{80}$ than replacement.}
	\label{figure3}
	\vspace{-5mm}
\end{figure*}
\subsection{Early Trajectories Shape Final Source Selection}
\label{sec:intervention}

To test whether early trajectory states causally influence final source selection, we perform corrective and corruptive prefix interventions with varying lengths $k$. For failed edits, we replace the first $k$ model-predicted velocities with oracle velocities constructed from the correct editing target. For successful edits, we instead replace the first $k$ velocities with those constructed from a deliberately corrupted target encoding incorrect source selection. In both cases, the model resumes its own denoising under the original instruction.

As shown in Figure~\ref{figure3}, short corrective prefixes can rescue failed edits, whereas short corruptive prefixes can induce source-selection failures. Across the evaluated tasks and models, extending these prefixes into later denoising stages yields limited additional effects. These results provide evidence that early trajectory states shape final source selection, while subsequent denoising often follows the editing direction established early, even when it is incorrect. Two properties of standard training may partly explain this behavior. SD3-style logit-normal sampling~\cite{esser2024scaling} emphasizes intermediate noise levels, leaving high-noise early states relatively underrepresented. Meanwhile, training conditions on noisy versions of correct targets, whereas inference states are self-generated, which may make recovery from early source-selection errors difficult.

Different editing tasks exhibit different sensitivities to early prefix replacement, reflecting differences among editing operations. Removal sharply contrasts target suppression with non-target preservation, making source selection apparent within only a few steps, whereas replacement requires more steps for the modification to become distinguishable, yielding broader but still early critical ranges. We therefore group operations into \textit{generation-oriented}, \textit{selective suppression}, and \textit{selective transformation} tasks, depending on whether they synthesize new content, apply strongly contrasting operations to the target and preserved sources, or modify the target's content or acoustic attributes.

\section{Unified Audio Editing with STEER}
\label{sec:method}
Motivated by the findings in Section~\ref{sec:analysis}, we aim to develop a unified audio editor that combines broad task coverage with reliable source selection. Section~\ref{sec:base} introduces our base editor, which supports six audio generation and editing tasks, including four selective editing tasks. Section~\ref{STEER} then presents STEER, a two-stage post-training algorithm for improving source-selection reliability.

\subsection{Unified Audio Editor}
\label{sec:base}
To develop a unified audio editor that combines broad task coverage with strong selective editing capabilities, we first build a multi-task base editor. Unlike existing editors, which typically support only a subset of the selective operations considered here, our base editor supports six tasks: two generation-oriented tasks (generation and addition) and four selective editing tasks (removal, replacement, pitch modification, and volume modification). This unified task coverage also enables systematic cross-task analysis of source selection within a single model. We pretrain the model on large-scale audio editing data to establish strong initial multi-task audio editing capabilities. Further details on our base editor are provided in Section~\ref{sec:setup}.

\begin{figure*}[t]
	\centering
	\noindent\makebox[\textwidth][s]{%
	
	\begin{minipage}[t]{0.466\textwidth}
		\vspace{0pt}
		\centering
		
		\includegraphics[width=\linewidth]{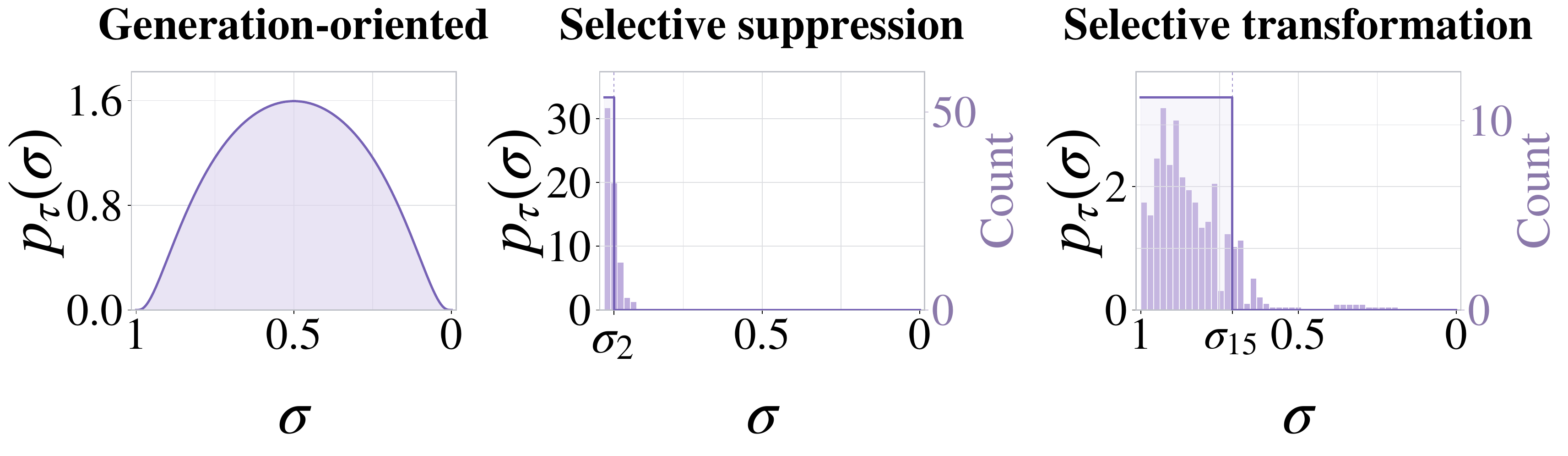}
		
		\vspace{-3mm}
		\captionof{figure}{Noise-level sampling distributions for different task groups in STEER-Emphasis.}
		\label{figure4}
		\vspace{1mm}
		
		\includegraphics[width=\linewidth]{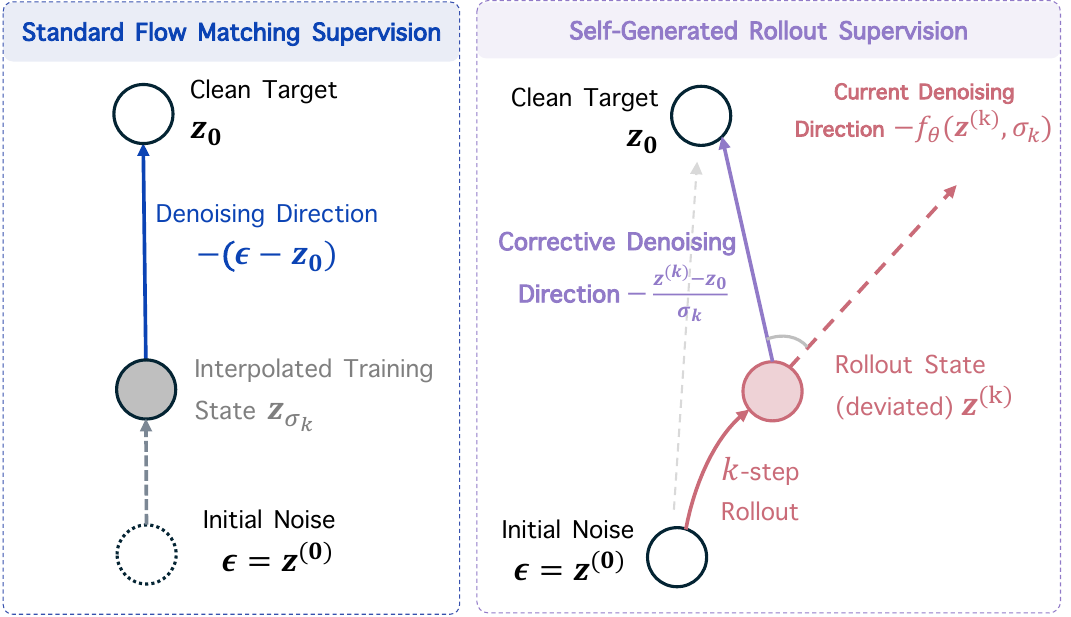}
		
		\vspace{-3mm}
		\caption{Comparison of standard Flow Matching supervision on interpolated states and STEER-Recovery supervision on self-generated rollout states.}
		\label{fig:ter_supervision}
	\end{minipage}%
	\hfill%
\begin{minipage}[t]{0.500\textwidth}
	\vspace{0pt}
	\refstepcounter{algorithm}
	\label{alg1}
	
	\noindent
	{\footnotesize\bfseries Algorithm~\thealgorithm: STEER}
	\par\vspace{2pt}
	\hrule height 0.5pt
	\vspace{3pt}
	
	\begingroup
	\fontsize{6.6pt}{8pt}\selectfont
	\algrenewcommand{\algorithmicindent}{0.95em}
	\algrenewcommand{\algorithmicrequire}{\textbf{Input:}}
	\algrenewcommand{\algorithmicensure}{\textbf{Output:}}
	\algrenewcommand{\alglinenumber}[1]{%
		\textcolor{black!50}{\tiny #1}}
	
	\begin{algorithmic}[1]
		\Require $\theta_0$, $\mathcal D_{\rm tr}$,
		$\mathcal D_{\rm val}$, $p_\tau$, $K$, $P$
		\Ensure $\theta^\star$
		
		\Statex \textbf{Stage I: STEER-Emphasis}
		\State Train $\theta_0$ with $\mathcal L_{\rm FM}$
		and $\sigma\sim p_\tau$
		\State $\theta^\star\gets$
		validation-best checkpoint
		
		\Statex \textbf{Stage II: STEER-Recovery}
		\State $k\gets1,\ q\gets0$
		\While{$q<P$ \textbf{and} $k<K$}
		\State $\theta\gets\theta^\star$;
		reset optimizer and schedule
		\For{each training update}
		\State Sample
		$(\mathbf z_0,c,\tau)\sim\mathcal D_{\rm tr}$
		\State Sample
		$\boldsymbol\epsilon\sim\mathcal N(\mathbf0,\mathbf I)$
		\If{$\tau\in\{\mathrm{Generate},\mathrm{Add}\}$}
		\State $\sigma\sim p_{\rm LN}$;
		$\mathcal L\gets\mathcal L_{\rm FM}$
		\Else
		\State $\mathbf z^{(k)}\gets
		\sg(\Call{Rollout}{\theta,\boldsymbol\epsilon,c,k})$
		\State $\mathcal L\gets
		\mathcal L_{\rm rec}^{(k)}$
		\EndIf
		\State Update $\theta$ using $\mathcal L$
		\EndFor
		\State $\hat\theta_k\gets$
		validation-best checkpoint at depth $k$
		\If{$S_{\rm val}(\hat\theta_k)>
			S_{\rm val}(\theta^\star)$}
		\State $\theta^\star\gets\hat\theta_k,\ q\gets0$
		\Else
		\State $q\gets q+1$
		\EndIf
		\State $k\gets k+1$
		\EndWhile
		\State \Return $\theta^\star$
	\end{algorithmic}
	\endgroup
	
	\vspace{2pt}
	\hrule height 0.5pt
\end{minipage}%
	}
	\vspace{-3mm}
\end{figure*}

\subsection{STEER}
\label{STEER}
STEER enhances source-selection reliability in pretrained audio editors through two complementary post-training stages. First, STEER-Emphasis prioritizes training on task-critical early denoising states to reduce the formation of source-selection errors. STEER-Recovery then learns corrective denoising from self-generated rollout states to recover from remaining early-trajectory deviations.

\subsubsection{Stage1: Early-Trajectory Emphasis}
\label{sec:focus}

Standard logit-normal sampling emphasizes intermediate noise levels, leaving early states critical to source selection relatively underrepresented during training. To address this, STEER-Emphasis prioritizes training on critical early denoising states to reduce source-selection errors. Since different editing tasks exhibit different early-trajectory sensitivities, we first identify their critical ranges and then increase sampling within the corresponding intervals.

To identify these ranges, we perform corrective prefix interventions on our base editor. For each failed edit, we determine the shortest early prefix whose correction leads to a successful edit. Let \(C_\tau(k)\) denote the cumulative fraction of failed edits for task \(\tau\) rescued by corrective prefix interventions of up to \(k\) denoising steps. Following the task grouping in Section~\ref{sec:intervention}, we define the critical range for each selective task group $\mathcal{G}$ as $k_{\mathcal{G}}^\rho=\max_{\tau\in\mathcal{G}}\min\{k:C_\tau(k)\geq\rho\}$, where $\rho$ specifies the fraction of rescued edits to be covered. Taking the maximum ensures that the range covers the critical prefix of each task in the group. With $\rho=0.8$, the resulting ranges cover the first 2 of 50 denoising steps for selective suppression and the first 15 for selective transformation.

Based on these ranges, we uniformly sample noise levels within the corresponding high-noise intervals for selective editing tasks, while retaining logit-normal sampling for generation-oriented tasks:
\begin{equation}
	p_\tau(\sigma)=
	\left\{
	\begin{array}{@{}l@{\quad}l@{\qquad}l@{}}
		p_{\mathrm{LN}}(\sigma),
		& \tau\in\{\mathrm{Generate},\mathrm{Add}\},
		& \text{\footnotesize\itshape generation-oriented}\\
		\mathcal{U}(\sigma_2,1),
		& \tau=\mathrm{Remove},
		& \text{\footnotesize\itshape selective suppression}\\
		\mathcal{U}(\sigma_{15},1),
		& \tau\in\{\mathrm{Replace},\mathrm{Pitch},\mathrm{Volume}\},
		& \text{\footnotesize\itshape selective transformation}
	\end{array}
	\right.
	\label{eq:focus}
\end{equation}
Here $p_{\mathrm{LN}}$ denotes the SD3-style logit-normal distribution~\cite{esser2024scaling}, defined by $\sigma=\operatorname{sigmoid}(u)$ with $u\sim\mathcal{N}(0,1)$. STEER-Emphasis increases sampling of critical early states, while leaving the Flow Matching objective unchanged. Figure~\ref{figure4} illustrates the sampling distributions, with additional intervention results provided in Appendix~\ref{AA}.

\begin{figure*}[t]
	\centering
	
	\begin{minipage}[t]{0.27\textwidth}
		\centering
		\vspace{0pt}
		\includegraphics[
		width=\linewidth,
		height=3.5cm,
		keepaspectratio
		]{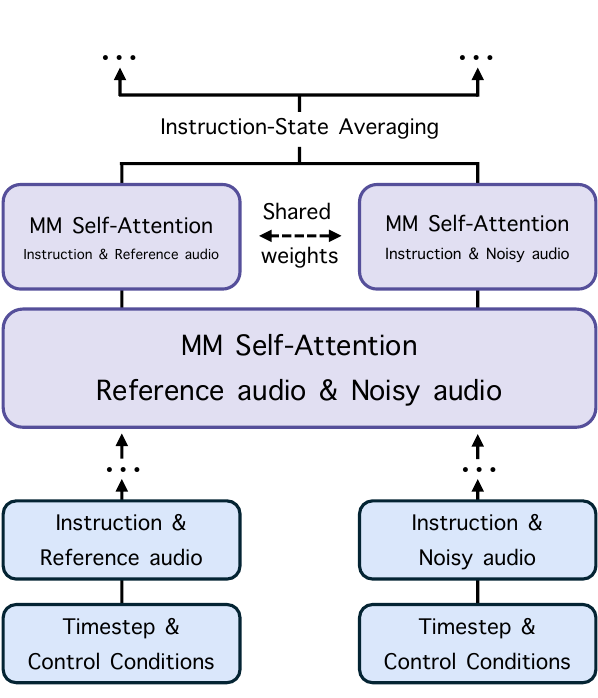}
		\vspace{-2mm}
		\captionof{figure}{Architecture of our base editor.}
		\label{figure6}
	\end{minipage}
	\hfill
	\begin{minipage}[t]{0.70\textwidth}
		\centering
		\vspace{0pt}
		\includegraphics[
		width=\linewidth,
		height=3.5cm,
		keepaspectratio
		]{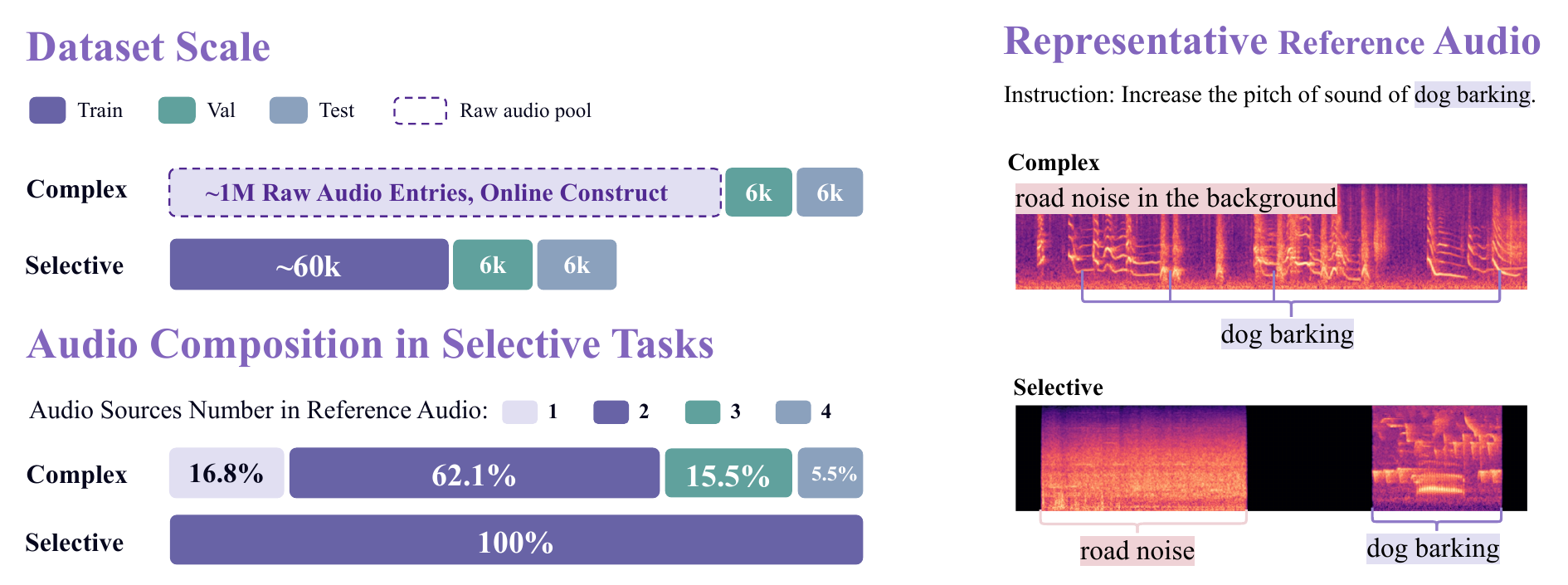}
		\vspace{-2mm}
		\caption{Statistics and representative examples of the two data settings, including dataset scale and source composition.}
		\label{figure7}
	\end{minipage}
	
	\vspace{-5mm}
\end{figure*}

\subsubsection{Stage 2: Early-Trajectory Recovery}
\label{sec:recover}

Although STEER-Emphasis reduces early source-selection errors, incorrect selections may still occur and persist through subsequent denoising. Standard Flow Matching supervises states interpolated between Gaussian noise and the clean target. Consequently, when early errors drive inference onto incorrect trajectories, subsequent denoising may continue along these trajectories rather than correcting them. To address this, STEER-Recovery introduces corrective supervision at self-generated rollout states, as illustrated in Figure~\ref{fig:ter_supervision}, enabling the model to learn to recover from its own trajectory deviations.

Starting from the validation-best STEER-Emphasis checkpoint, we perform a $k$-step denoising rollout from Gaussian noise $\boldsymbol{\epsilon}$ to obtain $\mathbf{z}^{(k)}$ for each selective editing example, without backpropagating through the rollout. We then supervise the predicted velocity with a corrective target determined by the rollout state and the desired clean edit $\mathbf{z}_0$:
\begin{equation}
	\mathcal L_{\mathrm{rec}}^{(k)}=
	\left\|f_\theta(\mathbf z^{(k)},\sigma_k,c)
	-\sg\!\left(\frac{\mathbf z^{(k)}-\mathbf z_0}{\sigma_k}\right)\right\|_2^2,
	\label{eq:recovery}
\end{equation}
where $\sg$ denotes stop-gradient. This target reduces to the standard Flow Matching velocity $\boldsymbol{\epsilon}-\mathbf z_0$ when $\mathbf{z}^{(k)}$ lies on the target--noise interpolation path, but provides corrective supervision when the rollout deviates from it. This enables the model to learn to recover from its own trajectory deviations during denoising.

We progressively increase the rollout depth $k$ from 1, initializing each depth from the best checkpoint found so far. Generate and Add retain the standard Flow Matching objective throughout this stage. Training stops after $P$ consecutive depths without improvement in the best validation score, and the best checkpoint is retained. Algorithm~\ref{alg1} summarizes the complete STEER procedure. We obtain SteerSound by applying both stages of STEER to our base editor.

\section{Experiments}
\label{sec:experiments}
We first describe the experimental configuration and compare unified audio editing performance across models. We then evaluate STEER's effectiveness, conduct ablation studies on its two post-training stages, and examine its robustness to alternative definitions of the critical early-trajectory range. Next, we assess STEER's generalizability across model architectures for audio editing. Finally, we investigate whether early-trajectory emphasis also benefits related selective-control tasks.

\begin{table*}[t]
	\centering
	\caption{Unified audio editing performance across models on the Complex Audio Editing Test Set. Our Base Editor denotes the pretrained model before STEER; SteerSound denotes our final model after both STEER stages. ``--'' in the \textit{Train.} column denotes training-free methods built upon existing pretrained audio generation models, while ``--'' in task columns indicates that the model or method does not support the corresponding task.}
	\label{table1}
	
	\vspace{-2mm}
	\scriptsize
	\setlength{\tabcolsep}{2.5pt}
	\renewcommand{\arraystretch}{1.10}
	
	\resizebox{\textwidth}{!}{
		\begin{tabular}{l*{17}{c}}
			\toprule
			
			& \multicolumn{2}{c}{\textbf{Params.}}
			& \multicolumn{2}{c}{\textbf{Generate}}
			& \multicolumn{2}{c}{\textbf{Add}}
			& \multicolumn{2}{c}{\textbf{Replace}}
			& \multicolumn{3}{c}{\textbf{Remove}}
			& \multicolumn{3}{c}{\textbf{Pitch}}
			& \multicolumn{3}{c}{\textbf{Volume}} \\
			
			\cmidrule(lr){2-3}
			\cmidrule(lr){4-5}
			\cmidrule(lr){6-7}
			\cmidrule(lr){8-9}
			\cmidrule(lr){10-12}
			\cmidrule(lr){13-15}
			\cmidrule(lr){16-18}
			
			\textbf{Model}
			& Total & Train.
			& CLAP$\uparrow$ & KL$\downarrow$
			& CLAP$\uparrow$ & KL$\downarrow$
			& CLAP$\uparrow$ & KL$\downarrow$
			& CLAP$\uparrow$ & LSD$\downarrow$ & KL$\downarrow$
			& MAE$\downarrow$ & LSD$\downarrow$ & KL$\downarrow$
			& MAE$\downarrow$ & LSD$\downarrow$ & KL$\downarrow$ \\
			
			\midrule
			
			Audio-Omni
			& 7.907B & 3.047B
			& \underline{0.231} & 2.057
			& 0.196 & 1.955
			& 0.154 & 2.561
			& 0.001 & 1.695 & 2.803
			& -- & -- & --
			& -- & -- & -- \\
			
			MMEdit
			& 9.057B & 0.616B
			& -- & --
			& 0.159 & 1.882
			& 0.085 & 2.706
			& 0.012 & 2.060 & 2.155
			& -- & -- & --
			& 14.043 & 1.042 & 0.427 \\
			
			AUDIT
			& 1.067B & 0.860B
			& -- & --
			& 0.151 & 1.417
			& 0.082 & 1.897
			& 0.038 & 1.603 & 2.691
			& -- & -- & --
			& -- & -- & -- \\
			
			PPAE
			& 1.318B & --
			& -- & --
			& 0.182 & 1.566
			& 0.094 & 1.501
			& 0.044 & 1.539 & 1.384
			& -- & -- & --
			& -- & -- & -- \\
			
			SDEdit
			& 1.118B & --
			& -- & --
			& 0.111 & 2.347
			& 0.129 & 2.222
			& 0.129 & 1.778 & 1.969
			& 0.923 & 0.942 & 1.963
			& 5.047 & 0.895 & 1.992 \\
			
			ZETA
			& 1.118B & --
			& -- & --
			& 0.135 & 2.060
			& 0.140 & 1.939
			& 0.135 & 1.688 & 1.690
			& 0.570 & 0.803 & 1.441
			& 4.199 & 0.722 & 1.432 \\
			
			SAM Audio(Small)
			&  0.714B & 0.492B
			& -- & --
			& -- & --
			& -- & --
			& 0.059 & 1.514 & 1.726
			& -- & -- & --
			& -- & -- & -- \\
			
			SAM Audio(Large)
			& 3.154B & 2.929B
			& -- & --
			& -- & --
			& -- & --
			& 0.056 & 1.465 & 1.629
			& -- & -- & --
			& -- & -- & -- \\
			
			UNISON
			& 8.086B & 0.732B
			& \textbf{0.247} & \textbf{1.845}
			& \underline{0.250} & \textbf{1.218}
			& 0.222 & 1.783
			& 0.112 & 1.576 & 1.564
			& -- & -- & --
			& -- & -- & -- \\
			
			Our Base Editor
			& 1.348B & 0.673B
			& 0.223 & \underline{1.915}
			& 0.249 & \underline{1.286}
			& \underline{0.230} & \underline{1.032}
			& \underline{0.137} & \underline{1.346} & \underline{1.140}
			& \underline{0.055} & \underline{0.493} & \underline{0.281}
			& \underline{0.702} & \underline{0.433} & \underline{0.200} \\
			
			\rowcolor{LightPurple}
			\textbf{SteerSound}
			& 1.348B & 0.673B
			& 0.220 & 2.133
			& \textbf{0.251} & 1.352
			& \textbf{0.235} & \textbf{1.028}
			& \textbf{0.144} & \textbf{1.164} & \textbf{0.788}
			& \textbf{0.047} & \textbf{0.491} & \textbf{0.280}
			& \textbf{0.634} & \textbf{0.433} & \textbf{0.200} \\
			
			\bottomrule
		\end{tabular}
	}
	\vspace{-3mm}
\end{table*}

\begin{table*}[t]
	\centering
	\caption{Effectiveness of STEER across models on the Complex Audio Editing Test Set. TBG scores are computed relative to the corresponding base model. ``--'' indicates that the corresponding training stage is not applied or the editing task is not supported.}
	\label{table2}
	\vspace{-2mm}
	
	\begingroup
	\fontsize{8pt}{9.4pt}\selectfont
	\setlength{\tabcolsep}{1.75pt}
	\renewcommand{\arraystretch}{1.16}
	
	\def\steertick{%
		\raisebox{-0.12ex}{\scalebox{1.4}{\ding{51}}}}
	\def\steerdash{\textcolor{black!45}{--}}
	
	\begin{adjustbox}{max width=\textwidth,center}
		\begin{tabular}{@{}lcc*{17}{c}@{}}
			\toprule
			
			\multirow[c]{2}{*}{\textbf{Model}}
			& \multirow[c]{2}{*}{
				\shortstack[c]{
					\textbf{STEER-}\\[-0.4ex]
					\textbf{Emphasis}
				}
			}
			& \multirow[c]{2}{*}{
				\shortstack[c]{
					\textbf{STEER-}\\[-0.4ex]
					\textbf{Recovery}
				}
			}
			& \multicolumn{2}{c}{\textbf{Generate}}
			& \multicolumn{2}{c}{\textbf{Add}}
			& \multicolumn{2}{c}{\textbf{Replace}}
			& \multicolumn{3}{c}{\textbf{Remove}}
			& \multicolumn{3}{c}{\textbf{Pitch}}
			& \multicolumn{3}{c}{\textbf{Volume}}
			& \textbf{Overall}
			& \textbf{Selective} \\
			
			\cmidrule(lr){4-5}
			\cmidrule(lr){6-7}
			\cmidrule(lr){8-9}
			\cmidrule(lr){10-12}
			\cmidrule(lr){13-15}
			\cmidrule(lr){16-18}
			\cmidrule(lr){19-19}
			\cmidrule(lr){20-20}
			
			& & &
			CLAP$\uparrow$ & KL$\downarrow$
			& CLAP$\uparrow$ & KL$\downarrow$
			& CLAP$\uparrow$ & KL$\downarrow$
			& CLAP$\uparrow$ & LSD$\downarrow$ & KL$\downarrow$
			& MAE$\downarrow$ & LSD$\downarrow$ & KL$\downarrow$
			& MAE$\downarrow$ & LSD$\downarrow$ & KL$\downarrow$
			& $\mathrm{TBG}_{\mathrm{all}}\uparrow$
			& $\mathrm{TBG}_{\mathrm{sel}}\uparrow$ \\
			
			\midrule
			
			
			\textbf{Our Base Editor}
			& \steerdash & \steerdash
			& \textbf{0.22} & \textbf{1.92}
			& \textbf{0.25} & \textbf{1.29}
			& \textbf{0.23} & \textbf{1.03}
			& \textbf{0.14} & 1.35 & 1.14
			& 0.06 & \textbf{0.49} & \textbf{0.28}
			& 0.70 & \textbf{0.43} & \textbf{0.20}
			& 0.00 & 0.00 \\
			
			+Emph. only
			& \steertick & \steerdash
			& \textbf{0.22} & 2.12
			& \textbf{0.25} & 1.32
			& \textbf{0.23} & 1.05
			& \textbf{0.14} & 1.22 & 0.90
			& \textbf{0.05} & \textbf{0.49} & \textbf{0.28}
			& 0.71 & \textbf{0.43} & \textbf{0.20}
			& 1.67 & 2.97 \\
			
			\rowcolor{LightPurple}
			\textbf{+Full STEER}
			& \steertick & \steertick
			& \textbf{0.22} & 2.13
			& \textbf{0.25} & 1.35
			& \textbf{0.23} & \textbf{1.03}
			& \textbf{0.14} & \textbf{1.16} & \textbf{0.79}
			& \textbf{0.05} & \textbf{0.49} & \textbf{0.28}
			& \textbf{0.63} & \textbf{0.43} & \textbf{0.20}
			& \textbf{5.36} & \textbf{8.25} \\
			
			\midrule
			
			
			\textbf{UNISON}
			& \steerdash & \steerdash
			& \textbf{0.25} & 1.85
			& 0.25 & \textbf{1.22}
			& 0.22 & 1.78
			& 0.11 & 1.58 & 1.56
			& \steerdash & \steerdash & \steerdash
			& \steerdash & \steerdash & \steerdash
			& 0.00 & 0.00 \\
			
			+Emph. only
			& \steertick & \steerdash
			& \textbf{0.25} & 1.84
			& 0.26 & 1.37
			& \textbf{0.24} & 1.16
			& \textbf{0.15} & 1.39 & 0.89
			& \steerdash & \steerdash & \steerdash
			& \steerdash & \steerdash & \steerdash
			& 10.27 & 17.85 \\
			
			\rowcolor{LightPurple}
			\textbf{+Full STEER}
			& \steertick & \steertick
			& 0.25 & \textbf{1.83}
			& \textbf{0.27} & 1.32
			& 0.23 & \textbf{1.15}
			& \textbf{0.15} & \textbf{1.36} & \textbf{0.82}
			& \steerdash & \steerdash & \steerdash
			& \steerdash & \steerdash & \steerdash
			& \textbf{10.81} & \textbf{18.42} \\
			
			\midrule
			
			
			\textbf{SAM Audio}
			& \steerdash & \steerdash
			& \steerdash & \steerdash
			& \steerdash & \steerdash
			& \steerdash & \steerdash
			& 0.06 & 1.51 & 1.73
			& \steerdash & \steerdash & \steerdash
			& \steerdash & \steerdash & \steerdash
			& 0.00 & 0.00 \\
			
			+Emph. only
			& \steertick & \steerdash
			& \steerdash & \steerdash
			& \steerdash & \steerdash
			& \steerdash & \steerdash
			& 0.12 & 1.43 & 1.14
			& \steerdash & \steerdash & \steerdash
			& \steerdash & \steerdash & \steerdash
			& 108.62 & 108.62 \\
			
			\rowcolor{LightPurple}
			\textbf{+Full STEER}
			& \steertick & \steertick
			& \steerdash & \steerdash
			& \steerdash & \steerdash
			& \steerdash & \steerdash
			& \textbf{0.13} & \textbf{1.27} & \textbf{0.97}
			& \steerdash & \steerdash & \steerdash
			& \steerdash & \steerdash & \steerdash
			& \textbf{124.53} & \textbf{124.53} \\
			
			\bottomrule
		\end{tabular}
	\end{adjustbox}
	
	\par\vspace{1mm}
	\fontsize{7.2pt}{8pt}\selectfont
	\endgroup
	\vspace{-1mm}
\end{table*}

\subsection{Experimental Configuration}
\label{sec:setup}

\paragraph{Base Editor.}
As shown in Figure~\ref{figure6}, our base editor adopts a SeedEdit-inspired dual-branch DiT~\cite{shi2024seededit}. The reference and generated audio branches exchange information through multimodal self-attention with shared attention parameters, and both receive timestep and task-specific conditioning. The model combines a 673.35M-parameter trainable backbone with a T5-Base~\cite{raffel2020exploring} text encoder, an audio VAE~\cite{rombach2022high} trained from scratch, and a HiFi-GAN~\cite{kong2020hifi} vocoder. Further training details are provided in Appendix~\ref{AB}.

\paragraph{Data, Benchmark, and Metrics.}
We use two complementary data settings. To better approximate real-world audio editing scenarios, we construct Complex Audio Editing Data with diverse multi-source compositions. However, source overlap in these scenes makes selective editing failures difficult to identify automatically. To support systematic analysis and large-scale experiments, we introduce the Selective Audio Editing Benchmark. Its deliberately designed source composition allows target manipulation and non-target preservation to be evaluated separately, enabling automatic identification and measurement of selective editing failures. We also construct training and validation sets from the same distribution. For experiments, results reported on each test set are obtained from models post-trained on its corresponding training set, with checkpoints selected using the corresponding validation set. Figure~\ref{figure7} presents statistics and representative examples.

On the complex test set, we report task-aware CLAP Score, pitch and volume MAE, LSD~\cite{wang2023audit}, and KL-softmax~\cite{liu2023audioldm}. To summarize overall unified editing performance, we introduce the Task-Balanced Gain ($\mathrm{TBG}_{\mathrm{all}}$) to measure a model's overall unified editing capability, and $\mathrm{TBG}_{\mathrm{sel}}$ to specifically quantify its selective editing capability. On the selective audio editing benchmark, we develop an evaluation protocol that measures overall editing success while identifying and categorizing different types of source-selection errors. Further details on the two datasets and their corresponding metrics are provided in Appendices~\ref{AC} and~\ref{AD} respectively.

\paragraph{Implementation Details.}
Across models, we use the critical early-trajectory ranges defined in Section~\ref{sec:focus} for STEER-Emphasis and set the patience hyperparameter to $P=2$ for STEER-Recovery. Unless otherwise specified, our base editor and its post-trained variants use classifier-free guidance (CFG) at inference, with guidance scales of 7.5 for source addition and 3.5 for the other tasks, while other models retain their official default configurations. Experiment-specific hyperparameters, training procedures, and learning dynamics are provided in the corresponding appendices.

\begin{table*}[t]
	\centering
	\caption{Effectiveness of STEER across models on the Selective Audio Editing Benchmark. Succ. and SErr. denote success and selective error rates, respectively. ``--'' indicates that the corresponding training stage is not applied or the editing task is not supported.}
	\label{table3}
	
	\vspace{-3mm}
	
	\begingroup
	\fontsize{6.2pt}{7.5pt}\selectfont
	\setlength{\tabcolsep}{3.2pt}
	\renewcommand{\arraystretch}{1.20}
	
	\def\steertick{%
		\raisebox{-0.15ex}{\scalebox{1.5}{\ding{51}}}}
	\def\steerdash{\textcolor{black!40}{--}}
	
	\resizebox{\textwidth}{!}{%
		\begin{tabular}{lcc*{12}{c}}
			\toprule
			
			\multirow[c]{2}{*}{\textbf{Model}}
			& \multirow[c]{2}{*}{
				\shortstack[c]{
					\textbf{STEER-}\\
					\textbf{Emphasis}
				}
			}
			& \multirow[c]{2}{*}{
				\shortstack[c]{
					\textbf{STEER-}\\
					\textbf{Recovery}
				}
			}
			& \textbf{Generate}
			& \textbf{Add}
			& \multicolumn{2}{c}{\textbf{Replace}}
			& \multicolumn{2}{c}{\textbf{Remove}}
			& \multicolumn{2}{c}{\textbf{Pitch}}
			& \multicolumn{2}{c}{\textbf{Volume}}
			& \multicolumn{2}{c}{\textbf{Overall}} \\
			
			\cmidrule(lr){4-4}
			\cmidrule(lr){5-5}
			\cmidrule(lr){6-7}
			\cmidrule(lr){8-9}
			\cmidrule(lr){10-11}
			\cmidrule(lr){12-13}
			\cmidrule(lr){14-15}
			
			& &
			& Succ.$\uparrow$
			& Succ.$\uparrow$
			& Succ.$\uparrow$ & SErr.$\downarrow$
			& Succ.$\uparrow$ & SErr.$\downarrow$
			& Succ.$\uparrow$ & SErr.$\downarrow$
			& Succ.$\uparrow$ & SErr.$\downarrow$
			& Succ.$\uparrow$ & SErr.$\downarrow$ \\
			
			\midrule
			
			
			\textbf{Our Base Editor}
			& \steerdash & \steerdash
			& 92.80 & 93.20
			& 67.90 & 24.10
			& 72.50 & 22.80
			& 51.20 & 48.40
			& 39.10 & 24.70
			& 70.42 & 28.48 \\
			
			+Emph. only
			& \steertick & \steerdash
			& 94.00 & \textbf{94.60}
			& 78.90 & 14.00
			& 88.10 & \textbf{7.90}
			& 87.90 & 11.70
			& 61.60 & 6.80
			& 84.02 & 9.99 \\
			
			\rowcolor{LightPurple}
			\textbf{+Full STEER}
			& \steertick & \steertick
			& \textbf{94.80} & 93.50
			& \textbf{79.10} & \textbf{12.30}
			& \textbf{88.20} & \textbf{7.90}
			& \textbf{92.80} & \textbf{6.80}
			& \textbf{68.00} & \textbf{6.40}
			& \textbf{85.78} & \textbf{8.45} \\
			
			\midrule
			
			
			\textbf{UNISON}
			& \steerdash & \steerdash
			& \textbf{96.40} & \textbf{90.10}
			& 33.40 & 50.90
			& 25.30 & 26.90
			& \steerdash & \steerdash
			& \steerdash & \steerdash
			& 61.30 & 38.90 \\
			
			+Emph. only
			& \steertick & \steerdash
			& 94.80 & 85.60
			& 78.40 & 13.60
			& 38.30 & 8.50
			& \steerdash & \steerdash
			& \steerdash & \steerdash
			& 74.28 & 11.05 \\
			
			\rowcolor{LightPurple}
			\textbf{+Full STEER}
			& \steertick & \steertick
			& 94.90 & 85.40
			& \textbf{79.60} & \textbf{13.10}
			& \textbf{41.10} & \textbf{8.00}
			& \steerdash & \steerdash
			& \steerdash & \steerdash
			& \textbf{75.25} & \textbf{10.55} \\
			
			\midrule
			
			
			\textbf{SAM Audio}
			& \steerdash & \steerdash
			& \steerdash & \steerdash
			& \steerdash & \steerdash
			& 37.60 & 55.90
			& \steerdash & \steerdash
			& \steerdash & \steerdash
			& 37.60 & 55.90 \\
			
			+Emph. only
			& \steertick & \steerdash
			& \steerdash & \steerdash
			& \steerdash & \steerdash
			& 80.60 & 19.10
			& \steerdash & \steerdash
			& \steerdash & \steerdash
			& 80.60 & 19.10 \\
			
			\rowcolor{LightPurple}
			\textbf{+Full STEER}
			& \steertick & \steertick
			& \steerdash & \steerdash
			& \steerdash & \steerdash
			& \textbf{84.10} & \textbf{15.70}
			& \steerdash & \steerdash
			& \steerdash & \steerdash
			& \textbf{84.10} & \textbf{15.70} \\
			
			\bottomrule
		\end{tabular}%
	}
	
	\endgroup
	
	\vspace{-2mm}
\end{table*}

\begin{figure*}[t]
	\centering
	
	\begin{minipage}[t]{0.46\textwidth}
		\centering
		\vspace{0pt}
		
		\parbox[c][3.0cm][t]{\linewidth}{%
			\centering
			\vspace{7.7pt}
			\includegraphics[
			width=\linewidth,
			height=3.0cm,
			keepaspectratio
			]{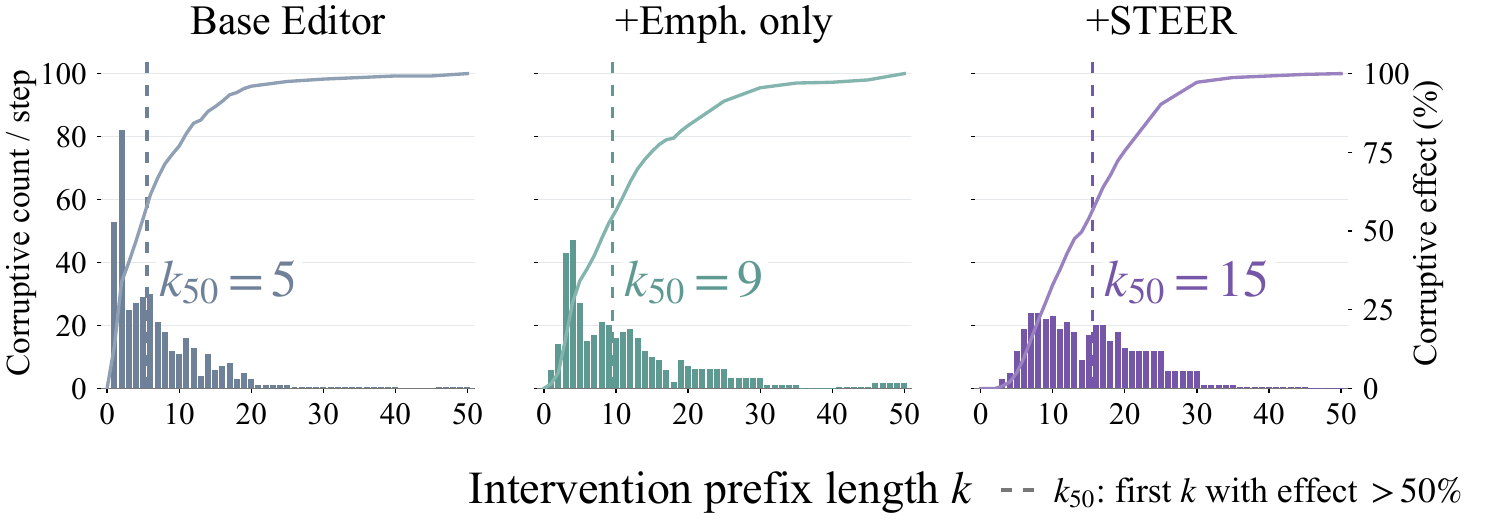}
		}
		\vspace{1mm}
		\caption{Recovery capability under early-trajectory corruption for the base editor, the STEER-Emphasis-only variant, and the full STEER variant.}
		\label{figure8}
	\end{minipage}%
	\hfill
	\begin{minipage}[t]{0.50\textwidth}
		\centering
		\vspace{0pt}
		
		\parbox[c][3.0cm][t]{\linewidth}{%
			\centering
			\vspace{5.1pt}
			\includegraphics[
			width=\linewidth,
			height=3.0cm,
			keepaspectratio
			]{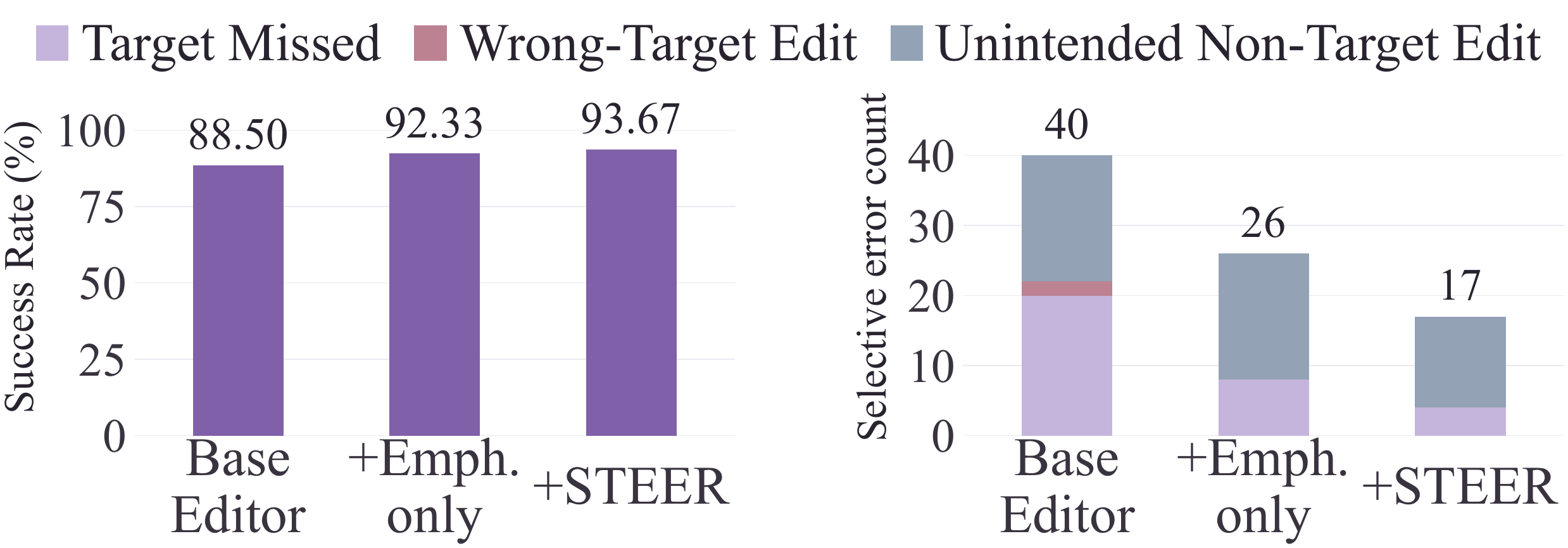}
		}
		\vspace{1mm}
		\caption{Human evaluation of our base editor, its Emphasis-only variant, and full STEER variant on the Complex Audio Editing Test Set, 100 random samples per task.}
		\label{figure9}
	\end{minipage}
\vspace{-6pt}
\end{figure*}

\subsection{Unified Audio Editing Performance}
\label{sec:unified_results}

We compare SteerSound with eight existing audio editors on the Complex Audio Editing Test Set, as shown in Table~\ref{table1}. SteerSound supports all six tasks, whereas existing editors cover only subsets, and achieves the best primary metric on five tasks, including all four selective editing tasks. Notably, it outperforms substantially larger models on removal and replacement while achieving the lowest pitch and volume errors. Compared with our already competitive base editor, STEER further improves removal and replacement CLAP scores from $0.137$ and $0.230$ to $0.144$ and $0.235$, while reducing pitch and volume MAEs from $0.055$ and $0.702$ to $0.047$ and $0.634$. These results demonstrate SteerSound's comprehensive task coverage and strong selective editing performance, further enhanced by STEER.

\subsection{Effectiveness of STEER across Models and Datasets}
\label{5.3}
We evaluate STEER on our base editor, UNISON, and SAM Audio using the Complex Audio Editing Test Set and the Selective Audio Editing Benchmark. As shown in Tables~\ref{table2} and~\ref{table3}, STEER improves aggregate editing performance across all three models in both settings. On the selective benchmark, it increases overall success rates while reducing selective error rates by over 70\% relative to each corresponding base model. STEER-Emphasis consistently improves performance, and adding STEER-Recovery yields further gains. These results demonstrate the effectiveness of STEER across different editing backbones under both controlled source compositions and complex audio scenes. Further experimental details and results are provided in Appendix~\ref{AE}.

\begin{table*}[t]
	\centering
	\begingroup
	\setlength{\tabcolsep}{3pt}
	\renewcommand{\arraystretch}{1.2}
	
	\fontsize{8pt}{9.5pt}\selectfont
	\setlength{\SteerBackboneWidth}{0pt}
	\setlength{\SteerStrategyWidth}{0pt}
	\setlength{\SteerComplexWidth}{0pt}
	\setlength{\SteerSelectiveWidth}{0pt}
	
	\SteerMaxWidth{\SteerBackboneWidth}{\textbf{Backbone}}
	\SteerMaxWidth{\SteerBackboneWidth}{\shortstack[c]{\textbf{Our Base}\\\textbf{Editor}}}
	\SteerMaxWidth{\SteerBackboneWidth}{\textbf{UNISON}}
	\SteerMaxWidth{\SteerBackboneWidth}{\textbf{SAM Audio}}
	
	\SteerMaxWidth{\SteerStrategyWidth}{\textbf{Strategy}}
	\SteerMaxWidth{\SteerStrategyWidth}{\textbf{STEER-Emphasis}}
	\SteerMaxWidth{\SteerStrategyWidth}{\textbf{STEER-Recovery}}
	\SteerMaxWidth{\SteerStrategyWidth}{Standard FM}
	\SteerMaxWidth{\SteerStrategyWidth}{Emph. (Cont.)}
	
	\SteerMaxWidth{\SteerComplexWidth}{$\mathrm{TBG}_{\mathrm{all}}\uparrow$}
	\SteerMaxWidth{\SteerComplexWidth}{$\mathrm{TBG}_{\mathrm{sel}}\uparrow$}
	\SteerMaxWidth{\SteerComplexWidth}{\textbf{124.53}}
	\SteerMaxWidth{\SteerComplexWidth}{\textbf{108.62}}
	
	\SteerMaxWidth{\SteerSelectiveWidth}{Succ.$\uparrow$}
	\SteerMaxWidth{\SteerSelectiveWidth}{SErr.$\downarrow$}
	\SteerMaxWidth{\SteerSelectiveWidth}{\textbf{84.10}}
	\normalsize
	
	\sbox{\SteerLeftCaption}{%
		\begin{minipage}[t]{0.49\textwidth}
			\caption{STEER-Emphasis versus standard Flow Matching post-training under matched updates on both sets.}
			\label{table4}
		\end{minipage}%
	}
	\sbox{\SteerRightCaption}{%
		\begin{minipage}[t]{0.49\textwidth}
			\caption{STEER-Recovery versus continued STEER-Emphasis under matched updates on both sets.}
			\label{table5}
		\end{minipage}%
	}
	
	\setlength{\SteerCaptionHeight}{%
		\dimexpr\ht\SteerLeftCaption+\dp\SteerLeftCaption\relax
	}
	\setlength{\SteerMeasureWidth}{%
		\dimexpr\ht\SteerRightCaption+\dp\SteerRightCaption\relax
	}
	\ifdim\SteerMeasureWidth>\SteerCaptionHeight
	\setlength{\SteerCaptionHeight}{\SteerMeasureWidth}
	\fi
	
	\begin{minipage}[t]{0.49\textwidth}
		\centering
		\vspace{0pt}
		\begin{minipage}[t][\SteerCaptionHeight][t]{\linewidth}
			\usebox{\SteerLeftCaption}
		\end{minipage}
		\vspace{-3mm}
		
		\fontsize{8pt}{9.5pt}\selectfont
		\resizebox{\linewidth}{!}{%
			\begin{tabular}{
					w{l}{\SteerBackboneWidth}
					w{l}{\SteerStrategyWidth}
					*{2}{w{c}{\SteerComplexWidth}}
					*{2}{w{c}{\SteerSelectiveWidth}}
				}
				\toprule[0.8pt]
				\multicolumn{1}{c}{\multirow{2}{*}{\textbf{Backbone}}}
				& \multicolumn{1}{c}{\multirow{2}{*}{\textbf{Strategy}}}
				& \multicolumn{2}{c}{\textbf{Complex}}
				& \multicolumn{2}{c}{\textbf{Selective}}\\
				\cmidrule(lr){3-4}\cmidrule(lr){5-6}
				& & $\mathrm{TBG}_{\mathrm{all}}\uparrow$
				& $\mathrm{TBG}_{\mathrm{sel}}\uparrow$
				& Succ.$\uparrow$ & SErr.$\downarrow$\\
				\midrule[0.4pt]
				
				& Standard FM & 0.92 & 0.66 & 81.05 & 13.58\\
				\rowcolor{LightPurple}
				\cellcolor{white}\multirow{-2}{*}{%
					\shortstack[c]{\textbf{Our Base}\\\textbf{Editor}}%
				}
				& \textbf{STEER-Emphasis}
				& \textbf{1.67} & \textbf{2.97}
				& \textbf{84.02} & \textbf{9.99}\\
				\midrule[0.4pt]
				
				& Standard FM & 6.97 & 12.89 & 68.18 & 18.95\\
				\rowcolor{LightPurple}
				\cellcolor{white}\multirow{-2}{*}{\textbf{UNISON}}
				& \textbf{STEER-Emphasis}
				& \textbf{10.27} & \textbf{17.85}
				& \textbf{74.28} & \textbf{11.05}\\
				\midrule[0.4pt]
				
				& Standard FM & 65.69 & 65.69 & 50.50 & 49.40\\
				\rowcolor{LightPurple}
				\cellcolor{white}\multirow{-2}{*}{\textbf{SAM Audio}}
				& \textbf{STEER-Emphasis}
				& \textbf{108.62} & \textbf{108.62}
				& \textbf{80.60} & \textbf{19.10}\\
				\bottomrule[0.8pt]
			\end{tabular}%
		}
	\end{minipage}%
	\hfill%
	\begin{minipage}[t]{0.49\textwidth}
		\centering
		\vspace{0pt}
		\begin{minipage}[t][\SteerCaptionHeight][t]{\linewidth}
			\usebox{\SteerRightCaption}
		\end{minipage}
		\vspace{-3mm}
		
		\fontsize{8pt}{9.5pt}\selectfont
		\resizebox{\linewidth}{!}{%
			\begin{tabular}{
					w{l}{\SteerBackboneWidth}
					w{l}{\SteerStrategyWidth}
					*{2}{w{c}{\SteerComplexWidth}}
					*{2}{w{c}{\SteerSelectiveWidth}}
				}
				\toprule[0.8pt]
				\multicolumn{1}{c}{\multirow{2}{*}{\textbf{Backbone}}}
				& \multicolumn{1}{c}{\multirow{2}{*}{\textbf{Strategy}}}
				& \multicolumn{2}{c}{\textbf{Complex}}
				& \multicolumn{2}{c}{\textbf{Selective}}\\
				\cmidrule(lr){3-4}\cmidrule(lr){5-6}
				& & $\mathrm{TBG}_{\mathrm{all}}\uparrow$
				& $\mathrm{TBG}_{\mathrm{sel}}\uparrow$
				& Succ.$\uparrow$ & SErr.$\downarrow$\\
				\midrule[0.4pt]
				
				& Emphasis (continue) & 3.61 & 5.50 & 84.51 & 8.67\\
				\rowcolor{LightPurple}
				\cellcolor{white}\multirow{-2}{*}{%
					\shortstack[c]{\textbf{Our Base}\\\textbf{Editor}}%
				}
				& \textbf{STEER-Recovery}
				& \textbf{5.36} & \textbf{8.25}
				& \textbf{85.78} & \textbf{8.45}\\
				\midrule[0.4pt]
				
				& Emphasis (continue) & 10.27 & 17.85 & 74.28 & 11.05\\
				\rowcolor{LightPurple}
				\cellcolor{white}\multirow{-2}{*}{\textbf{UNISON}}
				& \textbf{STEER-Recovery}
				& \textbf{10.81} & \textbf{18.42}
				& \textbf{75.25} & \textbf{10.55}\\
				\midrule[0.4pt]
				
				& Emphasis (continue) & 105.08 & 105.08 & 82.10 & 17.50\\
				\rowcolor{LightPurple}
				\cellcolor{white}\multirow{-2}{*}{\textbf{SAM Audio}}
				& \textbf{STEER-Recovery}
				& \textbf{124.53} & \textbf{124.53}
				& \textbf{84.10} & \textbf{15.70}\\
				\bottomrule[0.8pt]
			\end{tabular}%
		}
	\end{minipage}
	
	\endgroup
	\vspace{-2mm}
\end{table*}

\begin{figure*}[t]
	\centering
	
	\begin{minipage}[t]{0.45\textwidth}
		\centering
		\vspace{0pt}
		
		\parbox[c][3.2cm][t]{\linewidth}{%
			\centering
			\vspace{6.7pt}
			\includegraphics[
			width=\linewidth,
			height=3.2cm,
			keepaspectratio
			]{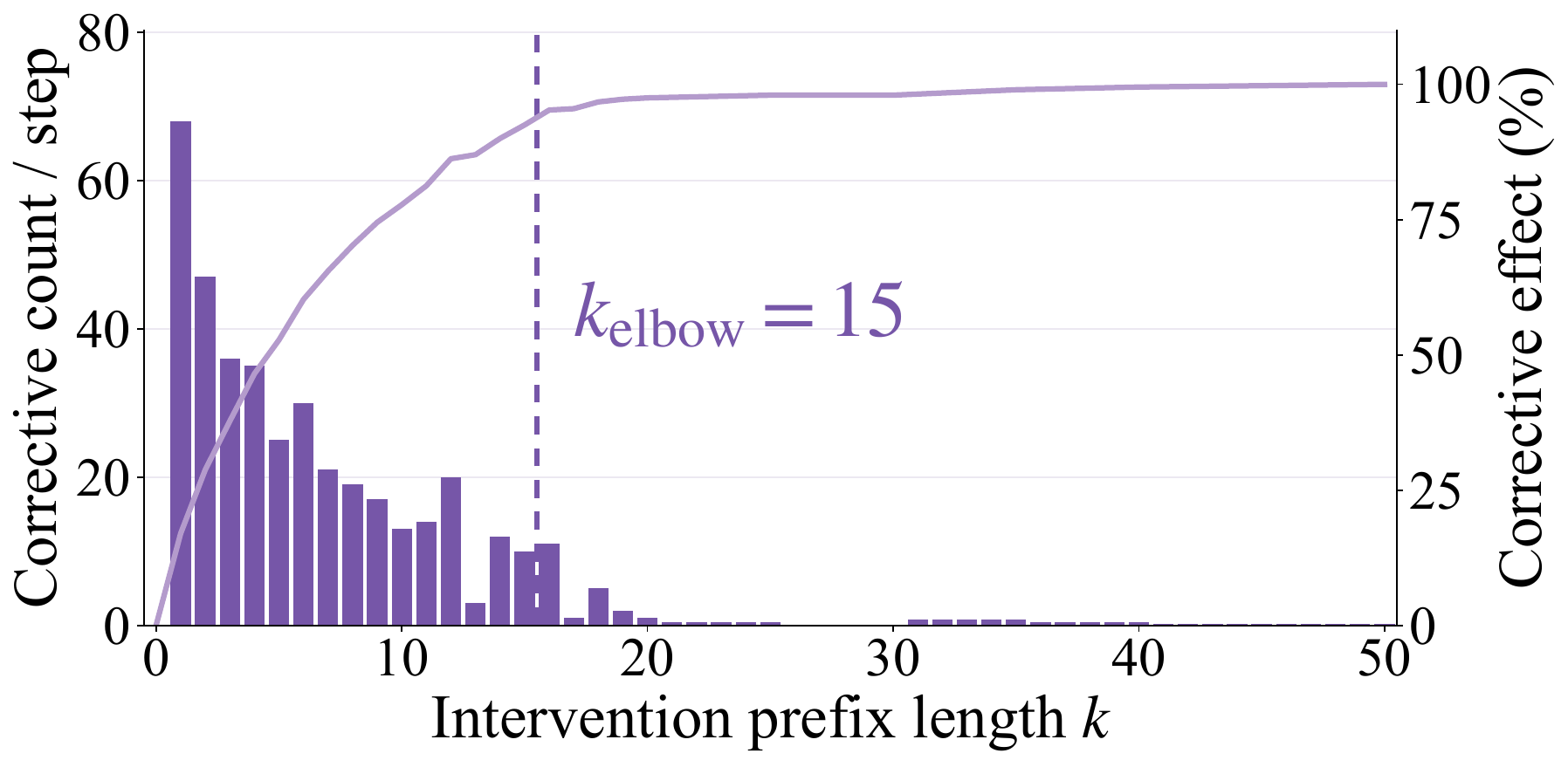}
		}
		\vspace{1mm}
		\caption{Corrective effect and per-step counts, with the elbow at $k=15$ defining the shared early range.}
		\label{figure10}
	\end{minipage}%
	\hfill
	\begin{minipage}[t]{0.50\textwidth}
		\centering
		\vspace{0pt}
		
		\parbox[c][3.2cm][t]{\linewidth}{%
			\centering
			\vspace{4.1pt}
			\includegraphics[
			width=\linewidth,
			height=3.2cm,
			keepaspectratio
			]{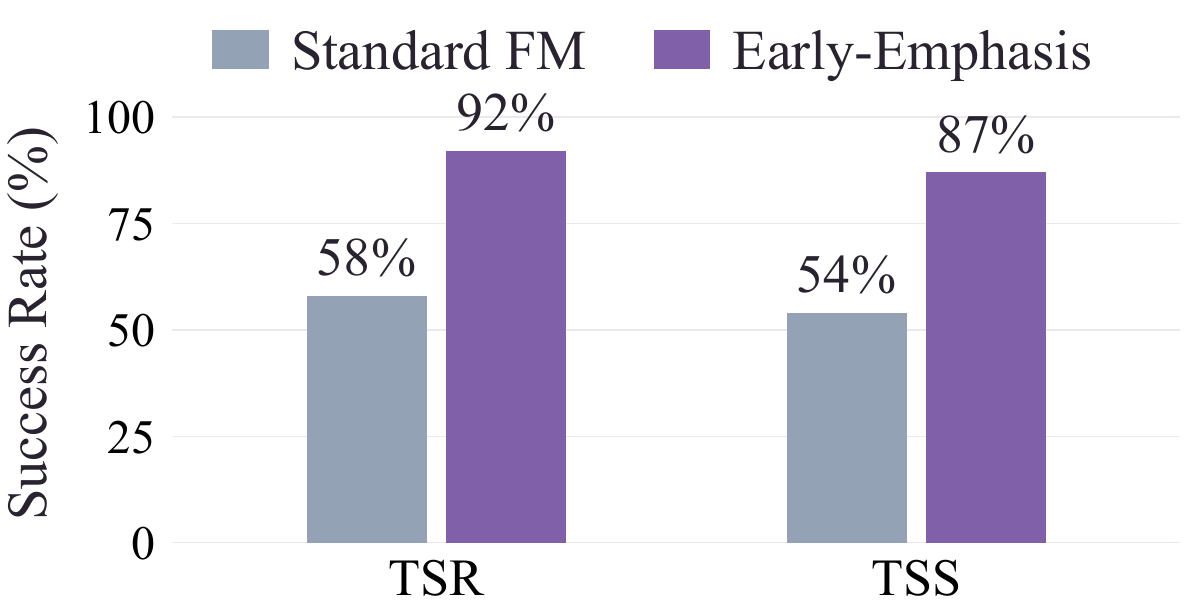}
		}
		\vspace{1mm}
		\caption{Human evaluation of standard Flow Matching and early-trajectory emphasis on TSR and TSS.}
		\label{figure11}
	\end{minipage}
\end{figure*}

\subsection{Analysis of Source-Selection Reliability}
\label{5.4}
\paragraph{Recovery from Early-Trajectory Errors.} Following the corruptive prefix intervention protocol in Section~\ref{sec:intervention}, we evaluate recovery capability across three model variants on the Selective Audio Editing Benchmark. As shown in Figure~\ref{figure8}, the prefix length required to induce a 50\% failure rate increases from 5 steps for our base editor to 9 after STEER-Emphasis and 15 after full STEER. These results demonstrate that STEER-Emphasis improves the model's ability to recover from early source-selection errors, while STEER-Recovery further strengthens this capability. Further experimental results are provided in Appendix~\ref{AF1}.

\paragraph{Human Evaluation.}
To assess source-selection reliability in complex audio scenes, we conduct human evaluation on 100 randomly sampled examples per task from the Complex Audio Editing Test Set. Annotators assess whether the intended source is correctly edited while non-target content is preserved. As shown in Figure~\ref{figure9}, overall success increases from 88.50\% to 92.33\% after STEER-Emphasis and 93.67\% after full STEER, while selective error counts decrease from 40 to 26 and 17, respectively. These results further confirm improved source-selection reliability in complex audio scenes. Detailed results are provided in Appendix~\ref{AF2}.

\subsection{Ablation Studies}
\label{5.5}
\paragraph{Effectiveness of the Two STEER Stages.} We compare STEER-Emphasis with standard Flow Matching post-training and STEER-Recovery with continued STEER-Emphasis, using matched numbers of optimizer updates within each comparison. Within each comparison, both runs start from the same checkpoint, with comparable training settings. Tables~\ref{table4} and~\ref{table5} show that both stages outperform their respective controls across all three models and both evaluation settings. These results demonstrate the effectiveness of both STEER stages beyond simply increasing the number of training updates. Further experimental details and results are provided in Appendix~\ref{AG1} and \ref{AG2}.

\paragraph{Alternative Early-Trajectory Range Selection.}
To examine whether the benefits of STEER-Emphasis stem from emphasizing critical early states rather than task-specific range selection, we evaluate an alternative that uses a single shared range across all selective tasks. Specifically, instead of using task-group-specific ranges determined by the 80\% cumulative-effect threshold, we average the cumulative corrective effects across selective tasks and identify the point beyond which longer corrective prefixes provide diminishing gains (the elbow). As shown in Figure~\ref{figure10}, this gives a shared range covering the first 15 denoising steps, which we use for uniform noise-level sampling across all selective tasks. Table~\ref{table6} shows that this alternative also improves editing performance on both datasets, increasing the overall success rate from $70.42\%$ to $83.98\%$ and reducing selective errors from $28.48\%$ to $9.68\%$ on the selective benchmark. Although our task-group-specific design achieves slightly higher task-balanced gains on the complex test set, both strategies provide substantial improvements. These results suggest that the benefits of STEER-Emphasis stem from emphasizing critical early states rather than task-specific range selection. Further details are provided in Appendix~\ref{AG3}.

\begin{table*}[t]
	\centering
	\begingroup
	\setlength{\tabcolsep}{1.5pt}
	\renewcommand{\arraystretch}{1.10}
	
	\def\SteerNum#1{{\fontsize{7.5pt}{8.6pt}\selectfont #1}}
	
	\sbox{\SteerLeftCaption}{%
		\begin{minipage}[t]{0.5\textwidth}
			\caption{Comparison of task-group-specific and elbow-based shared early trajectory ranges for STEER-Emphasis on both evaluation sets.}
			\label{table6}
		\end{minipage}%
	}
	\sbox{\SteerRightCaption}{%
		\begin{minipage}[t]{0.47\textwidth}
			\caption{Evaluation of standard Flow Matching and early-trajectory emphasis on TSR and TSS using our base editor and SAM Audio.}
			\label{table7}
		\end{minipage}%
	}
	
	\setlength{\SteerCaptionHeight}{%
		\dimexpr\ht\SteerLeftCaption+\dp\SteerLeftCaption\relax
	}
	\ifdim\dimexpr\ht\SteerRightCaption+\dp\SteerRightCaption
	\relax>\SteerCaptionHeight
	\setlength{\SteerCaptionHeight}{%
		\dimexpr\ht\SteerRightCaption+\dp\SteerRightCaption\relax
	}
	\fi
	
	\begin{minipage}[t]{0.50\textwidth}
		\centering
		\vspace{0pt}
		\begin{minipage}[t][\SteerCaptionHeight][t]{\linewidth}
			\usebox{\SteerLeftCaption}
		\end{minipage}
		\vspace{-3mm}
		
		\fontsize{6.5pt}{7.8pt}\selectfont
		\begin{tabularx}{\linewidth}{
				@{}ll*{4}{>{\centering\arraybackslash}X}@{}
			}
			\toprule[0.8pt]
			\multicolumn{1}{c}{\multirow{2}{*}{\textbf{Backbone}}}
			& \multicolumn{1}{c}{\multirow{2}{*}{\textbf{Strategy}}}
			& \multicolumn{2}{c}{\textbf{Complex}}
			& \multicolumn{2}{c}{\textbf{Selective}}\\
			\cmidrule(lr){3-4}\cmidrule(lr){5-6}
			& & $\mathrm{TBG}_{\mathrm{all}}\uparrow$
			& $\mathrm{TBG}_{\mathrm{sel}}\uparrow$
			& Succ.$\uparrow$
			& SErr.$\downarrow$\\
			\midrule[0.4pt]
			
			& Emph.$_{\mathrm{elb.}}$
			& \SteerNum{1.36}
			& \SteerNum{2.82}
			& \SteerNum{83.98}
			& \textbf{\SteerNum{9.68}}\\
			
			\rowcolor{LightPurple}
			\cellcolor{white}\multirow{-2}{*}{\shortstack[c]{\textbf{Our Base}\\\textbf{Editor}}}
			& \textbf{Emphasis}
			& \textbf{\SteerNum{1.67}}
			& \textbf{\SteerNum{2.97}}
			& \textbf{\SteerNum{84.02}}
			& \SteerNum{9.99}\\
			\bottomrule[0.8pt]
		\end{tabularx}
	\end{minipage}%
	\hfill%
	\begin{minipage}[t]{0.47\textwidth}
		\centering
		\vspace{0pt}
		\begin{minipage}[t][\SteerCaptionHeight][t]{\linewidth}
			\usebox{\SteerRightCaption}
		\end{minipage}
		\vspace{-3mm}
		
		\fontsize{6.5pt}{7.8pt}\selectfont
		\begin{tabularx}{\linewidth}{
				@{}l*{2}{>{\centering\arraybackslash}X}@{}
			}
			\toprule[0.8pt]
			\multicolumn{1}{c}{\multirow{2}{*}{\textbf{Method}}}
			& \shortstack[c]{\textbf{Our Base Editor}}
			& \textbf{SAM Audio}\\
			\cmidrule(lr){2-2}\cmidrule(lr){3-3}
			& $\mathrm{CLAP\text{-}AA}_{\mathrm{TSR}}\uparrow$
			& $\mathrm{CLAP\text{-}AA}_{\mathrm{TSS}}\uparrow$\\
			\midrule[0.4pt]
			
			Standard FM
			& \underline{\SteerNum{0.74}}
			& \underline{\SteerNum{0.00}}\\
			
			\rowcolor{LightPurple}
			\textbf{Early-Emphasis}
			& \textbf{\SteerNum{0.86}}
			& \textbf{\SteerNum{0.21}}\\
			\bottomrule[0.8pt]
		\end{tabularx}
	\end{minipage}
	
	\endgroup
\end{table*}

\subsection{Extension to Fine-Grained Selective Control}
\label{5.6}
We further examine whether early-trajectory emphasis extends beyond category-level audio editing to related selective-control tasks. To this end, we study transcript-conditioned target speaker removal (TSR) and target speaker separation (TSS), adapting our base editor for TSR and SAM Audio for TSS. These tasks require distinguishing competing speakers by speech content rather than sound category. We compare standard Flow Matching training with early-trajectory emphasis from the start of task adaptation. The quantitative results in Table~\ref{table7} and human evaluation on 100 randomly sampled test examples per task in Figure~\ref{figure11} show substantial gains on both tasks, with early-trajectory emphasis enabling effective target selection where standard training struggles. These findings support the effectiveness of early-trajectory emphasis for fine-grained source selection and its broader applicability to related selective-control tasks. Further details are provided in Appendix~\ref{AH}.
\section{Discussion}
Our analysis shows that source selection in audio editing is largely shaped during early denoising, and early selection errors often persist through subsequent steps. STEER addresses this behavior through early-trajectory emphasis and recovery training on self-generated rollout states. Extensive experiments across different model architectures demonstrate consistent improvements in selective editing reliability and overall editing performance. Furthermore, the effectiveness of early-trajectory emphasis on transcript-conditioned speaker removal and separation suggests its applicability to broader fine-grained selective-control tasks. 

These findings suggest that training for selective control should account for both when consequential decisions are made and how models respond to trajectory deviations. Preventing early errors and recovering from errors that remain are complementary needs. Beyond audio, our findings suggest a general training principle for iterative generative models: improving early decisions and learning to recover from trajectory deviations may enhance the reliability of final outputs. This perspective offers a promising direction for image and video generation and editing, where early decisions about scene structure and target selection may influence results.

\bibliographystyle{plainnat}
\bibliography{references}

@article{wang2023audit,
	title={Audit: Audio editing by following instructions with latent diffusion models},
	author={Wang, Yuancheng and Ju, Zeqian and Tan, Xu and He, Lei and Wu, Zhizheng and Bian, Jiang and others},
	journal={Advances in Neural Information Processing Systems},
	volume={36},
	pages={71340--71357},
	year={2023}
}

@article{pan2025transfer,
	title={Transfer between modalities with metaqueries},
	author={Pan, Xichen and Shukla, Satya Narayan and Singh, Aashu and Zhao, Zhuokai and Mishra, Shlok Kumar and Wang, Jialiang and Xu, Zhiyang and Chen, Jiuhai and Li, Kunpeng and Juefei-Xu, Felix and others},
	journal={arXiv preprint arXiv:2504.06256},
	year={2025}
}

@article{tian2026audio,
	title={Audio-Omni: Extending Multi-modal Understanding to Versatile Audio Generation and Editing},
	author={Tian, Zeyue and Yang, Binxin and Liu, Zhaoyang and Zhang, Jiexuan and Yuan, Ruibin and Yin, Hubery and Chen, Qifeng and Li, Chen and Lyu, Jing and Xue, Wei and others},
	journal={arXiv preprint arXiv:2604.10708},
	year={2026}
}

@article{yang2026moss,
	title={Moss-audio technical report},
	author={Yang, Chen and Yu, Chufan and Chen, Hanfu and Zhu, Jie and Chen, Jingqi and Chen, Ke and Wang, Wenxuan and Wang, Yang and Jiang, Yaozhou and Jiang, Yi and others},
	journal={arXiv preprint arXiv:2606.01802},
	year={2026}
}

@inproceedings{elizalde2023clap,
	title={Clap learning audio concepts from natural language supervision},
	author={Elizalde, Benjamin and Deshmukh, Soham and Al Ismail, Mahmoud and Wang, Huaming},
	booktitle={ICASSP 2023-2023 IEEE International Conference on Acoustics, Speech and Signal Processing (ICASSP)},
	pages={1--5},
	year={2023},
	organization={IEEE}
}

@inproceedings{pang2025aligning,
	title={Aligning generative denoising with discriminative objectives unleashes diffusion for visual perception},
	author={Pang, Ziqi and Xu, Xin and Wang, Yu-Xiong},
	booktitle={International Conference on Learning Representations},
	volume={2025},
	pages={101991--102013},
	year={2025}
}

@article{seki2025language,
	title={Language-queried target speech extraction using para-linguistic and non-linguistic prompts},
	author={Seki, Kentaro and Ito, Nobutaka and Yamauchi, Kazuki and Okamoto, Yuki and Yamaoka, Kouei and Saito, Yuki and Takamichi, Shinnosuke and Saruwatari, Hiroshi},
	journal={Acoustical Science and Technology},
	pages={e25--27},
	year={2025},
	publisher={ACOUSTICAL SOCIETY OF JAPAN}
}

@article{li2026unison,
	title={UNISON: A Unified Sound Generation and Editing Framework via Deep LLM Fusion},
	author={Li, Zhaoqing and Xu, Haoning and Su, Jingran and Liu, Yaofang and Rao, Zhefan and Wang, Huimeng and Deng, Jiajun and Wang, Tianzi and Jin, Zengrui and Liu, Rui and others},
	journal={arXiv preprint arXiv:2605.31530},
	year={2026}
}

@article{tao2025mmedit,
	title={Mmedit: A unified framework for multi-type audio editing via audio language model},
	author={Tao, Ye and Wu, Wen and Zhang, Chao and Wu, Mengyue and Wang, Shuai and Xu, Xuenan},
	journal={arXiv preprint arXiv:2512.20339},
	year={2025}
}

@article{xu2024prompt,
	title={Prompt-guided precise audio editing with diffusion models},
	author={Xu, Manjie and Li, Chenxing and Su, Dan and Liang, Wei and Yu, Dong and others},
	journal={arXiv preprint arXiv:2406.04350},
	year={2024}
}

@inproceedings{jia2025audioeditor,
	title={Audioeditor: A training-free diffusion-based audio editing framework},
	author={Jia, Yuhang and Chen, Yang and Zhao, Jinghua and Zhao, Shiwan and Zeng, Wenjia and Chen, Yong and Qin, Yong},
	booktitle={ICASSP 2025-2025 IEEE International Conference on Acoustics, Speech and Signal Processing (ICASSP)},
	pages={1--5},
	year={2025},
	organization={IEEE}
}

@article{balaji2022ediff,
	title={ediff-i: Text-to-image diffusion models with an ensemble of expert denoisers},
	author={Balaji, Yogesh and Nah, Seungjun and Huang, Xun and Vahdat, Arash and Song, Jiaming and Zhang, Qinsheng and Kreis, Karsten and Aittala, Miika and Aila, Timo and Laine, Samuli and others},
	journal={arXiv preprint arXiv:2211.01324},
	year={2022}
}

@inproceedings{hang2023efficient,
	title={Efficient diffusion training via min-snr weighting strategy},
	author={Hang, Tiankai and Gu, Shuyang and Li, Chen and Bao, Jianmin and Chen, Dong and Hu, Han and Geng, Xin and Guo, Baining},
	booktitle={2023 IEEE/CVF International Conference on Computer Vision (ICCV)},
	pages={7407--7417},
	year={2023},
	organization={IEEE}
}

@article{liu2023audioldm,
	title={Audioldm: Text-to-audio generation with latent diffusion models},
	author={Liu, Haohe and Chen, Zehua and Yuan, Yi and Mei, Xinhao and Liu, Xubo and Mandic, Danilo and Wang, Wenwu and Plumbley, Mark D},
	journal={arXiv preprint arXiv:2301.12503},
	year={2023}
}

@inproceedings{esser2024scaling,
	title={Scaling rectified flow transformers for high-resolution image synthesis},
	author={Esser, Patrick and Kulal, Sumith and Blattmann, Andreas and Entezari, Rahim and M{\"u}ller, Jonas and Saini, Harry and Levi, Yam and Lorenz, Dominik and Sauer, Axel and Boesel, Frederic and others},
	booktitle={Forty-first international conference on machine learning},
	year={2024}
}

@article{huang2026self,
	title={Self forcing: Bridging the train-test gap in autoregressive video diffusion},
	author={Huang, Xun and Li, Zhengqi and He, Guande and Zhou, Mingyuan and Shechtman, Eli},
	journal={Advances in Neural Information Processing Systems},
	volume={38},
	pages={167283--167308},
	year={2026}
}

@article{qin2026soar,
	title={SOAR: Self-Correction for Optimal Alignment and Refinement in Diffusion Models},
	author={Qin, You and Wang, Linqing and Fei, Hao and Zimmermann, Roger and Bo, Liefeng and Lu, Qinglin and Wang, Chunyu},
	journal={arXiv preprint arXiv:2604.12617},
	year={2026}
}

@article{lipman2022flow,
	title={Flow matching for generative modeling},
	author={Lipman, Yaron and Chen, Ricky TQ and Ben-Hamu, Heli and Nickel, Maximilian and Le, Matt},
	journal={arXiv preprint arXiv:2210.02747},
	year={2022}
}

@article{manor2024zero,
	title={Zero-shot unsupervised and text-based audio editing using ddpm inversion},
	author={Manor, Hila and Michaeli, Tomer},
	journal={arXiv preprint arXiv:2402.10009},
	year={2024}
}

@article{shi2024seededit,
	title={Seededit: Align image re-generation to image editing},
	author={Shi, Yichun and Wang, Peng and Huang, Weilin},
	journal={arXiv preprint arXiv:2411.06686},
	year={2024}
}

@article{raffel2020exploring,
	title={Exploring the limits of transfer learning with a unified text-to-text transformer},
	author={Raffel, Colin and Shazeer, Noam and Roberts, Adam and Lee, Katherine and Narang, Sharan and Matena, Michael and Zhou, Yanqi and Li, Wei and Liu, Peter J},
	journal={Journal of machine learning research},
	volume={21},
	number={140},
	pages={1--67},
	year={2020}
}

@inproceedings{rombach2022high,
	title={High-resolution image synthesis with latent diffusion models},
	author={Rombach, Robin and Blattmann, Andreas and Lorenz, Dominik and Esser, Patrick and Ommer, Bj{\"o}rn},
	booktitle={2022 IEEE/CVF conference on computer vision and pattern recognition (CVPR)},
	pages={10674--10685},
	year={2022},
	organization={ieee}
}

@article{kong2020hifi,
	title={Hifi-gan: Generative adversarial networks for efficient and high fidelity speech synthesis},
	author={Kong, Jungil and Kim, Jaehyeon and Bae, Jaekyoung},
	journal={Advances in neural information processing systems},
	volume={33},
	pages={17022--17033},
	year={2020}
}

@inproceedings{chen2020vggsound,
	title={Vggsound: A large-scale audio-visual dataset},
	author={Chen, Honglie and Xie, Weidi and Vedaldi, Andrea and Zisserman, Andrew},
	booktitle={ICASSP 2020-2020 IEEE International Conference on Acoustics, Speech and Signal Processing (ICASSP)},
	pages={721--725},
	year={2020},
	organization={IEEE}
}

@inproceedings{kim2019audiocaps,
	title={Audiocaps: Generating captions for audios in the wild},
	author={Kim, Chris Dongjoo and Kim, Byeongchang and Lee, Hyunmin and Kim, Gunhee},
	booktitle={Proceedings of the 2019 Conference of the North American Chapter of the Association for Computational Linguistics: Human Language Technologies, Volume 1 (Long and Short Papers)},
	pages={119--132},
	year={2019}
}

@article{fonseca2021fsd50k,
	title={Fsd50k: an open dataset of human-labeled sound events},
	author={Fonseca, Eduardo and Favory, Xavier and Pons, Jordi and Font, Frederic and Serra, Xavier},
	journal={IEEE/ACM Transactions on Audio, Speech, and Language Processing},
	volume={30},
	pages={829--852},
	year={2021},
	publisher={IEEE}
}

@inproceedings{drossos2020clotho,
	title={Clotho: An audio captioning dataset},
	author={Drossos, Konstantinos and Lipping, Samuel and Virtanen, Tuomas},
	booktitle={ICASSP 2020-2020 IEEE International Conference on Acoustics, Speech and Signal Processing (ICASSP)},
	pages={736--740},
	year={2020},
	organization={IEEE}
}

@article{morato2021macs,
	title={Macs-multi-annotator captioned soundscapes},
	author={Morato, Irene Martin and Mesaros, Annamaria},
	journal={MACS-Multi-Annotator Captioned Soundscapes},
	year={2021},
	publisher={Zenodo}
}

@inproceedings{peebles2023scalable,
	title={Scalable diffusion models with transformers},
	author={Peebles, William and Xie, Saining},
	booktitle={2023 IEEE/CVF International Conference on Computer Vision (ICCV)},
	pages={4172--4182},
	year={2023},
	organization={IEEE}
}

@article{ho2022classifier,
	title={Classifier-free diffusion guidance},
	author={Ho, Jonathan and Salimans, Tim},
	journal={arXiv preprint arXiv:2207.12598},
	year={2022}
}

@article{li2026semantically,
	title={A semantically consistent dataset for data-efficient query-based universal sound separation},
	author={Li, Kai and Cheng, Jintao and Zeng, Chang and Yan, Zijun and Wang, Helin and Su, Zixiong and Zheng, Bo and Hu, Xiaolin},
	journal={arXiv preprint arXiv:2601.22599},
	year={2026}
}

@article{koutini2021efficient,
	title={Efficient training of audio transformers with patchout},
	author={Koutini, Khaled and Schl{\"u}ter, Jan and Eghbal-Zadeh, Hamid and Widmer, Gerhard},
	journal={arXiv preprint arXiv:2110.05069},
	year={2021}
}

@article{chen2026lens,
	title={Lens: Rethinking Training Efficiency for Foundational Text-to-Image Models},
	author={Chen, Dong and Wei, Fangyun and Wan, Ziyu and Chen, Dongdong and Zhang, Jiawei and Zhao, Jinjing and Zhang, Sirui and Yue, Yang and Liang, Zhiyang and Guo, Baining and others},
	journal={arXiv preprint arXiv:2605.21573},
	year={2026}
}

@article{shi2025sam,
	title={Sam audio: Segment anything in audio},
	author={Shi, Bowen and Tjandra, Andros and Hoffman, John and Wang, Helin and Wu, Yi-Chiao and Gao, Luya and Richter, Julius and Le, Matt and Vyas, Apoorv and Chen, Sanyuan and others},
	journal={arXiv preprint arXiv:2512.18099},
	year={2025}
}

\clearpage
\beginappendix

\section{Early-trajectory experiments on our base editor}
\label{AA}
This appendix presents the early-trajectory analyses on our base editor. While Section~\ref{sec:analysis} establishes the early formation and causal influence of source selection on UNISON and SAM Audio, here we extend the same analyses to the four selective editing tasks supported by our base editor: removal, replacement, pitch modification, and volume modification. We first examine when erroneous source-selection tendencies become observable, and then use bidirectional prefix interventions to identify the corresponding source-selection-critical early ranges.
\subsection{Early-Trajectory Observation}
\label{AA1}
\begin{figure*}[!ht]
	\centering
	\includegraphics[width=\textwidth]{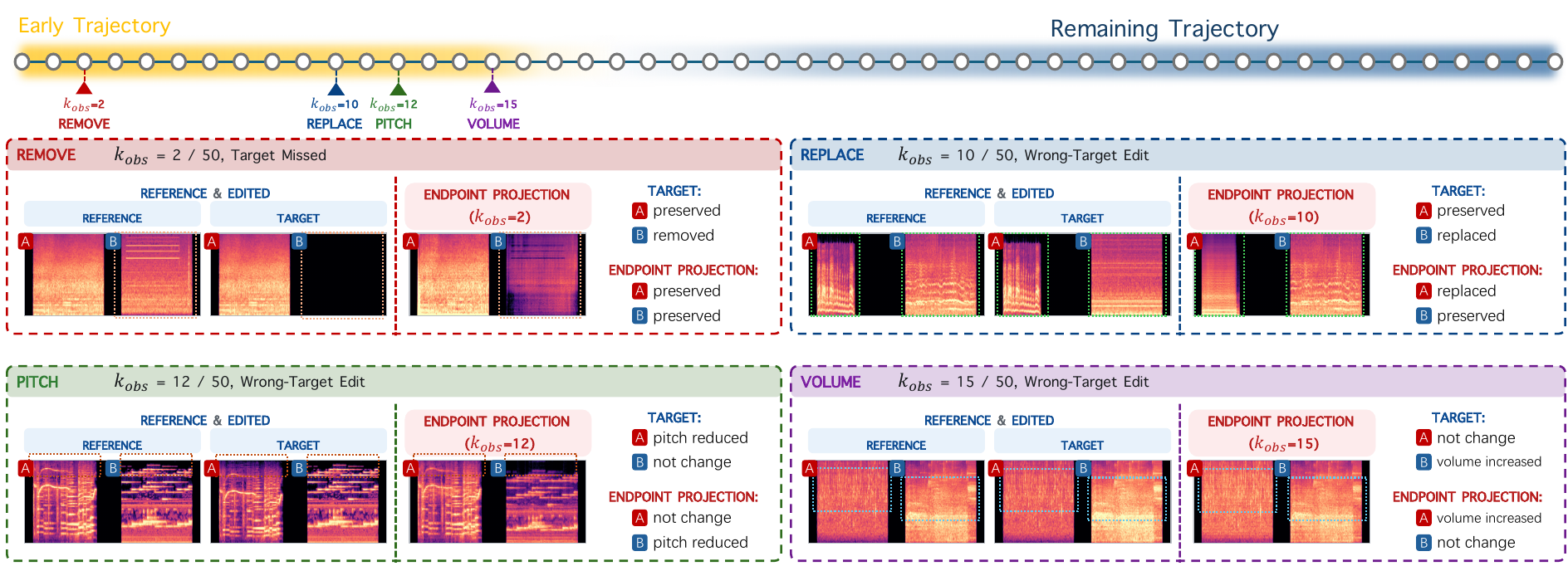}
	\vspace{-3mm}
	\caption{
		Early emergence of source-selection tendencies on our base editor.
		Intermediate denoising states are extrapolated to estimated zero-noise outputs.
		Representative failed edits from the four selective editing tasks are shown, with
		$k_{\mathrm{obs}}$ denoting the earliest step at which the erroneous source-selection tendency becomes perceptually distinguishable.
	}
	\label{figure12}
	\vspace{-3mm}
\end{figure*}

Figure~\ref{figure12} shows representative cases. Across all four tasks, erroneous source-selection tendencies become perceptually distinguishable during the early trajectory. Removal reveals the selected source within only a few steps, whereas replacement and source-attribute modifications require a broader, but still early, portion of the trajectory before their editing tendencies become clearly distinguishable. These observations are consistent with the task-dependent behavior discussed in Section~\ref{sec:observation}.

\subsection{Bidirectional Early-Trajectory Intervention}
\label{AA2}

\begin{figure*}[t]
	\centering
	\includegraphics[width=\textwidth]{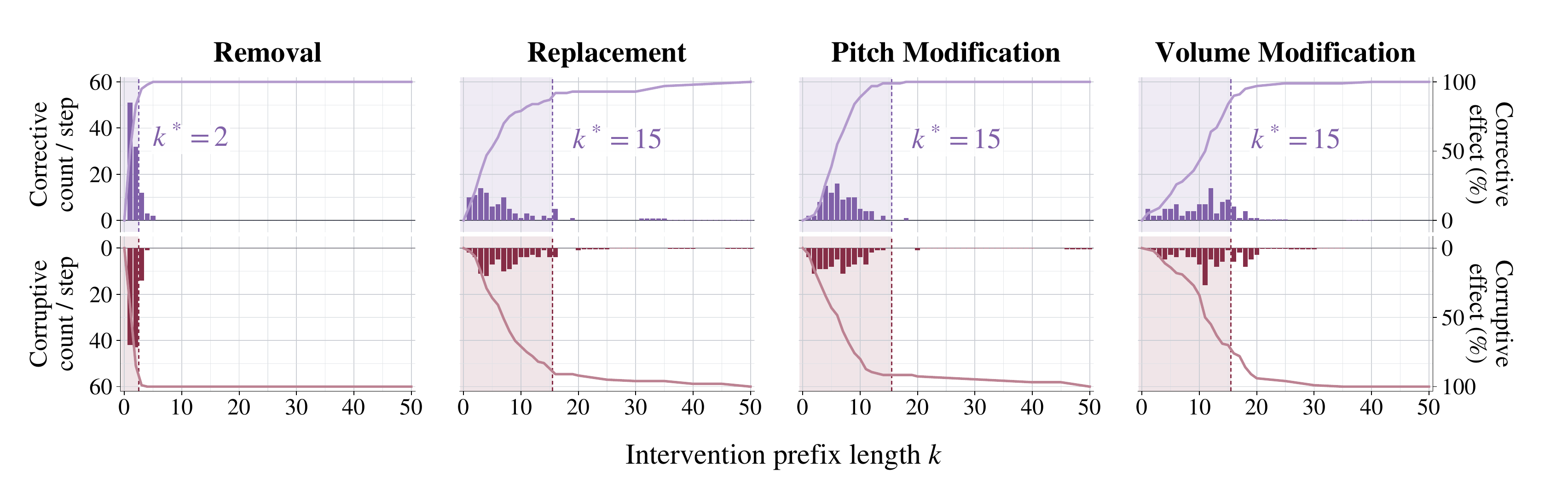}
	\vspace{-3mm}
	\caption{
		Bidirectional early-trajectory interventions on our base editor.
		Corrective (top) and corruptive (bottom) interventions are evaluated on 100 validation examples per selective editing task.
		Bars denote per-step intervention counts and curves denote cumulative effects.
		Dashed lines indicate the resulting critical ranges used by STEER-Emphasis:
		$k^*=2$ for selective suppression and $k^*=15$ for selective transformation.
	}
	\label{figure13}
	\vspace{-3mm}
\end{figure*}

Following the protocol in Section~\ref{sec:intervention}, we perform the same bidirectional prefix interventions on our base editor using 100 validation examples per selective editing task. As shown in Figure~\ref{figure13}, the intervention effects are likewise concentrated in the early trajectory, with removal exhibiting a substantially shorter sensitive range than replacement, pitch modification, and volume modification. Applying the cumulative-corrective-effect criterion defined in Section~\ref{sec:focus} with $\rho=0.8$ yields $k^*=2$ of 50 steps for selective suppression and $k^*=15$ for selective transformation. These ranges are used by STEER-Emphasis.

\section{Base editor architecture and training details}
\label{AB}

\begingroup
\raggedright
\setlength{\intextsep}{4pt}
\setlength{\columnsep}{12pt}

This appendix complements Section~\ref{sec:setup} with additional implementation details of our base editor. As introduced in the main paper, our base editor serves as the common backbone for our trajectory analysis and subsequent STEER post-training. Table~\ref{tab:model_parameters} summarizes the parameter composition of the complete system and its trainable editing backbone. 

\begin{wraptable}{r}{0.31\textwidth}
	\vspace{-13.5pt}
	\centering
	
	\setlength{\abovecaptionskip}{0pt}
	\setlength{\belowcaptionskip}{4pt}
	
	\caption{Parameter counts by module for our base editor.}
	\label{tab:model_parameters}
	
	\small
	\setlength{\tabcolsep}{4pt}
	\renewcommand{\arraystretch}{0.93}
	
	\begin{tabular}{@{}lr@{}}
		\toprule
		\textbf{Module} & \textbf{Params. (M)} \\
		\midrule
		DiT backbone & 673.35 \\
		T5-Base      & 109.63 \\
		Audio VAE    & 509.34 \\
		HiFi-GAN     & 55.26 \\
		\midrule
		Total        & 1,347.58 \\
		Trainable    & 673.35 \\
		\bottomrule
	\end{tabular}
	
	\vspace{-2pt}
\end{wraptable}
Here, we focus on the implementation details omitted from the main text, including the conditioning mechanism, base-model optimization, and task-dependent inference configuration. We first describe how task-specific scalar controls are incorporated into the denoising backbone. Then we provide the optimization and learning-rate configuration used to obtain our base editor, followed by the construction of task-dependent CFG reference prompts at inference time.

\par
\endgroup

\subsection{Task and Timestep Conditioning}
\label{AB1}

In addition to the text instruction, the backbone receives four scalar conditions: the noise level $\sigma$, task ID $\tau$, volume change $\Delta_v$, and pitch change $\Delta_p$. To combine denoising and editing information, each input is normalized to $[0,1]$ and passed through sinusoidal encoding followed by an independent MLP. The four equal-dimensional embeddings are then summed:
\begin{equation}
	\mathbf{c}_{\mathrm{global}}
	= g_{\sigma}\!\left(\phi(\bar\sigma)\right)
	+ g_{\tau}\!\left(\phi(\bar\tau)\right)
	+ g_v\!\left(\phi(\overline{\Delta_v})\right)
	+ g_p\!\left(\phi(\overline{\Delta_p})\right),
	\label{eq:base_global_condition}
\end{equation}
where bars denote normalization, $\phi$ is sinusoidal encoding, and each $g$ is an independent MLP. The resulting vector modulates intermediate features in both audio branches, following DiT and SD3~\citep{peebles2023scalable,esser2024scaling}, while T5 provides text-token conditioning separately. Notably, in our model, we set the pitch-change range $\Delta_p$ to $[-4,+4]$ semitones with a step size of $0.5$ semitones, and the volume-change range $\Delta_v$ to $[-5,+5]$~dB. The signs specify the direction of the requested change.

\Needspace{13\baselineskip}
\subsection{Training Configuration}
\label{AB2}

\begingroup
\setlength{\intextsep}{5pt}
\setlength{\columnsep}{12pt}

\begin{wrapfigure}{r}{0.44\textwidth}
	\vspace{-10pt}
	\centering
	\setlength{\parskip}{0pt}
	
	\includegraphics[
	width=\linewidth
	]{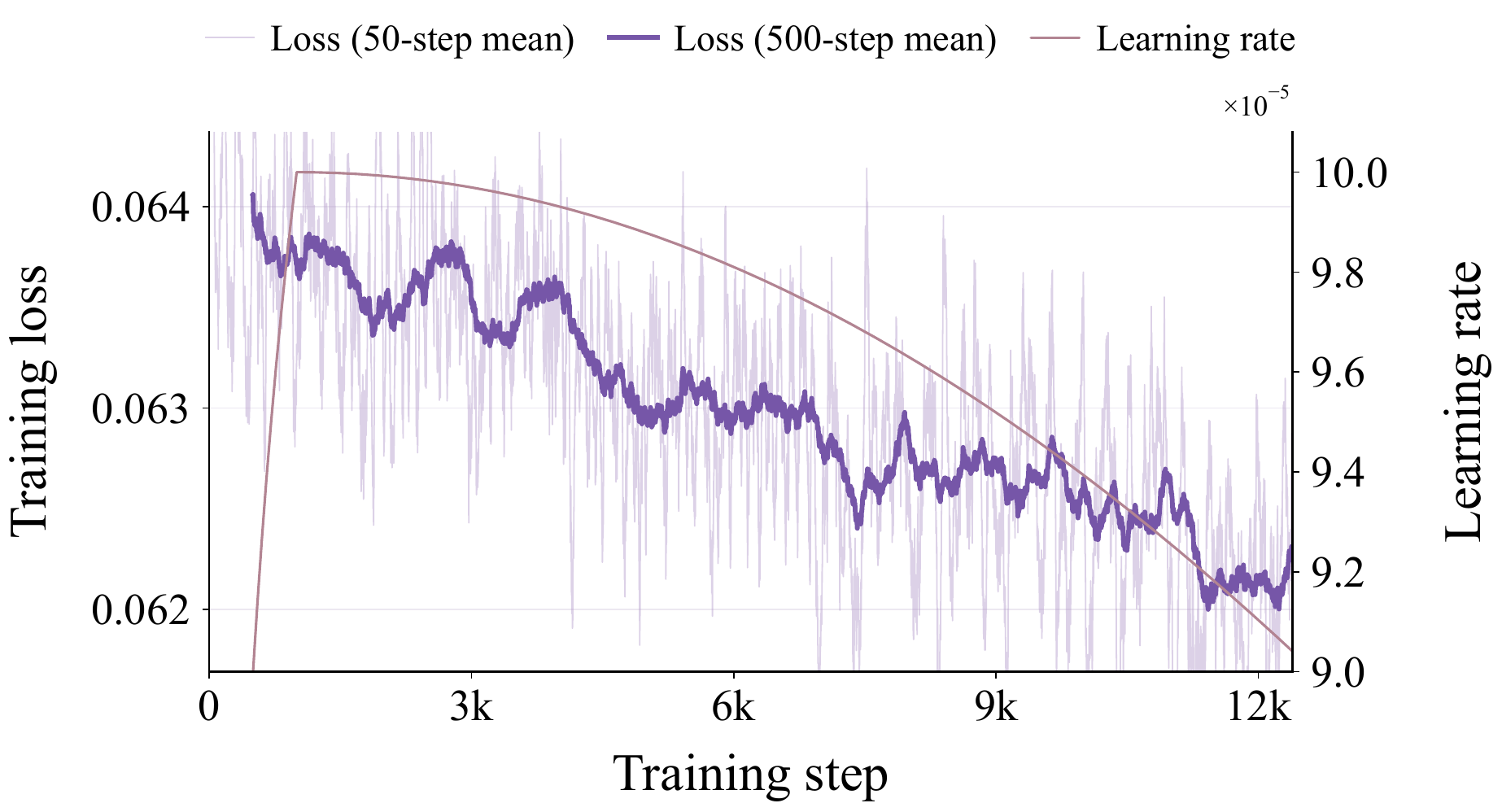}\par
	
	\vspace{-4pt}
	\setlength{\abovecaptionskip}{3pt}
	\setlength{\belowcaptionskip}{0pt}
	
	\caption{Training loss and learning rate.}
	\label{figure14}
	
	\vspace{-2pt}
\end{wrapfigure}

To train this conditioned backbone, we use the Flow Matching objective in Equation~\ref{eq:fm}, starting from random initialization. During pre-training, we adopt the standard logit-normal noise-level sampling strategy following SD3~\cite{esser2024scaling}. We use AdamW with $(\beta_1,\beta_2)=(0.9,0.999)$, $\epsilon=10^{-8}$, and zero weight decay. The learning rate reaches $10^{-4}$ after 1,000 warmup steps and subsequently follows a cosine decay schedule. Training uses BF16 precision and DeepSpeed~\mbox{ZeRO-2} for 15 epochs on our Complex Audio Editing Training Set and takes approximately 16 hours on 16 Ascend 910B NPUs. Figure~\ref{figure14} reports the training loss and learning-rate schedule.
\par
\endgroup

\subsection{Task-Dependent Inference Prompts}
\label{AB3}
For inference, we use the CFG configuration~\cite{ho2022classifier} with task-dependent reference prompts (Table~\ref{tab:base_cfg_prompts}). Generate, Add, Remove, and Replace use an empty string. In contrast, Volume and Pitch retain the direction and magnitude but omit the source description. Attribute magnitudes are absolute values with one decimal place, and Increase or Reduce specifies the direction. Units are implicit: dB for Volume and semitones for Pitch.

\par\smallskip
\noindent\begin{minipage}{\linewidth}
	\centering
	\setlength{\abovecaptionskip}{0pt}
	\setlength{\belowcaptionskip}{4pt}
	
	\captionof{table}{
		Example conditional and CFG reference prompts.
	}
	\label{tab:base_cfg_prompts}
	
	\small
	\setlength{\tabcolsep}{4pt}
	\renewcommand{\arraystretch}{1.03}
	
	\begin{tabularx}{\linewidth}{@{}
			>{\raggedright\arraybackslash}p{0.14\linewidth}
			>{\raggedright\arraybackslash}X
			>{\raggedright\arraybackslash}p{0.28\linewidth}
			@{}}
		\toprule
		\textbf{Task}
		& \textbf{Conditional prompt}
		& \textbf{CFG reference prompt} \\
		\midrule
		
		Generate
		& Generate the sound of rain.
		& \textit{Empty string} \\
		
		Add
		& Add the sound of rain.
		& \textit{Empty string} \\
		
		Remove
		& Remove the sound of rain.
		& \textit{Empty string} \\
		
		Replace
		& Replace the sound of rain by the sound of wind.
		& \textit{Empty string} \\
		
		\midrule
		
		Volume $\uparrow$
		& Increase the volume of the sound of rain by 3.0.
		& Increase the volume by 3.0. \\
		
		Volume $\downarrow$
		& Reduce the volume of the sound of rain by 3.0.
		& Reduce the volume by 3.0. \\
		
		Pitch $\uparrow$
		& Increase the pitch of the sound of rain by 2.0.
		& Increase the pitch by 2.0. \\
		
		Pitch $\downarrow$
		& Reduce the pitch of the sound of rain by 2.0.
		& Reduce the pitch by 2.0. \\
		
		\bottomrule
	\end{tabularx}
\end{minipage}\par

\section{Complex Audio Editing Data: Construction and Evaluation}
\label{AC}

\subsection{Audio Data Sources}
\label{AC1}
\paragraph{Task-Adaptive Caption Granularity.}
As described in Section~\ref{sec:base}, we consider six editing operations: generation, addition, removal, replacement, pitch modification, and volume modification. Through extensive experiments, we observe that different editing operations benefit from different granularities of textual supervision. Specifically, audio generation and source addition perform better with fine-grained audio descriptions, whereas source removal, pitch modification, and volume modification benefit more from class-level descriptions. This difference can be understood from their training and inference settings: fine-grained descriptions provide richer semantic constraints for generation-oriented operations, while concise class labels establish a clearer correspondence between semantic categories and audio sources, which better matches selective editing scenarios where the target source is typically specified by a category term. Similar benefits of information-rich captions have also been observed in image generation~\cite{chen2026lens}.

Based on this observation, we use two separate source pools constructed from different types of annotated audio data: the Fine-Grained Description Pool $\mathcal{P}$ and the Class-Label Pool $\mathcal{L}$. The former consists of audio paired with fine-grained natural-language descriptions, while the latter consists of single-source audio paired with class-level labels, where each clip contains only the labeled sound source. Samples from $\mathcal{P}$ are used for audio generation, source addition, and the generated content in source replacement, whereas samples from $\mathcal{L}$ are used for source removal, pitch modification, volume modification, and the source to be replaced in source replacement.
\par\smallskip
\noindent
\begin{minipage}[t]{0.62\linewidth}
	\vspace{0pt}
	The Fine-Grained Description Pool $\mathcal{P}$ is collected from VGGSound~\cite{chen2020vggsound}, AudioCaps~\cite{kim2019audiocaps}, FSD50K~\cite{fonseca2021fsd50k}, Clotho V2~\cite{drossos2020clotho}, and MACS~\cite{morato2021macs}. It contains 380,165 audio chunks with a total duration of 953.51 hours, with each sample paired with a fine-grained description of its acoustic content. The data composition across source datasets is summarized in Table~\ref{tab:source_pool}. The Class-Label Pool $\mathcal{L}$ is constructed from HIVE~\cite{li2026semantically} by retaining audio-label pairs with CLAP similarity above $0.1$. It contains 635,019 audio chunks with a total duration of 1,688.44 hours and covers 279 sound categories. 
	
\end{minipage}
\hfill
\begin{minipage}[t]{0.34\linewidth}
	\vspace{-8pt}
	\centering
	\small
	\setlength{\tabcolsep}{10pt}
	\renewcommand{\arraystretch}{1.10}
	
	\captionof{table}{Data sources for $\mathcal{P}$.}
	\label{tab:source_pool}
	
	\vspace{-1pt}
	\begin{tabular}{lr}
		\hline
		\textbf{Dataset} & \textbf{\# Chunks} \\
		\hline
		VGGSound  & 192,794 \\
		AudioCaps & 101,369 \\
		FSD50K    & 51,182 \\
		Clotho V2 & 17,545 \\
		MACS      & 17,275 \\
		\hline
		\textbf{Total} & \textbf{380,165} \\
		\hline
	\end{tabular}
\end{minipage}
\par
\medskip

\subsection{Online Data Constructor}
\label{AC2}
\begin{figure*}[t]
	\centering
	\includegraphics[width=\textwidth]{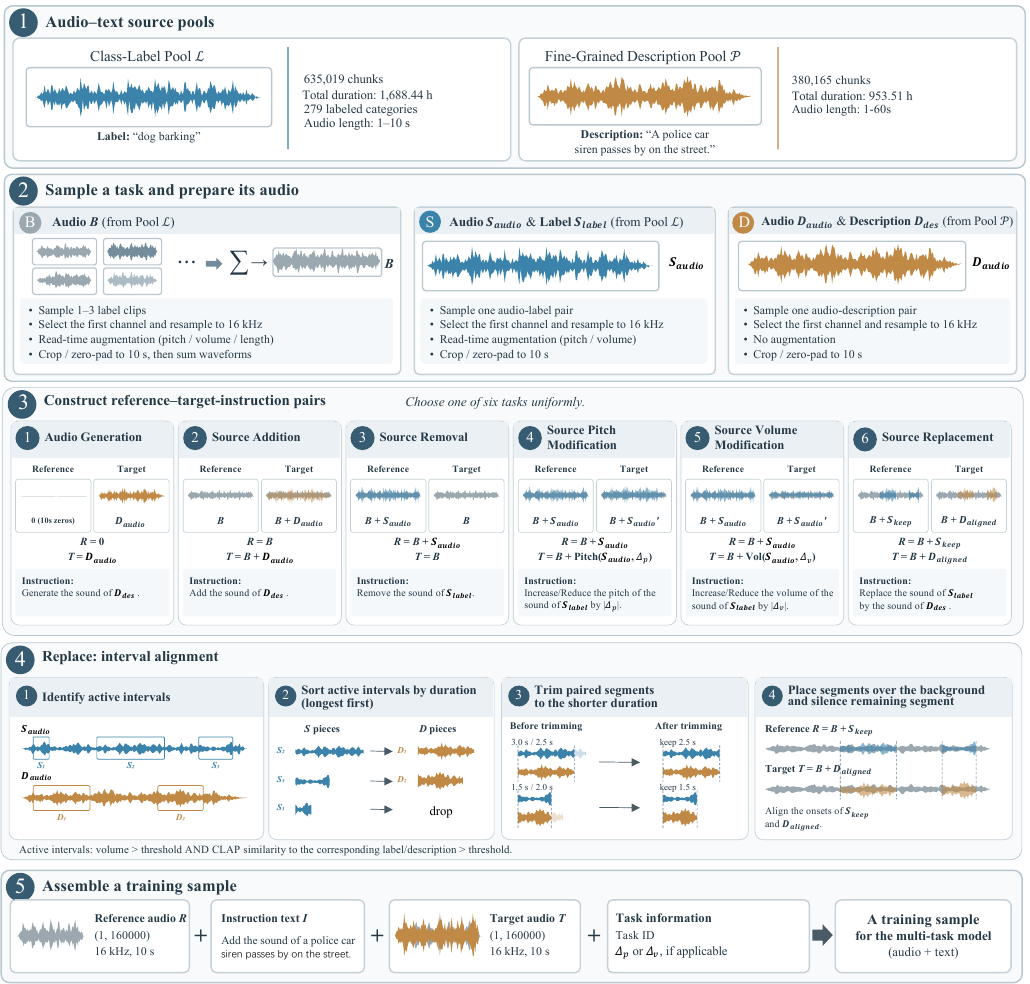}
	\caption{Overview of the online Complex Audio Editing Data constructor. Audio sources from the Class-Label Pool $\mathcal{L}$ and Fine-Grained Description Pool $\mathcal{P}$ are dynamically composed into reference--target--instruction pairs for six audio editing operations. $|\Delta_p|$ is capped at $4$ semitones, and $|\Delta_v|$ is capped at $5$~dB in the process.}
	\label{figure15}
\end{figure*}

We develop the online Complex Audio Editing Data constructor to generate training data for the overall model. For each editing task, audio sources are sampled from different source pools following the principles described in Appendix~\ref{AB1}, and are then composed into the reference audio, target audio, and editing instruction. The detailed training data construction process is illustrated in Fig.~\ref{figure15}.

Since training samples are constructed online, the training set is not restricted to a fixed collection of pre-generated editing pairs. We set its nominal length to approximately 600,000 samples. We additionally construct fixed validation and test sets, each containing 1,000 samples per operation and 6,000 samples in total. All audio sources appearing in the validation and test sets are excluded from training. Notably, during both training and evaluation, all audio samples are either cropped or padded to a fixed duration of 10 seconds.

\subsection{Evaluation Metrics}
\label{AC3}
We evaluate performance on Complex Audio Editing Data using task-aware semantic and attribute metrics, together with complementary audio-similarity metrics. For audio generation, source addition, source removal, and source replacement, we use a task-aware CLAP score as the primary metric. For source pitch and volume modification, the primary metrics are Pitch MAE and Volume MAE, respectively. We additionally report Log-Spectral Distance (LSD) and KL-softmax as complementary metrics.

\paragraph{Task-Aware CLAP.}
Task-aware CLAP is used to evaluate audio generation, source addition, source removal, and source replacement. These operations can naturally be characterized from three semantic aspects: whether the sources that should be preserved remain present, whether the sources that should be added are introduced, and whether the sources that should be removed are suppressed. We therefore design the task-aware CLAP score around these three aspects. Specifically, for each sample $i$, let $c_i^P$, $c_i^A$, and $c_i^R$ denote the average CLAP similarities between the generated audio and the labels corresponding to sources that should be preserved, added, and removed, respectively. The task-aware CLAP score is defined as
\begin{equation}
	s_i^{\mathrm{CLAP}}
	=
	\begin{cases}
		c_i^A, & \mathrm{Generate},\\
		c_i^P+c_i^A, & \mathrm{Add},\\
		c_i^P-c_i^R, & \mathrm{Remove},\\
		c_i^P+c_i^A-c_i^R, & \mathrm{Replace}.
	\end{cases}
\end{equation}
This formulation combines the semantic aspects involved in each operation to measure its overall task fulfillment. Higher values indicate better fulfillment of the task-specific semantic requirements, and the average score over all valid samples reflects the model's overall performance on the corresponding task.

\paragraph{Pitch and Volume MAE.}
For source pitch and volume modification, directly measuring the edited source would require first separating it from the generated mixture, which can introduce substantial measurement errors. We therefore evaluate these two operations indirectly by measuring the residual pitch and volume differences between the generated audio and the corresponding target audio without source separation. For each sample $i$, let $\widehat{\Delta}_i^{\mathrm{pitch}}$ and $\widehat{\Delta}_i^{\mathrm{volume}}$ denote the residual pitch and volume offsets, respectively. The pitch offset is estimated using bidirectional multi-resolution CQT alignment, while the volume offset is computed from the RMS-level difference between the generated and target audio. We define
\begin{equation}
	\mathrm{MAE}_{q}
	=
	\frac{1}{|\mathcal{V}_{q}|}
	\sum_{i\in\mathcal{V}_{q}}
	\left|\widehat{\Delta}_i^{q}\right|,
	\qquad
	q\in\{\mathrm{pitch},\mathrm{volume}\},
\end{equation}
where $\mathcal{V}_{q}$ denotes the set of valid samples for metric $q$. Pitch MAE is measured in semitones and Volume MAE in dB, with lower values indicating better editing performance. The task-level result is obtained by averaging over all valid samples. Unlike volume, pitch estimation depends on the measurement method, and the pitch of some audio samples cannot be measured. We therefore select the pitch estimator that provides the most accurate measurements through extensive validation of multiple estimation methods, and exclude pitch-unmeasurable samples from $\mathcal{V}_{\mathrm{pitch}}$. This measurability issue applies only to Pitch MAE.

\paragraph{LSD and KL-softmax.}
In addition to the task-related metrics above, we further evaluate the similarity between the generated and target audio from spectral and semantic perspectives. For source removal, pitch modification, and volume modification, the desired output is defined by a specific transformation of the input audio and is therefore expected to closely match the constructed target at the signal level. We thus report Log-Spectral Distance (LSD) for these three operations to measure their log-magnitude spectral discrepancy. For all six operations, we additionally report KL-softmax~\cite{liu2023audioldm}, which measures the KL divergence between the PaSST~\cite{koutini2021efficient} audio-event distributions of the target and generated audio and reflects their similarity in semantic audio-event content. Lower values indicate greater similarity for both metrics.

\paragraph{Overall Task-Balanced Gain.}
Since multiple editing tasks share a single model, improvements on one task may be accompanied by degradation on others. Moreover, their evaluation metrics differ in scale and optimization direction, making direct averaging inappropriate. We therefore introduce the overall Task-Balanced Gain, denoted as $\mathrm{TBG}_{\mathrm{all}}$. Let $\mathcal{T}_{\mathrm{all}}$ denote the set of all tasks supported by each model. Given a baseline model and one primary metric for each task, let $M_t$ and $M_t^{\mathrm{base}}$ denote the corresponding metrics of the evaluated and baseline models, respectively. Here, the baseline models refer to the corresponding pretrained models for each architecture. We define
\begin{equation}
	r_t =
	\begin{cases}
		\dfrac{M_t}{M_t^{\mathrm{base}}},
		& t\in\{\mathrm{Generate,Add,Remove,Replace}\},\\[5pt]
		\dfrac{M_t^{\mathrm{base}}}{M_t},
		& t\in\{\mathrm{Pitch,Volume}\},
	\end{cases}
\end{equation}
such that $r_t>1$ indicates improvement over the baseline. The overall score is computed from the equal-weight geometric mean across all supported tasks:
\begin{equation}
	\mathrm{TBG}_{\mathrm{all}}
	=100\left[
	\left(
	\prod_{t\in\mathcal{T}_{\mathrm{all}}}r_t
	\right)^{1/|\mathcal{T}_{\mathrm{all}}|}-1
	\right].
\end{equation}
The geometric mean prevents gains on a subset of tasks from overly masking degradation on others. For our base editor, $\mathcal{T}_{\mathrm{all}}$ includes all six tasks, while for UNISON and SAM Audio, it includes their four and one supported tasks, respectively. We use Task-Aware CLAP as the primary metric for audio generation, source addition, source removal, and source replacement, Pitch MAE for pitch modification, and Volume MAE for volume modification.

\paragraph{Selective Task-Balanced Gain.}
To more directly assess the model's relative selective editing capability, we report a separate gain score over its supported selective editing tasks. Let
$\mathcal{T}_{\mathrm{sel}}
=\mathcal{T}_{\mathrm{all}}\cap\allowbreak\{\mathrm{Remove},\mathrm{Replace},\mathrm{Pitch},\allowbreak\mathrm{Volume}\}$.
Using the same baseline and task-level ratios $r_t$ defined above, we compute
\begin{equation}
	\mathrm{TBG}_{\mathrm{sel}}
	=100\left[
	\left(
	\prod_{t\in\mathcal{T}_{\mathrm{sel}}}r_t
	\right)^{1/|\mathcal{T}_{\mathrm{sel}}|}-1
	\right].
\end{equation}
Like $\mathrm{TBG}_{\mathrm{all}}$, this score is zero for the corresponding pretrained model, while positive values indicate aggregate improvement over the baseline.

\section{Selective Audio Editing Benchmark: Construction and Evaluation}
\label{AD}
\subsection{Benchmark Construction}
\label{AD1}
As discussed in Section~\ref{sec:setup}, source overlap in natural acoustic scenes makes it difficult to automatically determine whether the intended source has been correctly edited while non-target content is preserved. We therefore construct the \textit{Selective Audio Editing Benchmark} with controlled source composition and develop a corresponding evaluation protocol for identifying editing failures and selective failure types.

The benchmark is designed to enable large-scale automatic evaluation of both editing success and selective editing failure types under a controlled setting. For audio generation, we follow the same construction strategy as the Complex Audio Editing Data. For source addition, we restrict the reference audio to a single source. For selective editing operations, the reference audio always contains exactly two temporally non-overlapping sources. One serves as the target source to be edited, while the other serves as the non-target source to be preserved. This simplified and controlled source composition facilitates the evaluation of audio generation and source addition while allowing the editing behaviors of the target and non-target sources in selective editing operations to be measured independently.

\begin{figure*}[t]
	\centering
	\includegraphics[width=\textwidth]{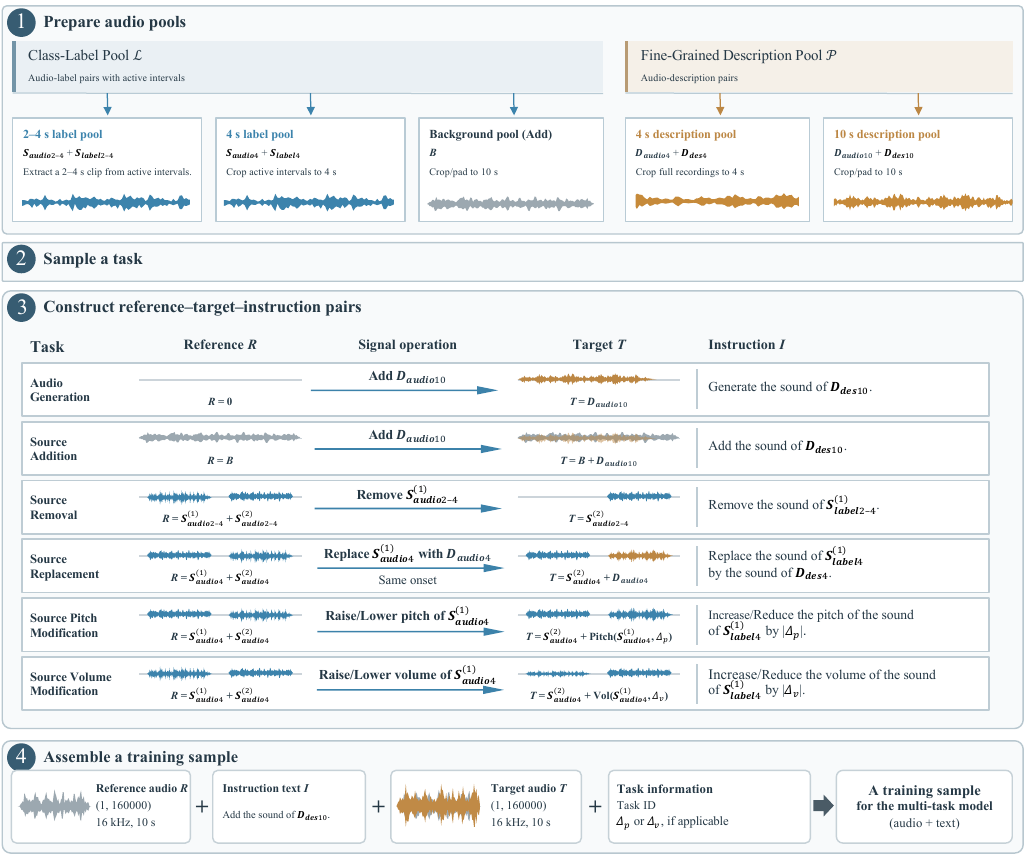}
	\caption{Overview of the Selective Audio Editing Benchmark constructor. Audio sources from the Class-Label Pool $\mathcal{L}$ and Fine-Grained Description Pool $\mathcal{P}$ are organized into task-specific pools and composed into reference--target--instruction pairs for six audio editing operations with controlled source layouts. $|\Delta_p|$ is fixed at $4$ semitones, and $|\Delta_v|$ is fixed at $5$~dB in this process.}
	\label{figure16}
\end{figure*}

We reuse the Class-Label Pool $\mathcal{L}$ and Fine-Grained Description Pool $\mathcal{P}$ introduced in Appendix~\ref{AB1}. We first construct task-specific source pools with different audio durations and then sample the required sources according to each editing task to form the reference audio $R$, target audio $T$, and editing instruction $I$. The overall construction procedure is illustrated in Fig.~\ref{figure16}.

In addition to the benchmark, we construct training and validation sets from the same controlled distribution to support controlled comparisons among different post-training strategies without introducing distribution differences. The training set contains 10,240 samples per task, resulting in 61,440 samples across the six tasks. The validation set and the benchmark each contain 1,000 samples per task. Throughout the construction process, we use the same audio source pools and the same source-level training/validation/test partition as the Complex Audio Editing data in Appendix~\ref{AB2}, ensuring that audio sources used for validation and benchmark evaluation do not appear in training.

\subsection{Automatic Evaluation Protocol}
\label{AD2}
The controlled source composition of the benchmark enables automatic evaluation of both editing success and selective editing failure types.
\paragraph{Task Success Identification and Selective Failure Categorization.}
Depending on the editing operation, we use CLAP similarity to measure source semantics, local RMS-level changes to measure removal and volume modification, and multi-resolution CQT alignment to measure relative pitch changes. The success criterion and failure categorization for each task are defined as follows:

\begin{table*}[t]
	\centering
	\caption{Automatic success criteria and selective failure categorization for the six editing tasks.}
	\label{tab:automatic_evaluation}
	
	\makebox[\textwidth][c]{%
		\begin{minipage}{0.94\textwidth}
			\fontsize{6.5pt}{7.1pt}\selectfont
			\setlength{\tabcolsep}{2.5pt}
			\renewcommand{\arraystretch}{0.96}
			
			\begin{tabularx}{\linewidth}{
					@{}
					>{\raggedright\arraybackslash}p{0.17\linewidth}
					>{\raggedright\arraybackslash}p{0.235\linewidth}
					>{\raggedright\arraybackslash}X
					>{\raggedright\arraybackslash}X
					@{}
				}
				\toprule
				\textbf{Task}
				& \textbf{Outcome}
				& \textbf{Target / edited source}
				& \textbf{Source to preserve} \\
				\midrule
				
				\multirow{2}{*}{\textbf{\shortstack[l]{Audio\\Generation}}}
				& \textbf{Success}
				& Target present (CLAP $>0.08$)
				& -- \\
				& \textit{Other}
				& Target absent (CLAP $\leq0.08$)
				& -- \\
				\midrule
				
				\multirow{2}{*}{\textbf{\shortstack[l]{Source\\Addition}}}
				& \textbf{Success}
				& Added source present (CLAP $>0.04$)
				& -- \\
				& \textit{Other}
				& Added source absent (CLAP $\leq0.04$)
				& -- \\
				\midrule
				
				\multirow{5}{*}{\textbf{\shortstack[l]{Source\\Removal}}}
				& \textbf{Success}
				& Removed ($\geq6$ dB decrease)
				& Preserved ($|\Delta v|<3$ dB) \\
				& \textit{Target Missed}
				& Not removed ($<6$ dB decrease)
				& Preserved ($|\Delta v|<3$ dB) \\
				& \mbox{\textit{Wrong-Target Edit}}
				& Not removed ($<6$ dB decrease)
				& Removed ($\geq6$ dB decrease) \\
				& \mbox{\textit{Unintended Non-Target Edit}}
				& Removed ($\geq6$ dB decrease)
				& Removed ($\geq6$ dB decrease) \\
				& \textit{Other}
				& \multicolumn{2}{l}{Remaining cases} \\
				\midrule
				
				\multirow{5}{*}{\textbf{\shortstack[l]{Source\\Replacement}}}
				& \textbf{Success}
				& Replacement source
				& Original preserved source \\
				& \textit{Target Missed}
				& Original target source
				& Original preserved source \\
				& \mbox{\textit{Wrong-Target Edit}}
				& Original target source
				& Replacement source \\
				& \mbox{\textit{Unintended Non-Target Edit}}
				& Replacement source
				& Replacement source \\
				& \textit{Other}
				& \multicolumn{2}{l}{Ambiguous or unmatched cases} \\
				\midrule
				
				\multirow{5}{*}{\textbf{\shortstack[l]{Source Pitch\\Modification}}}
				& \textbf{Success}
				& Modified ($r\in[0.8,1.2]$)
				& Preserved ($|\Delta p|<1$ semitone) \\
				& \textit{Target Missed}
				& Unchanged ($|\Delta p|<1$ semitone)
				& Unchanged ($|\Delta p|<1$ semitone) \\
				& \mbox{\textit{Wrong-Target Edit}}
				& Unchanged ($|\Delta p|<1$ semitone)
				& Modified ($r\in[0.8,1.2]$) \\
				& \mbox{\textit{Unintended Non-Target Edit}}
				& Modified ($r\in[0.8,1.2]$)
				& Modified ($r\in[0.8,1.2]$) \\
				& \textit{Other}
				& \multicolumn{2}{l}{Remaining measurable cases} \\
				\midrule
				
				\multirow{5}{*}{\textbf{\shortstack[l]{Source Volume\\Modification}}}
				& \textbf{Success}
				& Modified ($r\in[0.8,1.2]$)
				& Preserved ($|\Delta v|<1.5$ dB) \\
				& \textit{Target Missed}
				& Unchanged ($|\Delta v|<1.5$ dB)
				& Unchanged ($|\Delta v|<1.5$ dB) \\
				& \mbox{\textit{Wrong-Target Edit}}
				& Unchanged ($|\Delta v|<1.5$ dB)
				& Modified ($r\in[0.8,1.2]$) \\
				& \mbox{\textit{Unintended Non-Target Edit}}
				& Modified ($r\in[0.8,1.2]$)
				& Modified ($r\in[0.8,1.2]$) \\
				& \textit{Other}
				& \multicolumn{2}{l}{Remaining cases} \\
				\bottomrule
			\end{tabularx}
			
			\vspace{2pt}
			
			\raggedright
			\fontsize{6.2pt}{6.8pt}\selectfont
			For source replacement, a category is assigned only when its local CLAP similarity exceeds $0.08$; if both candidate categories pass the threshold, the one with higher similarity is selected. For pitch and volume modification, $\Delta p$ and $\Delta v$ denote the measured pitch and volume changes, respectively, while $r$ denotes the measured signed change divided by the requested signed change. Samples for which the pitch change cannot be measured for one or both sources are excluded from the final statistics.
			
		\end{minipage}%
	}
\end{table*}

\begin{itemize}
	
	\item \textbf{Audio Generation.}
	We compute the CLAP similarity between the generated audio and the target label. The generation is considered successful when the CLAP similarity is greater than $0.08$; otherwise, the sample is categorized as \textit{Other}.
	
	\item \textbf{Source Addition.}
	We compute the CLAP similarity between the edited audio and the label of the source to be added. The addition is considered successful when the CLAP similarity is greater than $0.04$; otherwise, the sample is categorized as \textit{Other}. The preservation of the original source is not included in the success criterion for source addition, as such failures are rare and additional preservation constraints can introduce false negatives.
	
	\item \textbf{Source Removal.}
	For source removal, the reference audio contains a source to be removed and a temporally non-overlapping source to be preserved. We measure the local volume changes of the two sources before and after editing. The edit is considered successful when the source to be removed decreases by at least $6$~dB, while the absolute volume change of the source to be preserved remains below $3$~dB. This design is also motivated by our observation that cases where the source to be preserved remains perceptually present but undergoes substantial waveform-level changes are rare. If neither source is removed, the result is categorized as \textit{Target Missed}; if the source to be preserved is removed instead of the target source, it is categorized as \textit{Wrong-Target Edit}; and if both sources are removed, it is categorized as \textit{Unintended Non-Target Edit}. All remaining failures are categorized as \textit{Other}.
	\item \textbf{Source Replacement.}
	For source replacement, the reference audio contains a source to be replaced and a temporally non-overlapping source to be preserved. We separately measure the local CLAP similarities within the two source regions. For the source to be replaced, we compare its original label with the replacement-source label; for the source to be preserved, we compare its original label with the replacement-source label. A source category is considered present only when its CLAP similarity is greater than $0.08$, and the category with the higher similarity is selected when both satisfy the threshold. The edit is considered successful when the source to be replaced is identified as the replacement source while the source to be preserved retains its original identity. If both sources retain their original identities, the result is categorized as \textit{Target Missed}; if the source to be replaced remains unchanged while the source to be preserved is identified as the replacement source, it is categorized as \textit{Wrong-Target Edit}; and if both regions are identified as the replacement source, it is categorized as \textit{Unintended Non-Target Edit}. All ambiguous or otherwise unmatched cases are categorized as \textit{Other}.
	\item \textbf{Source Pitch Modification.}
	For source pitch modification, the reference audio contains a source to be modified and a temporally non-overlapping source to be preserved. We separately estimate the relative pitch changes within the two source regions using multi-resolution CQT alignment. For the source to be modified, we compare its measured signed pitch change with the requested change; for the source to be preserved, we measure its pitch change relative to the original source. A source is considered successfully modified when its measured change reaches $80\%$--$120\%$ of the requested signed change, and is considered unchanged when its absolute pitch change is below $1$ semitone. The edit is considered successful when the source to be modified is successfully modified while the source to be preserved remains unchanged. If both sources remain unchanged, the result is categorized as \textit{Target Missed}; if the source to be modified remains unchanged while the source to be preserved undergoes the requested change, it is categorized as \textit{Wrong-Target Edit}; and if both sources undergo the requested change, it is categorized as \textit{Unintended Non-Target Edit}. All remaining failures are categorized as \textit{Other}. Samples for which the pitch change cannot be measured for either source are excluded.
	
	\item \textbf{Source Volume Modification.}
	For source volume modification, the reference audio contains a source to be modified and a temporally non-overlapping source to be preserved. We separately measure the local volume changes within the two source regions. For the source to be modified, we compare its measured signed volume change with the requested change; for the source to be preserved, we measure its volume change relative to the original source. A source is considered successfully modified when its measured change reaches $80\%$--$120\%$ of the requested signed change, and is considered unchanged when its absolute volume change is below $1.5$~dB. The edit is considered successful when the source to be modified is successfully modified while the source to be preserved remains unchanged. If both sources remain unchanged, the result is categorized as \textit{Target Missed}; if the source to be modified remains unchanged while the source to be preserved undergoes the requested change, it is categorized as \textit{Wrong-Target Edit}; and if both sources undergo the requested change, it is categorized as \textit{Unintended Non-Target Edit}. All remaining failures are categorized as \textit{Other}.
\end{itemize}
The criteria are summarized in Table~\ref{tab:automatic_evaluation}. The task-level success rate and overall success rate are computed over valid samples.

\section{Details on the Effectiveness of STEER across Models and Datasets}
\label{AE}

This section provides the training configurations and learning dynamics of STEER across different models on the complex and selective audio editing datasets corresponding to Section~\ref{5.3}. 

\subsection{STEER Adaptation on SAM Audio}
\label{AE1}
Since SAM Audio adopts a different timestep formulation, we first describe how STEER is adapted to SAM Audio. Specifically, SAM Audio uses the opposite time parameterization from our base editor: $t=0$ corresponds to noise, while $t=1$ corresponds to clean audio. We therefore map the training strategy from the $\sigma$ parameterization adopted by STEER to $t$ through
$$
t = 1-\sigma.
$$
Therefore, for STEER-Emphasis, the critical early range for the removal task, $\sigma\in[\sigma_2,1]$, is correspondingly mapped to $t\in[0,1-\sigma_2]$. This preserves the same training semantics of emphasizing the high-noise, early-denoising region. For STEER-Recovery, we retain the 50-step noise-position grid and convert each position according to $t_j=1-\sigma_j$. Specifically, starting from the initial noise, we perform a $k$-step explicit Euler rollout using the conditional velocity field of the model:
$$
\mathbf{z}_{j+1}
=
\mathbf{z}_j
+
(t_{j+1}-t_j)
\mathbf{v}_\theta(\mathbf{z}_j,t_j,c).
$$
After reaching the rollout state $\mathbf{z}_k$, we construct the corrective velocity toward the clean target latent as
$$
\mathbf{v}_k^*
=
\frac{\mathbf{z}_{\mathrm{GT}}-\mathbf{z}_k}{1-t_k},
$$
where $\mathbf{z}_{\mathrm{GT}}$ denotes the clean latent obtained by concatenating the encoded target and residual tracks. The Euler rollout is used only to implement the STEER-Recovery post-training procedure and is not used for validation or inference. For all validation and test evaluations of SAM Audio, we follow its default midpoint ODE solver and integrate the trajectory using \texttt{torchdiffeq.odeint} with the default SAM Audio integration step size $\Delta t = \frac{2}{32}=0.0625.$ We also follow the default SAM Audio setting and do not use CFG during inference. 

In addition, SAM Audio is a separation model that jointly outputs the target source and the residual audio. We post-train it only on the removal task, treating the source specified for removal in the editing instruction as the separation target and using the predicted residual audio as the final removal output. During training, both the target and residual branches are jointly supervised.

\subsection{Training Details on The Complex Dataset across Models}
We describe the training configurations and learning dynamics of STEER across different models on the complex audio editing datasets in this section. 
\label{AE2}

\begin{figure*}[t]
	\centering
	\includegraphics[width=\textwidth]{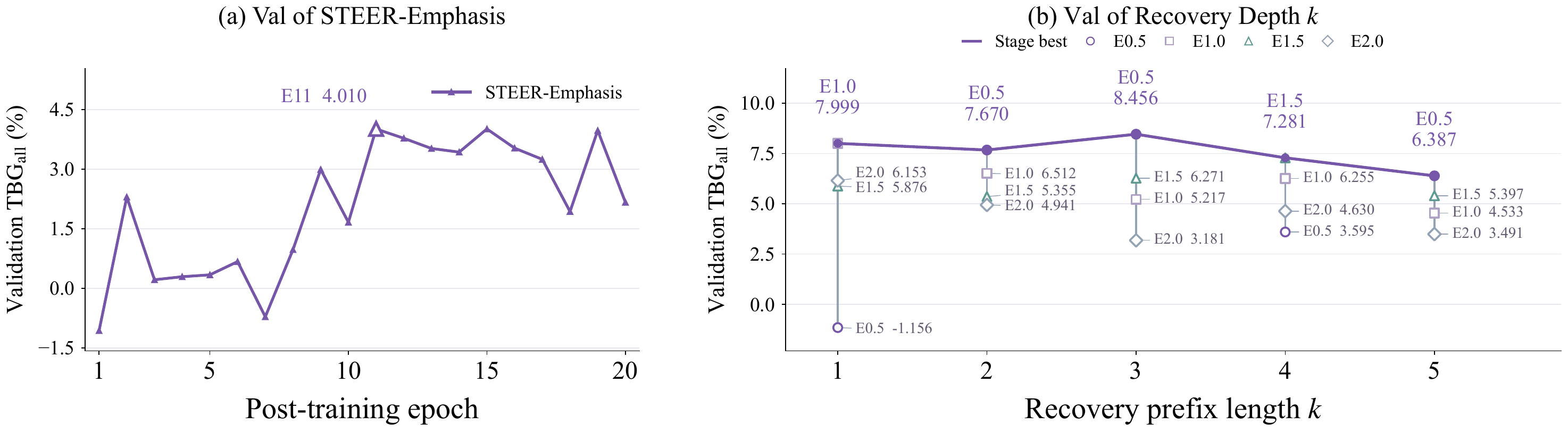}
	\caption{Validation dynamics of STEER on our base editor using the Complex Audio Editing validation set. (a) Validation $\mathrm{TBG}_{\mathrm{all}}$ across STEER-Emphasis training epochs, with the selected checkpoint highlighted. (b) Validation $\mathrm{TBG}_{\mathrm{all}}$ across recovery depths $k$. Markers denote checkpoints evaluated every 0.5 epoch at each depth, while the purple curve connects the validation-best checkpoint for each $k$.}
	\label{figure17}
\end{figure*}
\begin{figure*}[t]
	\centering
	\includegraphics[width=\textwidth]{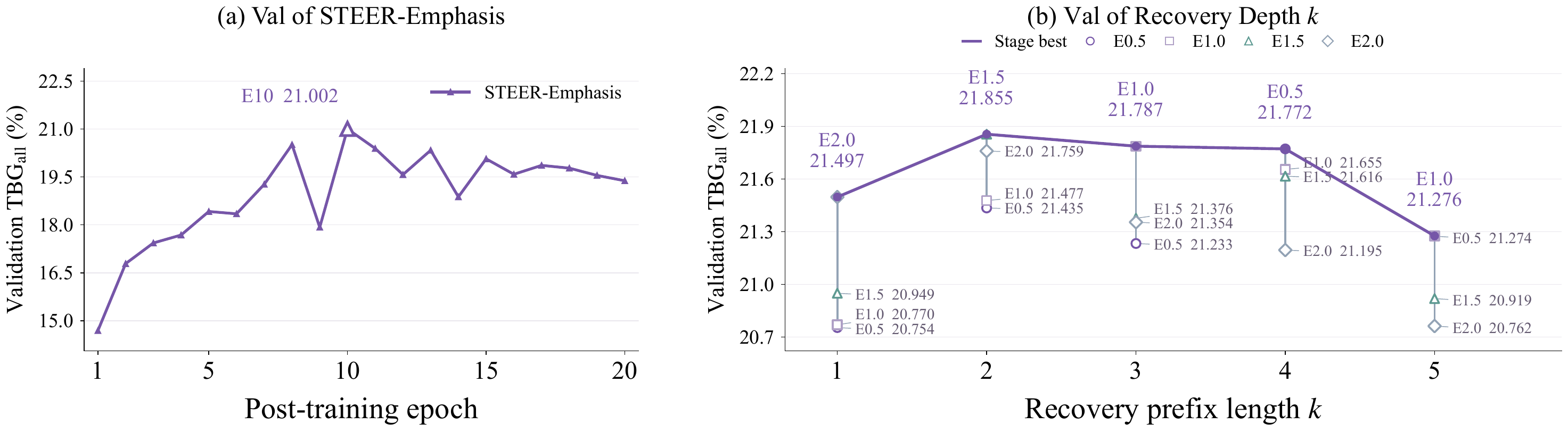}
	\caption{Validation dynamics of STEER on UNISON using the Complex Audio Editing validation set. (a) Validation $\mathrm{TBG}_{\mathrm{all}}$ across STEER-Emphasis training epochs, with the selected checkpoint highlighted. (b) Validation $\mathrm{TBG}_{\mathrm{all}}$ across recovery depths $k$. Markers denote checkpoints evaluated every 0.5 epoch at each depth, while the purple curve connects the validation-best checkpoint for each $k$.}
	\label{figure18}
\end{figure*}
\begin{figure*}[t]
	\centering
	\includegraphics[width=\textwidth]{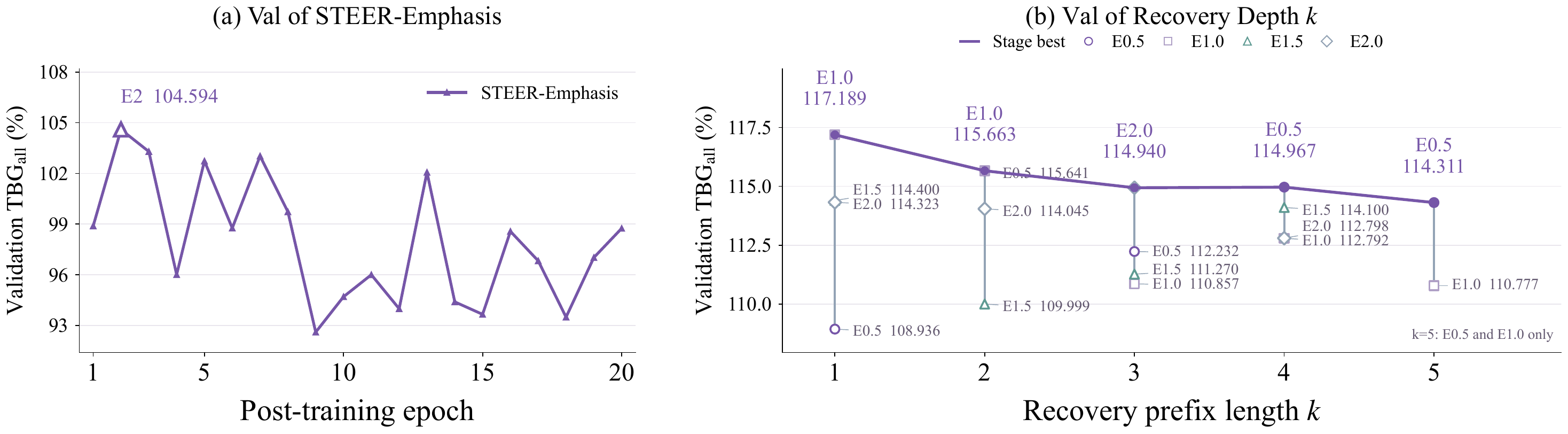}
	\caption{Validation dynamics of STEER on SAM Audio using the Complex Audio Editing validation set. (a) Validation $\mathrm{TBG}_{\mathrm{all}}$ across STEER-Emphasis training epochs, with the selected checkpoint highlighted. (b) Validation $\mathrm{TBG}_{\mathrm{all}}$ across recovery depths $k$. Markers denote checkpoints evaluated every 0.5 epoch at each depth, while the purple curve connects the validation-best checkpoint for each $k$.}
	\label{figure19}
\end{figure*}
\begin{figure*}[h]
	\centering
	\includegraphics[width=\textwidth]{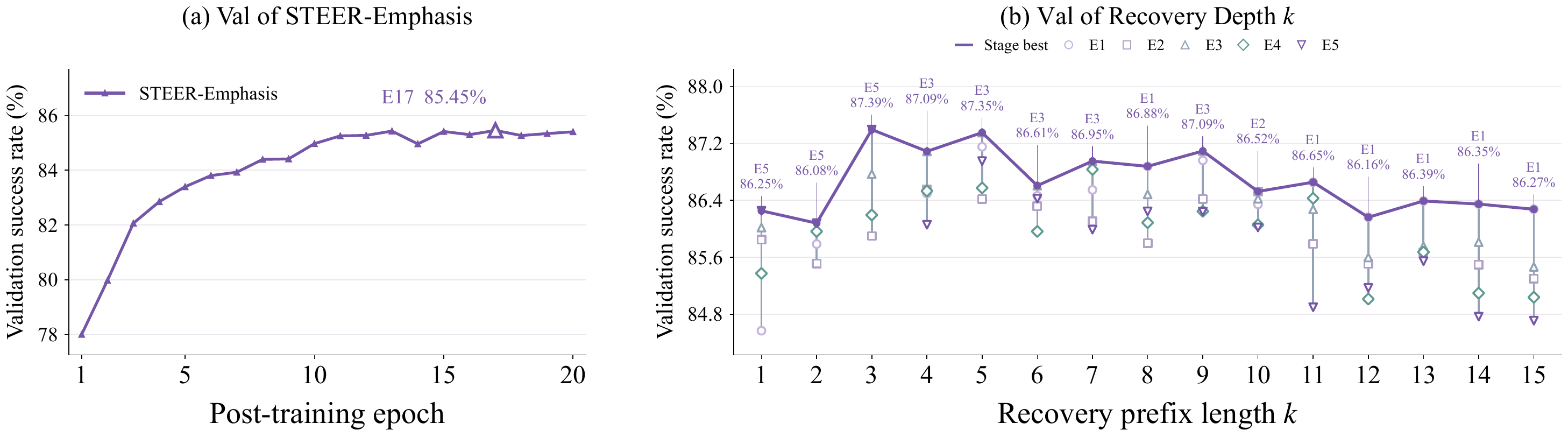}
	\caption{Validation dynamics of STEER on our base editor using the Selective Audio Editing validation set. (a) Overall validation success rate across STEER-Emphasis training epochs, with the selected checkpoint highlighted. (b) Overall validation success rate across recovery depths $k$. Markers denote checkpoints from the five training epochs at each depth, while the purple curve connects the validation-best checkpoint for each $k$.}
	\label{figure20}
\end{figure*}
\begin{figure*}[t]
	\centering
	\includegraphics[width=\textwidth]{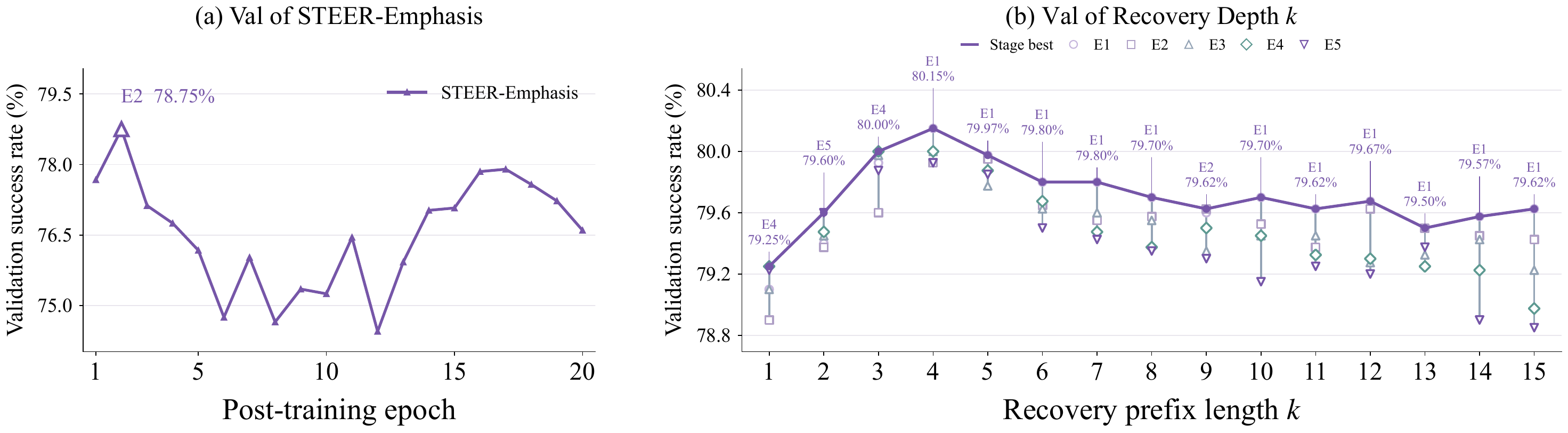}
	\caption{Validation dynamics of STEER on UNISON using the Selective Audio Editing validation set. (a) Overall validation success rate across STEER-Emphasis training epochs, with the selected checkpoint highlighted. (b) Overall validation success rate across recovery depths $k$. Markers denote checkpoints from the five training epochs at each depth, while the purple curve connects the validation-best checkpoint for each $k$.}
	\label{figure21}
\end{figure*}
\begin{figure*}[t]
	\centering
	\includegraphics[width=\textwidth]{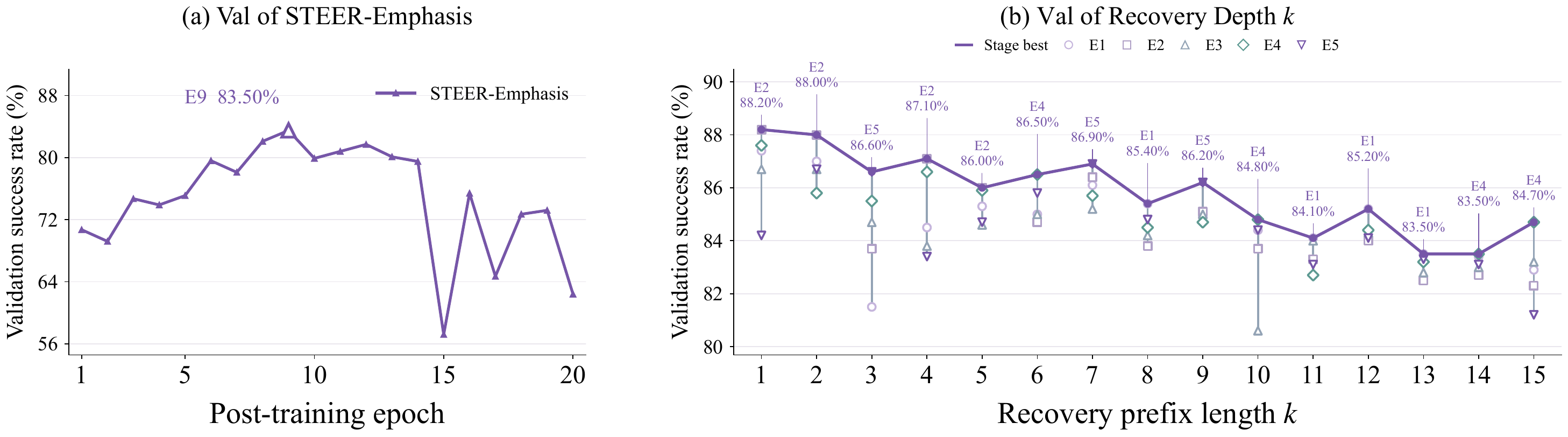}
	\caption{Validation dynamics of STEER on SAM Audio using the Selective Audio Editing validation set. (a) Overall validation success rate across STEER-Emphasis training epochs, with the selected checkpoint highlighted. (b) Overall validation success rate across recovery depths $k$. Markers denote checkpoints from the five training epochs at each depth, while the purple curve connects the validation-best checkpoint for each $k$.}
	\label{figure22}
\end{figure*}

For all three models, STEER-Emphasis starts from the corresponding pretrained checkpoint and runs for 20 epochs. We select the checkpoint with the highest validation $\mathrm{TBG}_{\mathrm{all}}$, which summarizes overall performance across the tasks supported by each model relative to its pretrained baseline. STEER-Recovery starts from the selected STEER-Emphasis checkpoint and is trained for 2 epochs at each recovery depth $k$, with validation performed every 0.5 epoch. At each subsequent depth, training starts from the best checkpoint found so far according to validation $\mathrm{TBG}_{\mathrm{all}}$, with the optimizer state reinitialized. We set the patience parameter to $P=2$. For SAM Audio, we follow the model adaptation procedure described in Section~\ref{AE1}. Across all models, we train only the DiT backbone, keeping all other components, including the language encoders and audio VAEs or codecs, frozen.

To account for differences in model size and pretrained editing capabilities, we use model-specific learning rates and batch sizes. Specifically, our base model uses learning rates of $2\times10^{-6}$ for STEER-Emphasis and $1\times10^{-5}$ for STEER-Recovery, with a global batch size of 480 in both stages. UNISON uses learning rates of $1\times10^{-5}$ and $1\times10^{-6}$ for the two stages, respectively, with a global batch size of 128 throughout. For SAM Audio, we use the Small variant with a learning rate of $1\times10^{-4}$ and a global batch size of 1,024 in both stages. The validation curves for different models on the complex validation set are shown in Figures~\ref{figure17},~\ref{figure18}, and~\ref{figure19}, respectively. And the final test results are shown in Table~\ref{table2}

\subsection{Training Details on The Selective Dataset across Models}
\label{AE3}
For all three models, STEER-Emphasis starts from the corresponding pretrained checkpoint and runs for 20 epochs. We select the checkpoint with the highest overall validation success rate, which summarizes editing performance across the tasks supported by each model. STEER-Recovery starts from the selected STEER-Emphasis checkpoint and is trained for 5 epochs at each recovery depth $k$, with validation performed once per epoch. At each subsequent depth, training starts from the best checkpoint found so far according to the overall validation success rate, with the optimizer state reinitialized. We set the patience parameter to $P=2$. For SAM Audio, we follow the model adaptation procedure described in Section~\ref{AE1}. Across all models, we train only the DiT backbone, keeping all other components, including the language encoders and audio VAEs or codecs, frozen.

To account for differences in model size and pretrained editing capabilities, we use model-specific learning rates and batch sizes. Specifically, our base model uses peak learning rates of $2\times10^{-6}$ for STEER-Emphasis and $1\times10^{-5}$ for STEER-Recovery, with a global batch size of 480 in both stages. UNISON uses learning rate of $1\times10^{-5}$ and a global batch size of 32 in both stages. For SAM Audio, we use the Small variant with a learning rate of $1\times10^{-4}$ and a global batch size of 1,024 in both stages. The validation curves for different models on the selective validation set are shown in Figures~\ref{figure20},~\ref{figure21}, and~\ref{figure22}, respectively. And the final test results are shown in Table~\ref{table3}

Additionally, on the selective dataset, we continue STEER-Recovery training beyond the stopping point prescribed by $P=2$, up to $k=15$, to examine whether deeper recovery yields further gains. Across different models, none of the subsequently evaluated depths surpasses the previous validation best, further supporting the effectiveness of the patience-based stopping rule.

\section{Details on Analysis of Source-Selection Reliability}
\label{AF}
This section provides additional results supporting the source-selection reliability analysis in Section~\ref{5.4}. Specifically, we present task-wise evaluations of recovery from early-trajectory errors and detailed human evaluation results across different editing tasks.

\subsection{Additional Results on Recovery from Early-Trajectory Errors}
\label{AF1}
\begin{figure*}[t]
	\centering
	\includegraphics[width=\textwidth]{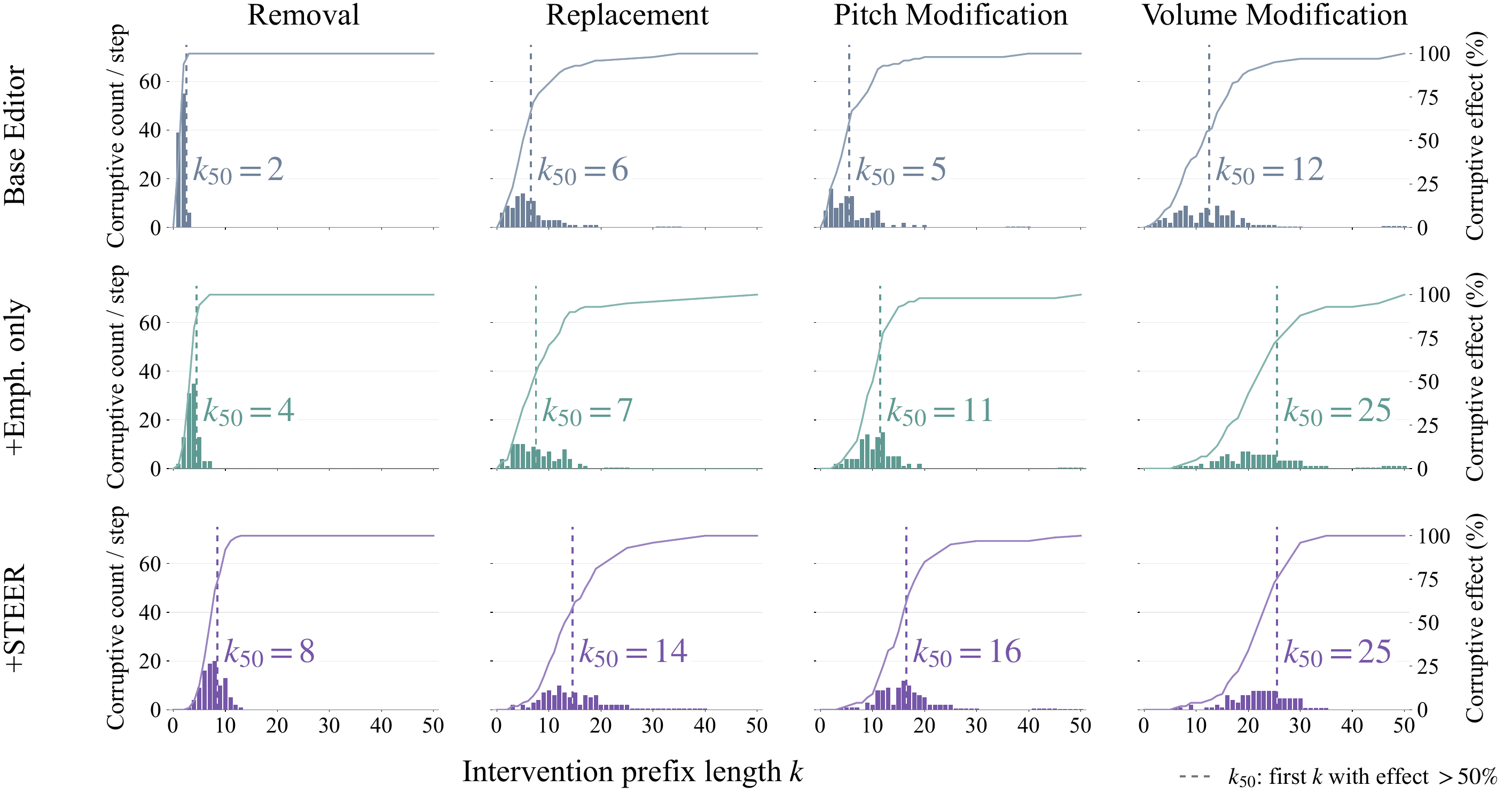}
	\caption{Task-wise recovery from early-trajectory corruption across model variants on the Selective Audio Editing Benchmark. Larger $k_{50}$ indicates stronger robustness to early-trajectory errors.}
	\label{figure23}
\end{figure*}
Figure \ref{figure23} provides the task-wise results corresponding to Figure~\ref{figure8}. Across all four selective editing tasks, STEER-Emphasis improves robustness to early-trajectory corruption. STEER-Recovery further increases $k_{50}$ substantially for removal, replacement, and pitch modification, while maintaining the gain on volume modification. These results further confirm that STEER-Recovery improves recovery from early-trajectory errors across selective editing tasks

\subsection{Additional results on human evaluation.}
\label{AF2}
\begin{figure*}[t]
	\centering
	\includegraphics[width=\textwidth]{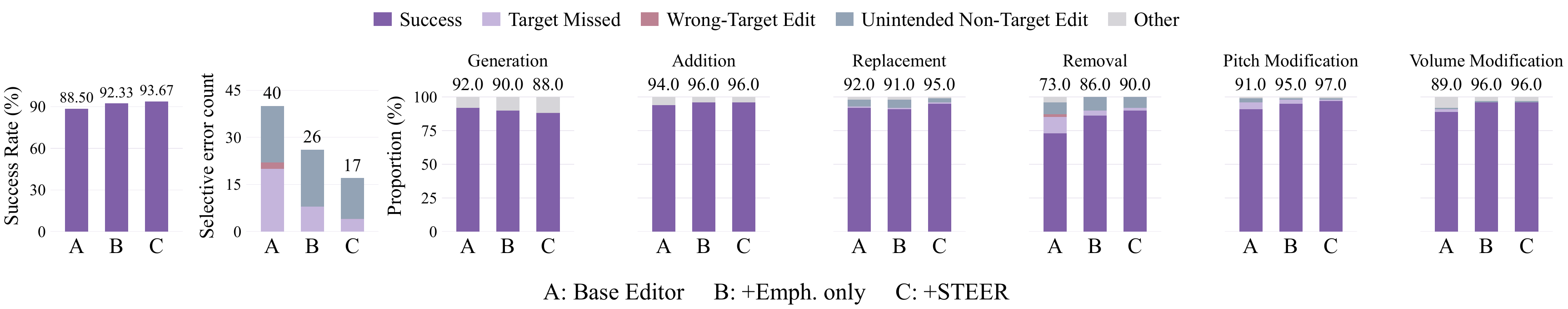}
	\caption{Human evaluation of our base editor, its STEER-Emphasis-only variant, and full STEER variant on the Complex Audio Editing Test Set, using 100 randomly sampled examples per task. Left: overall success rates and selective error counts. Right: task-wise distributions of successful edits and different failure types.}
	\label{figure24}
	\vspace{-5mm}
\end{figure*}
Figure~\ref{figure24} provides task-wise results corresponding to Figure~\ref{figure9}. The task-wise results provide a more detailed view of these improvements across different variants.

\section{Details on Ablation Studies}
\label{AG}
This subsection provides the training configurations, learning dynamics and additional results for the ablation studies presented in Section~\ref{5.5}.
\begin{figure*}[t]
	\centering
	
	\begin{minipage}[t]{0.49\textwidth}
		\centering
		\parbox[c][4.75cm][b]{\linewidth}{%
			\centering
			\includegraphics[
			width=\linewidth,
			height=4.75cm,
			keepaspectratio
			]{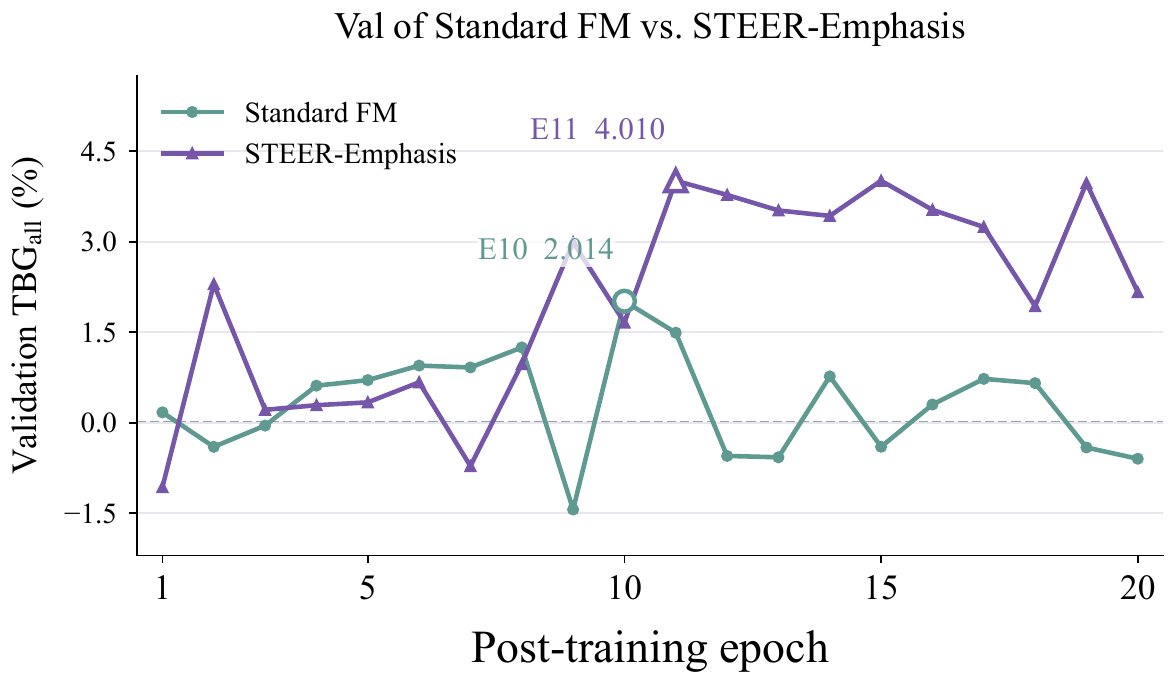}
		}
		\caption{Validation dynamics of standard Flow Matching and STEER-Emphasis on our base editor using the Complex Audio Editing validation set. Curves show validation $\mathrm{TBG}_{\mathrm{all}}$ across post-training epochs, with the validation-best checkpoint of each method highlighted.}

		\label{figure25}
	\end{minipage}%
	\hfill
	\begin{minipage}[t]{0.49\textwidth}
		\centering
		\parbox[c][4.75cm][b]{\linewidth}{%
			\centering
			\includegraphics[
			width=\linewidth,
			height=4.75cm,
			keepaspectratio
			]{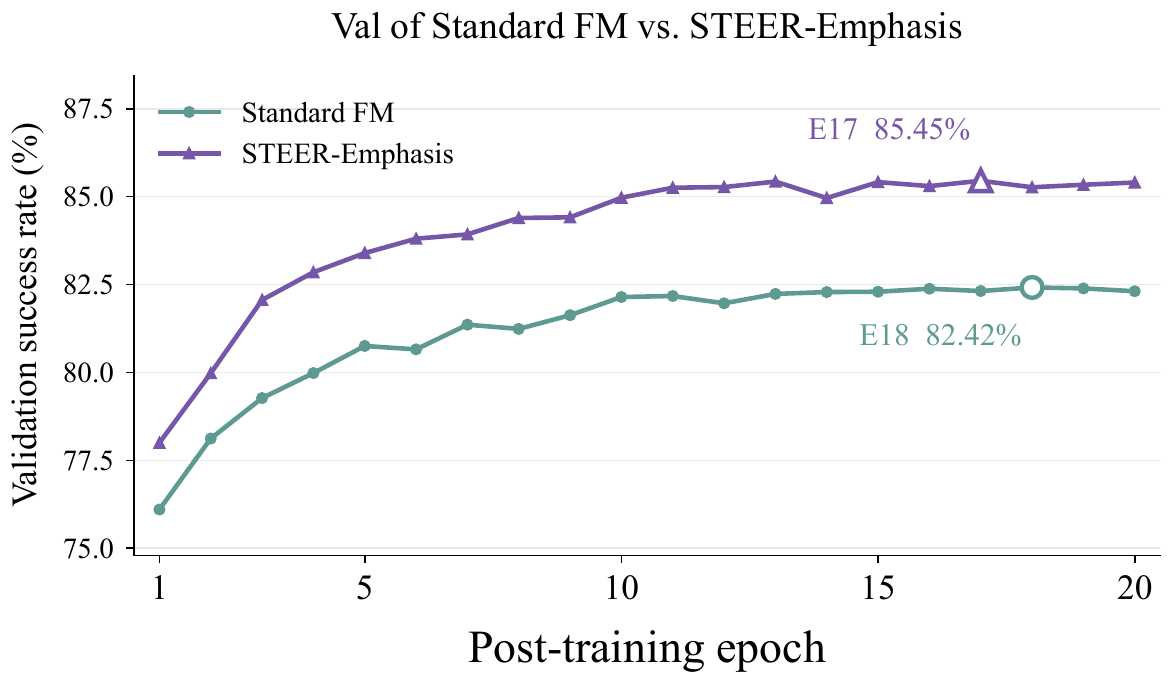}
		}
		\caption{Validation dynamics of standard Flow Matching and STEER-Emphasis on our base editor using the Selective Audio Editing validation set. Curves show overall validation success rates across post-training epochs, with the validation-best checkpoint of each method highlighted.}
		\label{figure26}
	\end{minipage}
\end{figure*}

\begin{figure*}[t]
	\centering
	
	\begin{minipage}[t]{0.49\textwidth}
		\centering
		\parbox[c][4.85cm][b]{\linewidth}{%
			\centering
			\includegraphics[
			width=\linewidth,
			height=4.85cm,
			keepaspectratio
			]{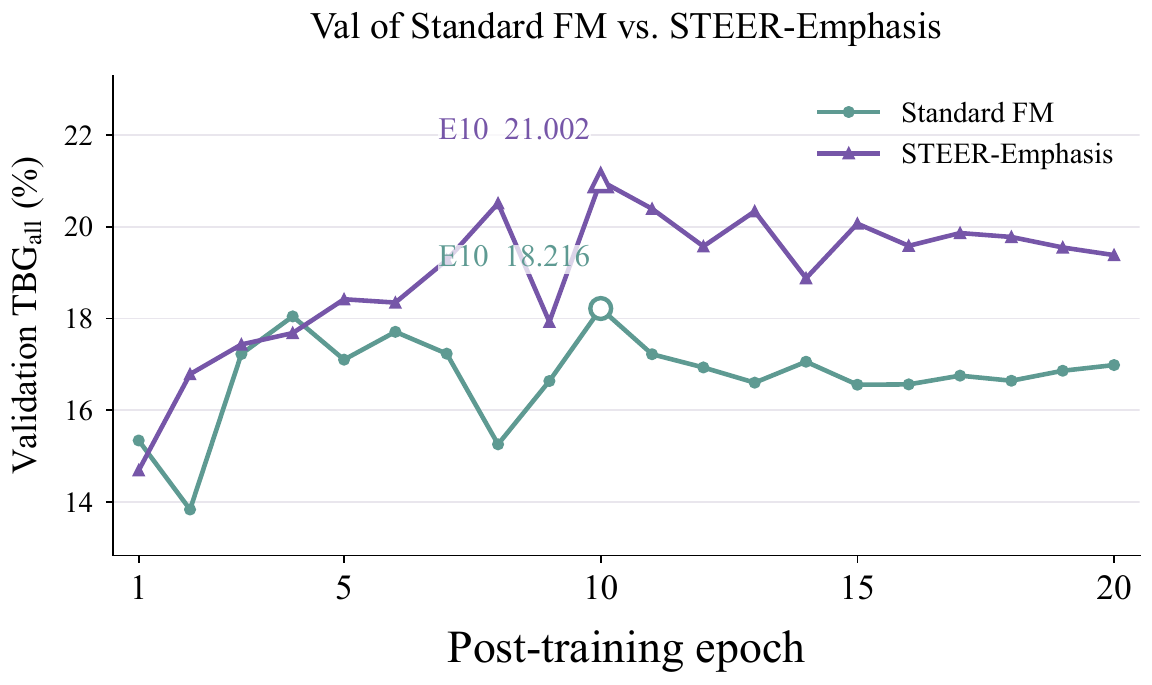}
		}
		\caption{Validation dynamics of standard Flow Matching and STEER-Emphasis on UNISON using the Complex Audio Editing validation set. Curves show validation $\mathrm{TBG}_{\mathrm{all}}$ across post-training epochs, with the validation-best checkpoint of each method highlighted.}
		\label{figure27}
	\end{minipage}%
	\hfill
	\begin{minipage}[t]{0.49\textwidth}
		\centering
		\parbox[c][4.85cm][b]{\linewidth}{%
			\centering
			\includegraphics[
			width=\linewidth,
			height=4.85cm,
			keepaspectratio
			]{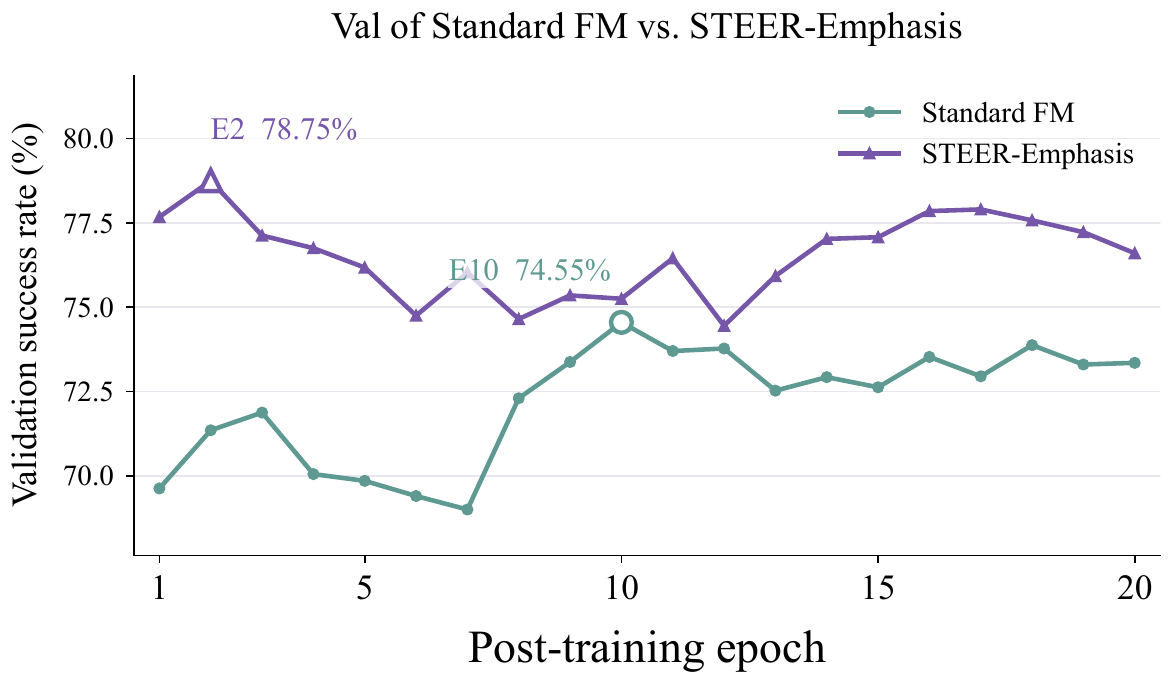}
		}
		\caption{Validation dynamics of standard Flow Matching and STEER-Emphasis on UNISON using the Selective Audio Editing validation set. Curves show overall validation success rates across post-training epochs, with the validation-best checkpoint of each method highlighted.}
		\label{figure28}
	\end{minipage}
\end{figure*}

\begin{figure*}[t]
	\centering
	
	\begin{minipage}[t]{0.49\textwidth}
		\centering
		\parbox[c][4.85cm][b]{\linewidth}{%
			\centering
			\includegraphics[
			width=\linewidth,
			height=4.85cm,
			keepaspectratio
			]{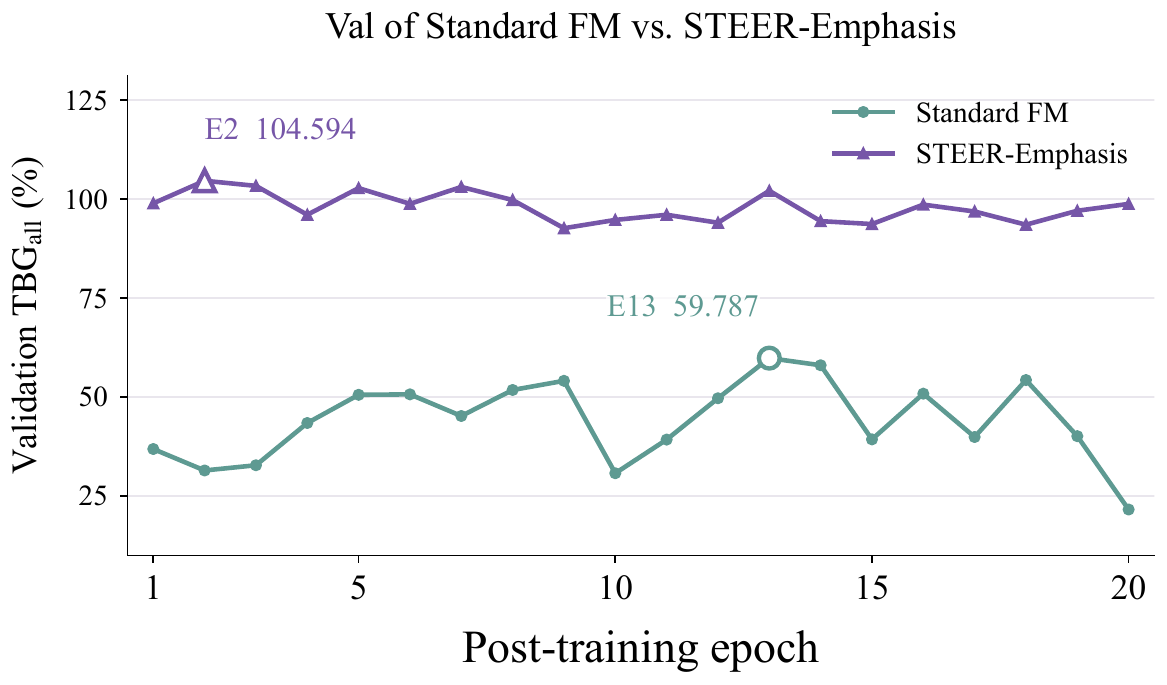}
		}
		\caption{Validation dynamics of standard Flow Matching and STEER-Emphasis on SAM Audio using the Complex Audio Editing validation set. Curves show validation $\mathrm{TBG}_{\mathrm{all}}$ across post-training epochs, with the validation-best checkpoint of each method highlighted.}
		\label{figure29}
	\end{minipage}%
	\hfill
	\begin{minipage}[t]{0.49\textwidth}
		\centering
		\parbox[c][4.85cm][b]{\linewidth}{%
			\centering
			\includegraphics[
			width=\linewidth,
			height=4.85cm,
			keepaspectratio
			]{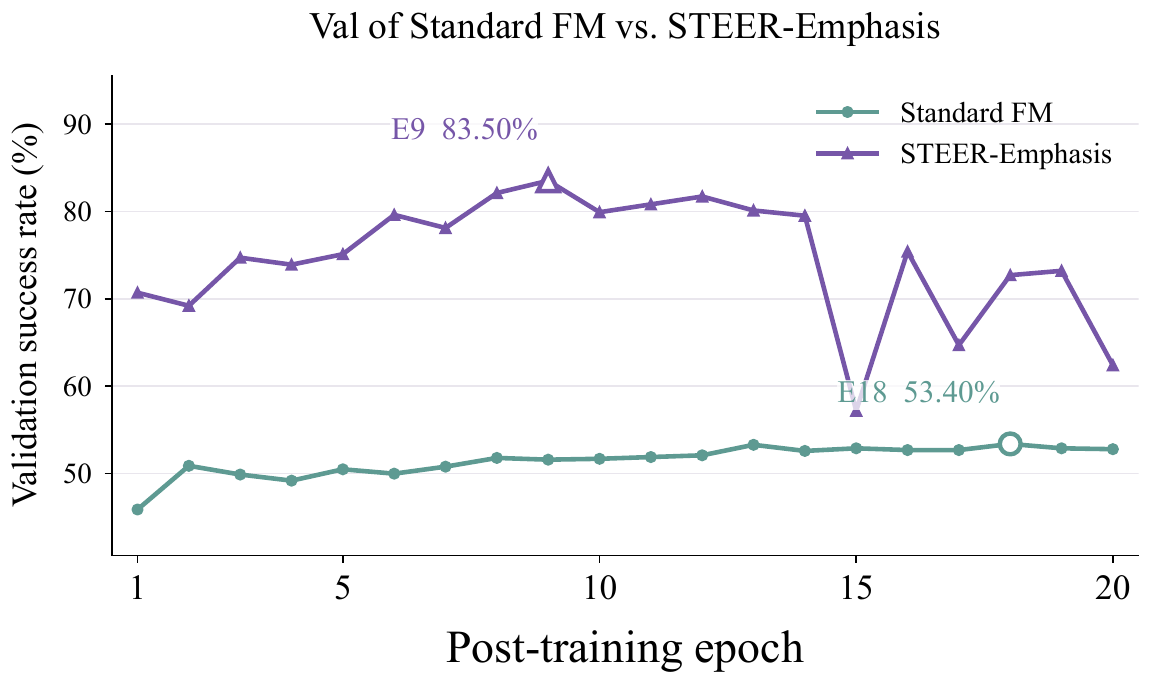}
		}
		\caption{Validation dynamics of standard Flow Matching and STEER-Emphasis on SAM Audio using the Selective Audio Editing validation set. Curves show validation removal success rates across post-training epochs, with the validation-best checkpoint of each method highlighted.}
		\label{figure30}
	\end{minipage}
\end{figure*}

\begin{figure*}[t]
	\centering
	
	\begin{minipage}[t]{0.49\textwidth}
		\centering
		\parbox[c][4.65cm][b]{\linewidth}{%
			\centering
			\includegraphics[
			width=\linewidth,
			height=4.65cm,
			keepaspectratio
			]{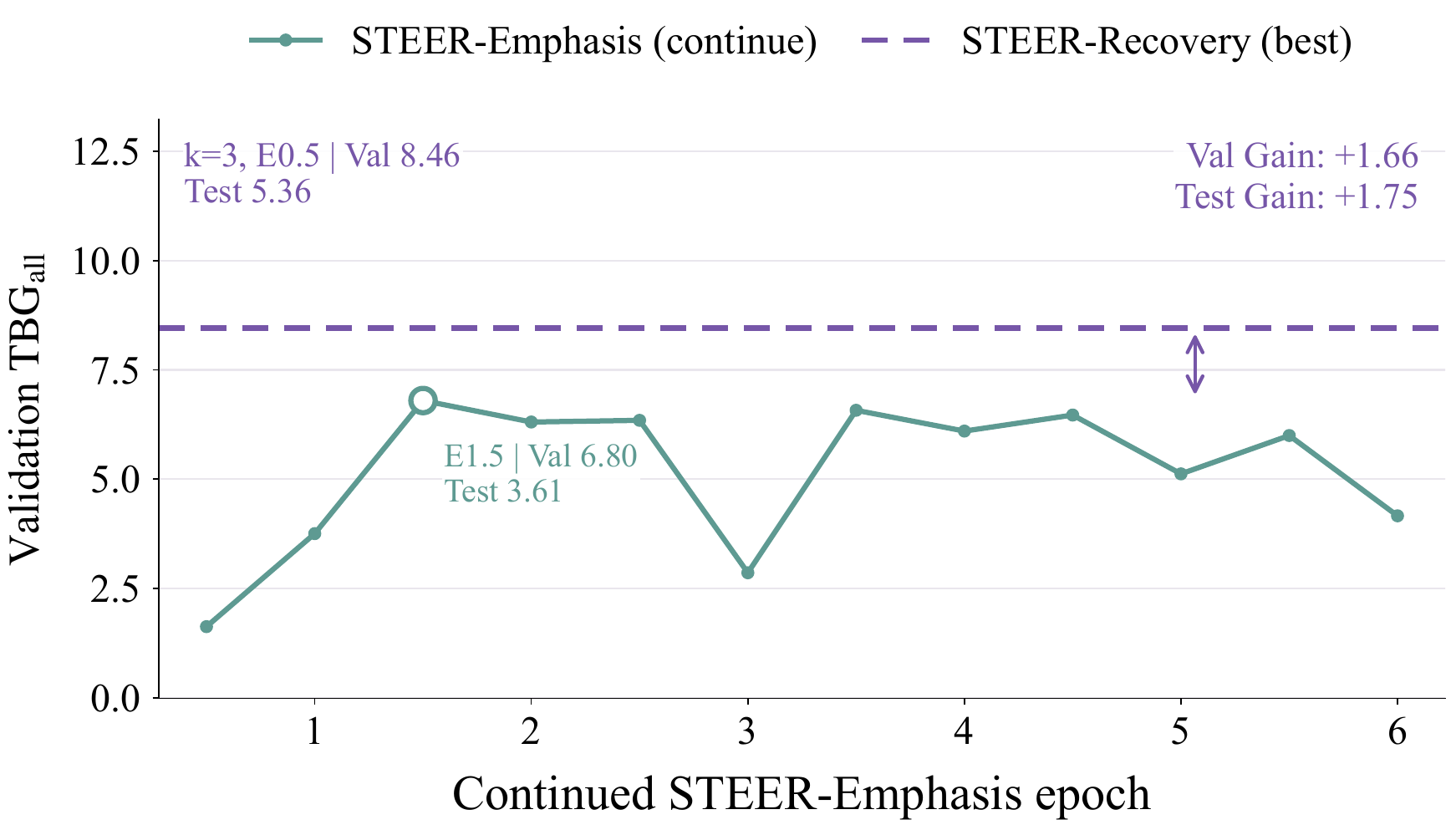}
		}
		\caption{Comparison of continued STEER-Emphasis and STEER-Recovery on our base editor using the Complex Audio Editing validation set under matched optimizer updates. The solid curve shows validation $\mathrm{TBG}_{\mathrm{all}}$ during continued STEER-Emphasis training, while the dashed line indicates the validation performance of the selected STEER-Recovery checkpoint. The highlighted marker denotes the validation-best continuation checkpoint. Validation and test gains are annotated.}
		\label{figure31}
	\end{minipage}%
	\hfill
	\begin{minipage}[t]{0.49\textwidth}
		\centering
		\parbox[c][4.65cm][b]{\linewidth}{%
			\centering
			\includegraphics[
			width=\linewidth,
			height=4.65cm,
			keepaspectratio
			]{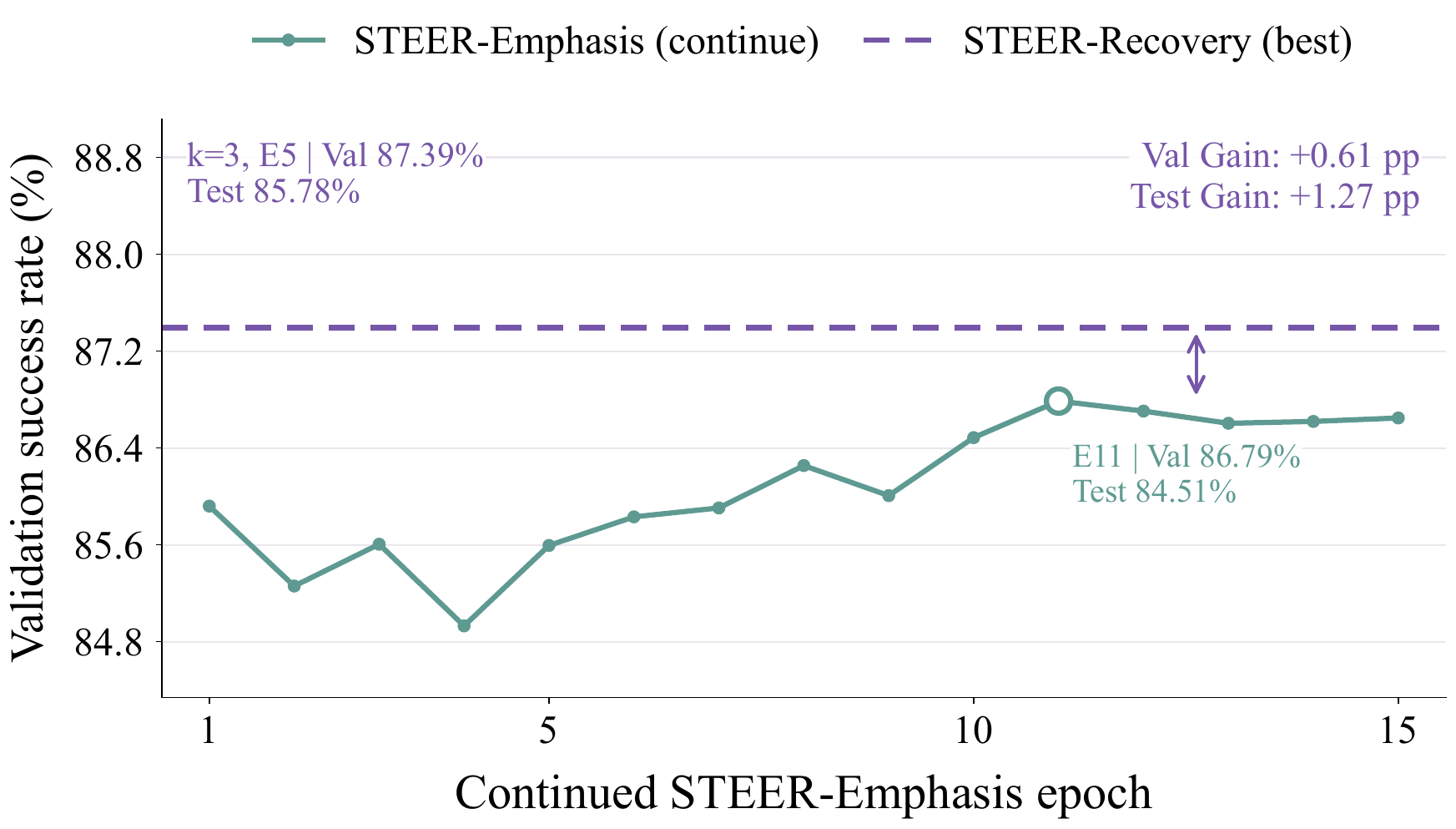}
		}
		\caption{Comparison of continued STEER-Emphasis and STEER-Recovery on our base editor using the Selective Audio Editing validation set under matched optimizer updates. The solid curve shows overall validation success rates during continued STEER-Emphasis training, while the dashed line indicates the validation performance of the selected STEER-Recovery checkpoint. The highlighted marker denotes the validation-best continuation checkpoint. Validation and test gains are annotated.}
		\label{figure32}
	\end{minipage}
\end{figure*}

\begin{figure*}[t]
	\centering
	
	\begin{minipage}[t]{0.49\textwidth}
		\centering
		\parbox[c][4.75cm][b]{\linewidth}{%
			\centering
			\includegraphics[
			width=\linewidth,
			height=4.75cm,
			keepaspectratio
			]{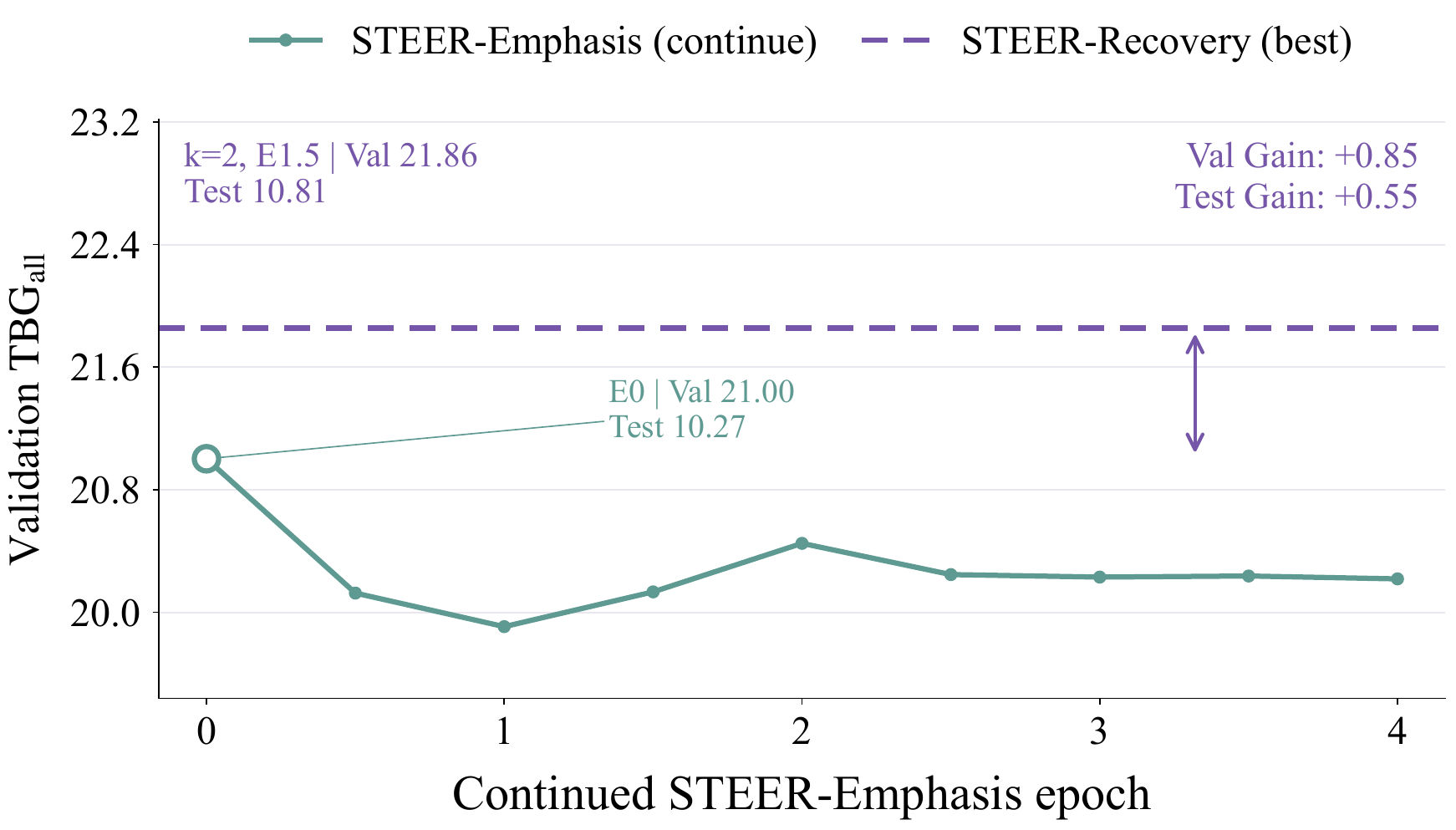}
		}
		\caption{Comparison of continued STEER-Emphasis and STEER-Recovery on UNISON using the Complex Audio Editing validation set under matched optimizer updates. The solid curve shows validation $\mathrm{TBG}_{\mathrm{all}}$ during continued STEER-Emphasis training, while the dashed line indicates the validation performance of the selected STEER-Recovery checkpoint. The highlighted marker denotes the validation-best continuation checkpoint. Validation and test gains are annotated.}
		\label{figure33}
	\end{minipage}%
	\hfill
	\begin{minipage}[t]{0.49\textwidth}
		\centering
		\parbox[c][4.75cm][b]{\linewidth}{%
			\centering
			\includegraphics[
			width=\linewidth,
			height=4.75cm,
			keepaspectratio
			]{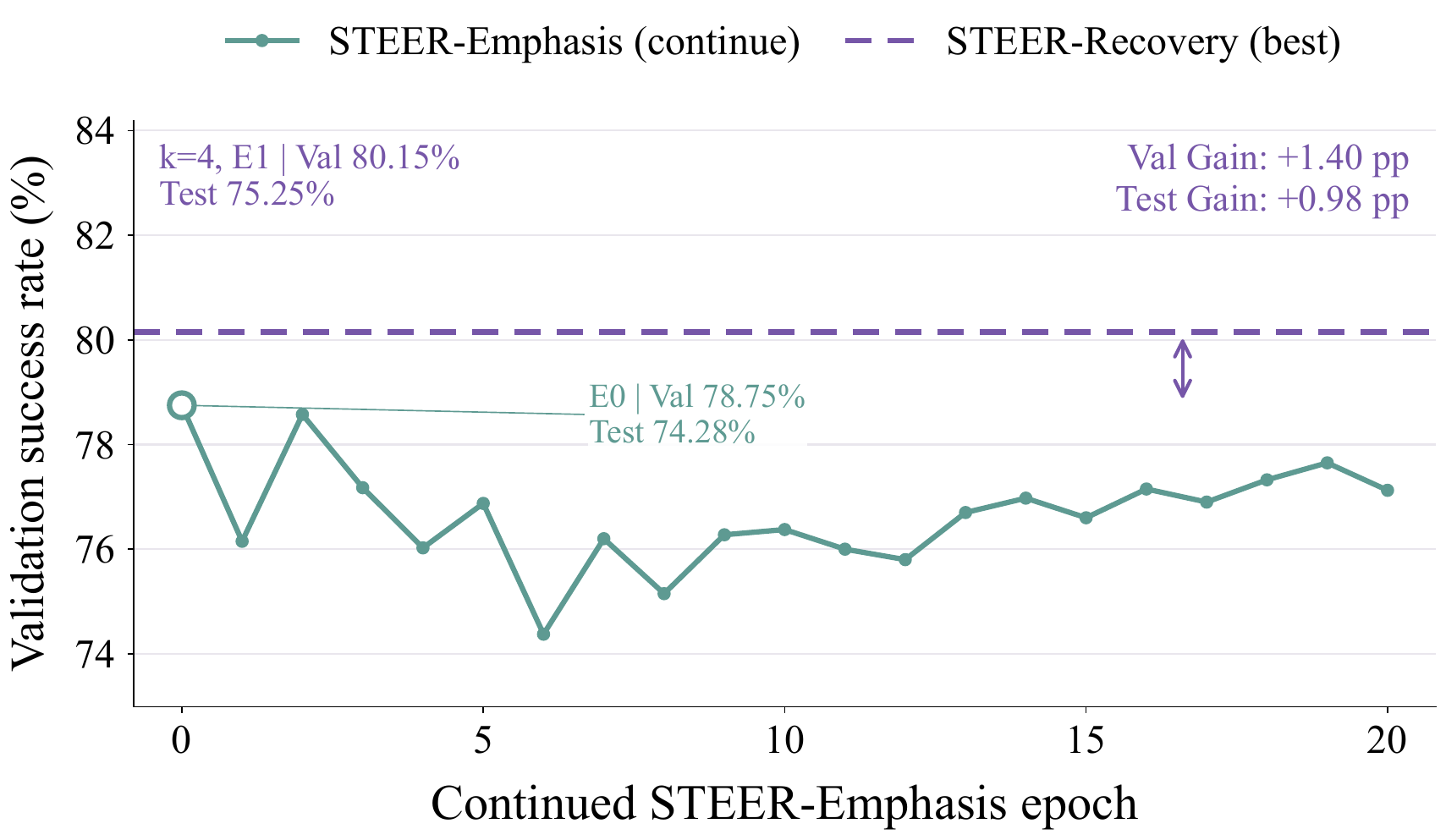}
		}
		\caption{Comparison of continued STEER-Emphasis and STEER-Recovery on UNISON using the Selective Audio Editing validation set under matched optimizer updates. The solid curve shows overall validation success rates during continued STEER-Emphasis training, while the dashed line indicates the validation performance of the selected STEER-Recovery checkpoint. The highlighted marker denotes the validation-best continuation checkpoint. Validation and test gains are annotated.}
		\label{figure34}
	\end{minipage}
\end{figure*}

\begin{figure*}[t]
	\centering
	
	\begin{minipage}[t]{0.49\textwidth}
		\centering
		\parbox[c][4.70cm][b]{\linewidth}{%
			\centering
			\includegraphics[
			width=\linewidth,
			height=4.70cm,
			keepaspectratio
			]{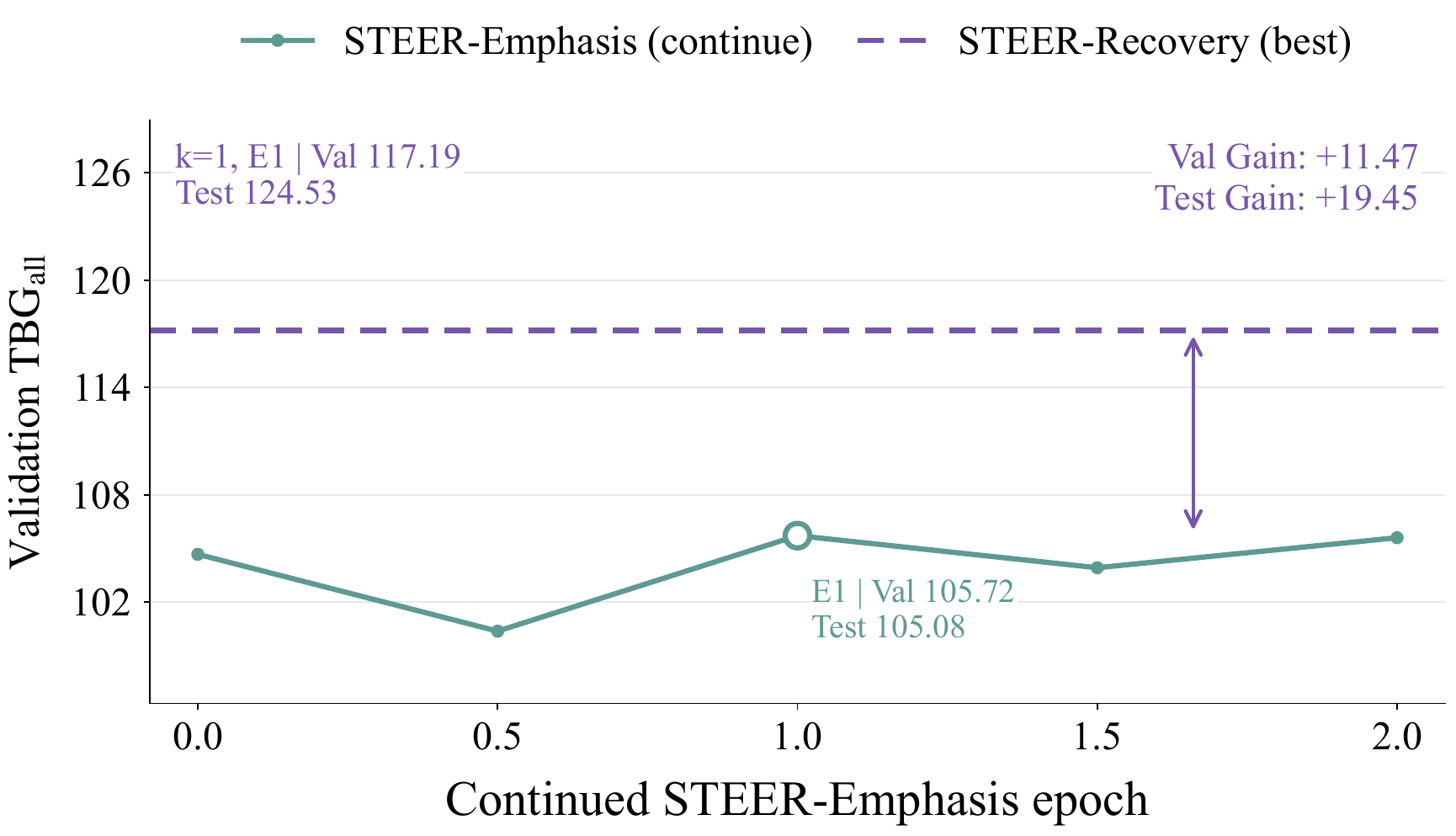}
		}
		\caption{Comparison of continued STEER-Emphasis and STEER-Recovery on SAM Audio using the Complex Audio Editing validation set under matched optimizer updates. The solid curve shows validation $\mathrm{TBG}_{\mathrm{all}}$ during continued STEER-Emphasis training, while the dashed line indicates the validation performance of the selected STEER-Recovery checkpoint. The highlighted marker denotes the validation-best continuation checkpoint. Validation and test gains are annotated.}
		\label{figure35}
	\end{minipage}%
	\hfill
	\begin{minipage}[t]{0.49\textwidth}
		\centering
		\parbox[c][4.70cm][b]{\linewidth}{%
			\centering
			\includegraphics[
			width=\linewidth,
			height=4.70cm,
			keepaspectratio
			]{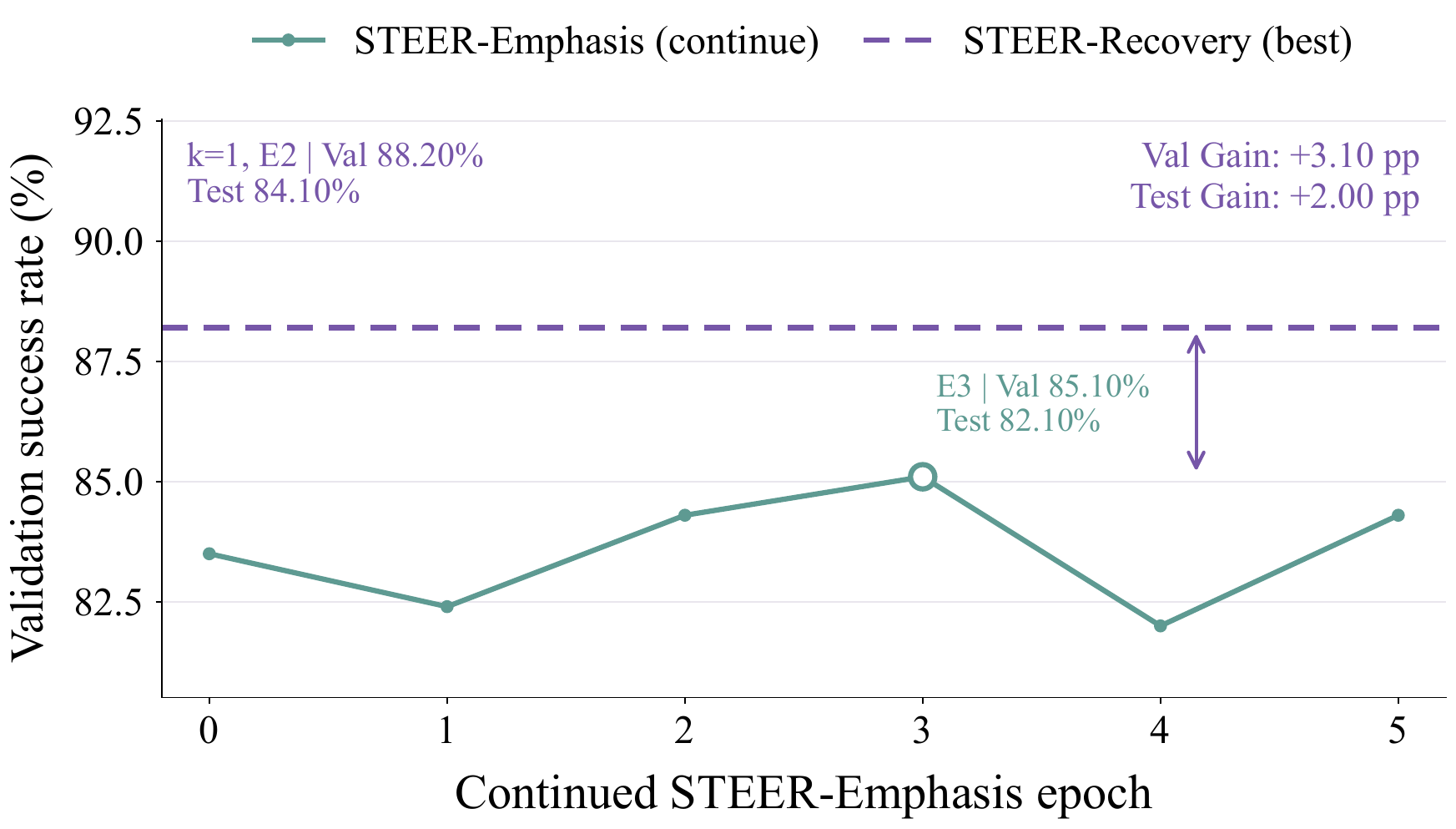}
		}
		\caption{Comparison of continued STEER-Emphasis and STEER-Recovery on SAM Audio using the Selective Audio Editing validation set under matched optimizer updates. The solid curve shows validation removal success rates during continued STEER-Emphasis training, while the dashed line indicates the validation performance of the selected STEER-Recovery checkpoint. The highlighted marker denotes the validation-best continuation checkpoint. Validation and test gains are annotated.}
		\label{figure36}
	\end{minipage}
\end{figure*}

\begin{table*}[t]
	\centering
	\caption{Detailed comparison of STEER-Emphasis and standard Flow Matching post-training under matched updates on the Complex Audio Editing Test Set. TBG scores are computed relative to the corresponding pretrained model.}
	\label{table12}
	
	\vspace{0mm}
	
	\begingroup
	\fontsize{6.2pt}{7.5pt}\selectfont
	\setlength{\tabcolsep}{2.6pt}
	\renewcommand{\arraystretch}{1.20}
	
	\def\steerdash{\textcolor{black!40}{--}}
	
	\resizebox{\textwidth}{!}{%
		\begin{tabular}{ll*{17}{c}}
			\toprule
			
			\multirow[c]{2}{*}{\textbf{Model}}
			& \multirow[c]{2}{*}{\textbf{Strategy}}
			& \multicolumn{2}{c}{\textbf{Generate}}
			& \multicolumn{2}{c}{\textbf{Add}}
			& \multicolumn{2}{c}{\textbf{Replace}}
			& \multicolumn{3}{c}{\textbf{Remove}}
			& \multicolumn{3}{c}{\textbf{Pitch}}
			& \multicolumn{3}{c}{\textbf{Volume}}
			& \multicolumn{2}{c}{\textbf{Overall}} \\
			
			\cmidrule(lr){3-4}
			\cmidrule(lr){5-6}
			\cmidrule(lr){7-8}
			\cmidrule(lr){9-11}
			\cmidrule(lr){12-14}
			\cmidrule(lr){15-17}
			\cmidrule(lr){18-19}
			
			& & CLAP$\uparrow$ & KL$\downarrow$
			& CLAP$\uparrow$ & KL$\downarrow$
			& CLAP$\uparrow$ & KL$\downarrow$
			& CLAP$\uparrow$ & LSD$\downarrow$ & KL$\downarrow$
			& MAE$\downarrow$ & LSD$\downarrow$ & KL$\downarrow$
			& MAE$\downarrow$ & LSD$\downarrow$ & KL$\downarrow$
			& $\mathrm{TBG}_{\mathrm{all}}\uparrow$
			& $\mathrm{TBG}_{\mathrm{sel}}\uparrow$ \\
			
			\midrule
			
			& Standard FM
			& \textbf{0.23} & \textbf{1.99}
			& \textbf{0.25} & \textbf{1.31}
			& \textbf{0.24} & 1.08
			& \textbf{0.14} & 1.33 & 1.06
			& 0.05 & 0.49 & \textbf{0.27}
			& 0.73 & \textbf{0.43} & \textbf{0.20}
			& 0.92 & 0.66 \\
			
			\rowcolor{LightPurple}
			\cellcolor{white}\multirow{-2}{*}{%
				\shortstack[l]{\textbf{Our Base}\\\textbf{Editor}}%
			}
			& \textbf{STEER-Emphasis}
			& 0.22 & 2.12
			& 0.25 & 1.33
			& 0.23 & \textbf{1.05}
			& 0.14 & \textbf{1.22} & \textbf{0.90}
			& \textbf{0.05} & \textbf{0.49} & 0.28
			& \textbf{0.71} & \textbf{0.43} & 0.20
			& \textbf{1.67} & \textbf{2.97} \\
			
			\midrule
			
			& Standard FM
			& 0.24 & 1.87
			& 0.26 & \textbf{1.26}
			& 0.23 & 1.24
			& 0.14 & 1.49 & 1.17
			& \steerdash & \steerdash & \steerdash
			& \steerdash & \steerdash & \steerdash
			& 6.97 & 12.89 \\
			
			\rowcolor{LightPurple}
			\cellcolor{white}\multirow{-2}{*}{\textbf{UNISON}}
			& \textbf{STEER-Emphasis}
			& \textbf{0.25} & \textbf{1.84}
			& \textbf{0.26} & 1.37
			& \textbf{0.24} & \textbf{1.16}
			& \textbf{0.15} & \textbf{1.39} & \textbf{0.89}
			& \steerdash & \steerdash & \steerdash
			& \steerdash & \steerdash & \steerdash
			& \textbf{10.27} & \textbf{17.85} \\
			
			\midrule
			
			& Standard FM
			& \steerdash & \steerdash
			& \steerdash & \steerdash
			& \steerdash & \steerdash
			& 0.10 & 1.47 & 1.45
			& \steerdash & \steerdash & \steerdash
			& \steerdash & \steerdash & \steerdash
			& 65.69 & 65.69 \\
			
			\rowcolor{LightPurple}
			\cellcolor{white}\multirow{-2}{*}{\textbf{SAM Audio}}
			& \textbf{STEER-Emphasis}
			& \steerdash & \steerdash
			& \steerdash & \steerdash
			& \steerdash & \steerdash
			& \textbf{0.12} & \textbf{1.43} & \textbf{1.14}
			& \steerdash & \steerdash & \steerdash
			& \steerdash & \steerdash & \steerdash
			& \textbf{108.62} & \textbf{108.62} \\
			
			\bottomrule
		\end{tabular}%
	}
	
	\endgroup
	\vspace{-2mm}
\end{table*}

\begin{table*}[t]
	\centering
	\caption{Detailed comparison of STEER-Emphasis and standard Flow Matching post-training under matched updates on the Selective Audio Editing Benchmark. Succ. and SErr. denote success and selective error rates, respectively.}
	\label{table13}
	
	\vspace{0mm}
	
	\begingroup
	\fontsize{6.2pt}{7.5pt}\selectfont
	\setlength{\tabcolsep}{3.2pt}
	\renewcommand{\arraystretch}{1.20}
	
	\def\steerdash{\textcolor{black!40}{--}}
	
	\resizebox{\textwidth}{!}{%
		\begin{tabular}{ll*{12}{c}}
			\toprule
			
			\multirow[c]{2}{*}{\textbf{Model}}
			& \multirow[c]{2}{*}{\textbf{Strategy}}
			& \textbf{Generate}
			& \textbf{Add}
			& \multicolumn{2}{c}{\textbf{Replace}}
			& \multicolumn{2}{c}{\textbf{Remove}}
			& \multicolumn{2}{c}{\textbf{Pitch}}
			& \multicolumn{2}{c}{\textbf{Volume}}
			& \multicolumn{2}{c}{\textbf{Overall}} \\
			
			\cmidrule(lr){3-3}
			\cmidrule(lr){4-4}
			\cmidrule(lr){5-6}
			\cmidrule(lr){7-8}
			\cmidrule(lr){9-10}
			\cmidrule(lr){11-12}
			\cmidrule(lr){13-14}
			
			& & Succ.$\uparrow$
			& Succ.$\uparrow$
			& Succ.$\uparrow$ & SErr.$\downarrow$
			& Succ.$\uparrow$ & SErr.$\downarrow$
			& Succ.$\uparrow$ & SErr.$\downarrow$
			& Succ.$\uparrow$ & SErr.$\downarrow$
			& Succ.$\uparrow$ & SErr.$\downarrow$ \\
			
			\midrule
			
			& Standard FM
			& 93.70
			& 92.90
			& 76.20 & 16.40
			& 83.00 & 13.40
			& 82.44 & 16.91
			& 58.40 & 8.40
			& 81.05 & 13.58 \\
			
			\rowcolor{LightPurple}
			\cellcolor{white}\multirow{-2}{*}{%
				\shortstack[l]{\textbf{Our Base}\\\textbf{Editor}}%
			}
			& \textbf{STEER-Emphasis}
			& \textbf{94.00}
			& \textbf{94.60}
			& \textbf{78.90} & \textbf{14.00}
			& \textbf{88.10} & \textbf{7.90}
			& \textbf{87.90} & \textbf{11.70}
			& \textbf{61.60} & \textbf{6.80}
			& \textbf{84.02} & \textbf{9.99} \\
			
			\midrule
			
			& Standard FM
			& \textbf{95.20}
			& 82.40
			& 67.40 & 23.30
			& 27.70 & 14.60
			& \steerdash & \steerdash
			& \steerdash & \steerdash
			& 68.18 & 18.95 \\
			
			\rowcolor{LightPurple}
			\cellcolor{white}\multirow{-2}{*}{\textbf{UNISON}}
			& \textbf{STEER-Emphasis}
			& 94.80
			& \textbf{85.60}
			& \textbf{78.40} & \textbf{13.60}
			& \textbf{38.30} & \textbf{8.50}
			& \steerdash & \steerdash
			& \steerdash & \steerdash
			& \textbf{74.28} & \textbf{11.05} \\
			
			\midrule
			
			& Standard FM
			& \steerdash
			& \steerdash
			& \steerdash & \steerdash
			& 50.50 & 49.40
			& \steerdash & \steerdash
			& \steerdash & \steerdash
			& 50.50 & 49.40 \\
			
			\rowcolor{LightPurple}
			\cellcolor{white}\multirow{-2}{*}{\textbf{SAM Audio}}
			& \textbf{STEER-Emphasis}
			& \steerdash
			& \steerdash
			& \steerdash & \steerdash
			& \textbf{80.60} & \textbf{19.10}
			& \steerdash & \steerdash
			& \steerdash & \steerdash
			& \textbf{80.60} & \textbf{19.10} \\
			
			\bottomrule
		\end{tabular}%
	}
	
	\endgroup
	\vspace{-2mm}
\end{table*}

\begin{table*}[t]
	\centering
	\caption{Detailed comparison of STEER-Recovery and continued STEER-Emphasis under matched updates on the Complex Audio Editing Test Set. TBG scores are computed relative to the corresponding pretrained model.}
	\label{table14}
	
	\vspace{0mm}
	
	\begingroup
	\fontsize{6.2pt}{7.5pt}\selectfont
	\setlength{\tabcolsep}{2.6pt}
	\renewcommand{\arraystretch}{1.20}
	
	\def\steerdash{\textcolor{black!40}{--}}
	
	\resizebox{\textwidth}{!}{%
		\begin{tabular}{ll*{17}{c}}
			\toprule
			
			\multirow[c]{2}{*}{\textbf{Model}}
			& \multirow[c]{2}{*}{\textbf{Strategy}}
			& \multicolumn{2}{c}{\textbf{Generate}}
			& \multicolumn{2}{c}{\textbf{Add}}
			& \multicolumn{2}{c}{\textbf{Replace}}
			& \multicolumn{3}{c}{\textbf{Remove}}
			& \multicolumn{3}{c}{\textbf{Pitch}}
			& \multicolumn{3}{c}{\textbf{Volume}}
			& \multicolumn{2}{c}{\textbf{Overall}} \\
			
			\cmidrule(lr){3-4}
			\cmidrule(lr){5-6}
			\cmidrule(lr){7-8}
			\cmidrule(lr){9-11}
			\cmidrule(lr){12-14}
			\cmidrule(lr){15-17}
			\cmidrule(lr){18-19}
			
			& & CLAP$\uparrow$ & KL$\downarrow$
			& CLAP$\uparrow$ & KL$\downarrow$
			& CLAP$\uparrow$ & KL$\downarrow$
			& CLAP$\uparrow$ & LSD$\downarrow$ & KL$\downarrow$
			& MAE$\downarrow$ & LSD$\downarrow$ & KL$\downarrow$
			& MAE$\downarrow$ & LSD$\downarrow$ & KL$\downarrow$
			& $\mathrm{TBG}_{\mathrm{all}}\uparrow$
			& $\mathrm{TBG}_{\mathrm{sel}}\uparrow$ \\
			
			\midrule
			
			& Emphasis (continue)
			& 0.22 & \textbf{2.08}
			& \textbf{0.25} & \textbf{1.32}
			& \textbf{0.24} & 1.04
			& 0.14 & 1.25 & 0.93
			& \textbf{0.05} & \textbf{0.49} & \textbf{0.28}
			& 0.68 & \textbf{0.43} & \textbf{0.20}
			& 3.61 & 5.50 \\
			
			\rowcolor{LightPurple}
			\cellcolor{white}\multirow{-2}{*}{%
				\shortstack[l]{\textbf{Our Base}\\\textbf{Editor}}%
			}
			& \textbf{STEER-Recovery}
			& \textbf{0.22} & 2.13
			& 0.25 & 1.35
			& \textbf{0.24} & \textbf{1.03}
			& \textbf{0.14} & \textbf{1.16} & \textbf{0.79}
			& \textbf{0.05} & \textbf{0.49} & 0.28
			& \textbf{0.63} & \textbf{0.43} & \textbf{0.20}
			& \textbf{5.36} & \textbf{8.25} \\
			
			\midrule
			
			& Emphasis (continue)
			& 0.25 & 1.84
			& 0.26 & 1.37
			& \textbf{0.24} & 1.16
			& 0.15 & 1.39 & 0.89
			& \steerdash & \steerdash & \steerdash
			& \steerdash & \steerdash & \steerdash
			& 10.27 & 17.85 \\
			
			\rowcolor{LightPurple}
			\cellcolor{white}\multirow{-2}{*}{\textbf{UNISON}}
			& \textbf{STEER-Recovery}
			& \textbf{0.25} & \textbf{1.83}
			& \textbf{0.27} & \textbf{1.32}
			& 0.23 & \textbf{1.16}
			& \textbf{0.15} & \textbf{1.36} & \textbf{0.82}
			& \steerdash & \steerdash & \steerdash
			& \steerdash & \steerdash & \steerdash
			& \textbf{10.81} & \textbf{18.42} \\
			
			\midrule
			
			& Emphasis (continue)
			& \steerdash & \steerdash
			& \steerdash & \steerdash
			& \steerdash & \steerdash
			& 0.12 & 1.36 & 1.06
			& \steerdash & \steerdash & \steerdash
			& \steerdash & \steerdash & \steerdash
			& 105.08 & 105.08 \\
			
			\rowcolor{LightPurple}
			\cellcolor{white}\multirow{-2}{*}{\textbf{SAM Audio}}
			& \textbf{STEER-Recovery}
			& \steerdash & \steerdash
			& \steerdash & \steerdash
			& \steerdash & \steerdash
			& \textbf{0.13} & \textbf{1.27} & \textbf{0.97}
			& \steerdash & \steerdash & \steerdash
			& \steerdash & \steerdash & \steerdash
			& \textbf{124.53} & \textbf{124.53} \\
			
			\bottomrule
		\end{tabular}%
	}
	
	\endgroup
	\vspace{-2mm}
\end{table*}

\begin{table*}[t]
	\centering
	\caption{Detailed comparison of STEER-Recovery and continued STEER-Emphasis under matched updates on the Selective Audio Editing Benchmark. Succ. and SErr. denote success and selective error rates, respectively.}
	\label{table15}
	
	\vspace{0mm}
	
	\begingroup
	\fontsize{6.2pt}{7.5pt}\selectfont
	\setlength{\tabcolsep}{3.2pt}
	\renewcommand{\arraystretch}{1.20}
	
	\def\steerdash{\textcolor{black!40}{--}}
	
	\resizebox{\textwidth}{!}{%
		\begin{tabular}{ll*{12}{c}}
			\toprule
			
			\multirow[c]{2}{*}{\textbf{Model}}
			& \multirow[c]{2}{*}{\textbf{Strategy}}
			& \textbf{Generate}
			& \textbf{Add}
			& \multicolumn{2}{c}{\textbf{Replace}}
			& \multicolumn{2}{c}{\textbf{Remove}}
			& \multicolumn{2}{c}{\textbf{Pitch}}
			& \multicolumn{2}{c}{\textbf{Volume}}
			& \multicolumn{2}{c}{\textbf{Overall}} \\
			
			\cmidrule(lr){3-3}
			\cmidrule(lr){4-4}
			\cmidrule(lr){5-6}
			\cmidrule(lr){7-8}
			\cmidrule(lr){9-10}
			\cmidrule(lr){11-12}
			\cmidrule(lr){13-14}
			
			& & Succ.$\uparrow$
			& Succ.$\uparrow$
			& Succ.$\uparrow$ & SErr.$\downarrow$
			& Succ.$\uparrow$ & SErr.$\downarrow$
			& Succ.$\uparrow$ & SErr.$\downarrow$
			& Succ.$\uparrow$ & SErr.$\downarrow$
			& Succ.$\uparrow$ & SErr.$\downarrow$ \\
			
			\midrule
			
			& Emphasis (continue)
			& 93.90
			& 93.20
			& \textbf{80.70} & 12.60
			& \textbf{88.30} & 8.00
			& 91.05 & 8.55
			& 61.50 & \textbf{5.50}
			& 84.51 & 8.67 \\
			
			\rowcolor{LightPurple}
			\cellcolor{white}\multirow{-2}{*}{%
				\shortstack[l]{\textbf{Our Base}\\\textbf{Editor}}%
			}
			& \textbf{STEER-Recovery}
			& \textbf{94.80}
			& \textbf{93.50}
			& 79.10 & \textbf{12.30}
			& 88.20 & \textbf{7.90}
			& \textbf{92.83} & \textbf{6.77}
			& \textbf{68.00} & 6.40
			& \textbf{85.78} & \textbf{8.45} \\
			
			\midrule
			
			& Emphasis (continue)
			& 94.80
			& 85.60
			& 78.40 & 13.60
			& 38.30 & 8.50
			& \steerdash & \steerdash
			& \steerdash & \steerdash
			& 74.28 & 11.05 \\
			
			\rowcolor{LightPurple}
			\cellcolor{white}\multirow{-2}{*}{\textbf{UNISON}}
			& \textbf{STEER-Recovery}
			& \textbf{94.90}
			& 85.40
			& \textbf{79.60} & \textbf{13.10}
			& \textbf{41.10} & \textbf{8.00}
			& \steerdash & \steerdash
			& \steerdash & \steerdash
			& \textbf{75.25} & \textbf{10.55} \\
			
			\midrule
			
			& Emphasis (continue)
			& \steerdash
			& \steerdash
			& \steerdash & \steerdash
			& 82.10 & 17.50
			& \steerdash & \steerdash
			& \steerdash & \steerdash
			& 82.10 & 17.50 \\
			
			\rowcolor{LightPurple}
			\cellcolor{white}\multirow{-2}{*}{\textbf{SAM Audio}}
			& \textbf{STEER-Recovery}
			& \steerdash
			& \steerdash
			& \steerdash & \steerdash
			& \textbf{84.10} & \textbf{15.70}
			& \steerdash & \steerdash
			& \steerdash & \steerdash
			& \textbf{84.10} & \textbf{15.70} \\
			
			\bottomrule
		\end{tabular}%
	}
	
	\endgroup
	\vspace{-2mm}
\end{table*}

\subsection{Training Details for the STEER-Emphasis Ablation}
\label{AG1}
{\widowpenalty=10000 \clubpenalty=10000 \displaywidowpenalty=10000
We compare STEER-Emphasis with standard Flow Matching post-training under matched numbers of optimizer updates on different datasets across different models. Within each model--dataset pair, both runs start from the same pretrained checkpoint and use the same training data, trainable components, batch size, and optimization settings as STEER-Emphasis described in Section~\ref{AE2} and \ref{AE3}. The standard Flow Matching baseline uses logit-normal noise-level sampling for all supported tasks, whereas STEER-Emphasis adopts the task-dependent sampling distributions in Equation~\ref{eq:focus}. The corresponding validation performance curves for the two strategies are shown in Figure~\ref{figure25}, \ref{figure26}, \ref{figure27}, \ref{figure28}, \ref{figure29} and \ref{figure30}. Detailed task-wise results corresponding to Table~\ref{table4} are provided in Table~\ref{table12} and \ref{table13}.\par}

\subsection{Training Details for the STEER-Recovery Ablation}
\label{AG2}
We compare STEER-Recovery with continued STEER-Emphasis training under matched numbers of optimizer updates. The continuation baseline, denoted STEER-Emphasis$_{\mathrm{cont}}$, retains the STEER-Emphasis noise-level sampling distribution and the standard Flow Matching objective, rather than introducing recovery supervision on self-generated trajectory states.

For a fair comparison, both strategies are initialized from the same validation-best STEER-Emphasis checkpoint, ensuring identical starting model weights. STEER-Emphasis$_{\mathrm{cont}}$ uses the same peak learning rate and global batch size as the corresponding STEER-Recovery run. However, since resetting the optimizer during continuous fine-tuning would disrupt optimization and is not part of standard training practice, we retain the optimizer state throughout continuation training, without checkpoint rollback or optimizer restarts. We also follow the same validation frequency as STEER-Recovery described in Sections~\ref{AE2} and~\ref{AE3}, evaluating every 0.5 epoch on the complex dataset and once per epoch on the selective dataset.

Across all models and datasets, we match the continuation training budget to the total number of optimizer updates performed through completion of the recovery depth containing the selected STEER-Recovery checkpoint. Specifically, for our base editor, we continue training for 6 epochs on the complex dataset and 15 epochs on the selective dataset. For UNISON, we continue training for 4 and 20 epochs, respectively. For SAM Audio, we continue training for 2 and 5 epochs, respectively.

The validation trajectories of continued STEER-Emphasis across training epochs are shown in Figures~\ref{figure31}, \ref{figure32}, \ref{figure33}, \ref{figure34}, \ref{figure35} and \ref{figure36}. Detailed task-wise results corresponding to Table~\ref{table5} are provided in Tables~\ref{table14} and~\ref{table15}.

\subsection{Training Details for the Alternative Early-Trajectory Range Selection.}
\label{AG3}
\begin{table*}[t]
	\centering
	\caption{Task-wise comparison of task-group-specific and elbow-based shared early-trajectory ranges for STEER-Emphasis on the Complex Audio Editing Test Set. TBG scores are computed relative to our base editor.}
	\label{table16}
	
	\vspace{-2mm}
	
	\begingroup
	\fontsize{6.2pt}{7.5pt}\selectfont
	\setlength{\tabcolsep}{2.6pt}
	\renewcommand{\arraystretch}{1.20}
	
	\resizebox{\textwidth}{!}{%
		\begin{tabular}{ll*{17}{c}}
			\toprule
			
			\multirow[c]{2}{*}{\textbf{Dataset}}
			& \multirow[c]{2}{*}{\textbf{Strategy}}
			& \multicolumn{2}{c}{\textbf{Generate}}
			& \multicolumn{2}{c}{\textbf{Add}}
			& \multicolumn{2}{c}{\textbf{Replace}}
			& \multicolumn{3}{c}{\textbf{Remove}}
			& \multicolumn{3}{c}{\textbf{Pitch}}
			& \multicolumn{3}{c}{\textbf{Volume}}
			& \multicolumn{2}{c}{\textbf{Overall}} \\
			
			\cmidrule(lr){3-4}
			\cmidrule(lr){5-6}
			\cmidrule(lr){7-8}
			\cmidrule(lr){9-11}
			\cmidrule(lr){12-14}
			\cmidrule(lr){15-17}
			\cmidrule(lr){18-19}
			
			& & CLAP$\uparrow$ & KL$\downarrow$
			& CLAP$\uparrow$ & KL$\downarrow$
			& CLAP$\uparrow$ & KL$\downarrow$
			& CLAP$\uparrow$ & LSD$\downarrow$ & KL$\downarrow$
			& MAE$\downarrow$ & LSD$\downarrow$ & KL$\downarrow$
			& MAE$\downarrow$ & LSD$\downarrow$ & KL$\downarrow$
			& $\mathrm{TBG}_{\mathrm{all}}\uparrow$
			& $\mathrm{TBG}_{\mathrm{sel}}\uparrow$ \\
			
			\midrule
			
			
			& Emph. (elbow)
			& 0.22 & 2.14
			& 0.25 & 1.35
			& 0.23 & 1.06
			& \textbf{0.14} & 1.25 & 0.95
			& \textbf{0.05} & \textbf{0.49} & \textbf{0.27}
			& 0.74 & \textbf{0.43} & \textbf{0.20}
			& 1.36 & 2.82 \\
			
			\rowcolor{LightPurple}
			\cellcolor{white}\multirow{-2}{*}{\textbf{Complex}}
			& \textbf{STEER-Emphasis}
			& \textbf{0.22} & \textbf{2.12}
			& \textbf{0.25} & \textbf{1.32}
			& \textbf{0.23} & \textbf{1.05}
			& 0.14 & \textbf{1.22} & \textbf{0.90}
			& 0.05 & 0.49 & 0.28
			& \textbf{0.71} & 0.43 & 0.20
			& \textbf{1.67} & \textbf{2.97} \\
			
			\bottomrule
		\end{tabular}%
	}
	
	\endgroup
	\vspace{-2mm}
\end{table*}

\begin{table*}[t]
	\centering
	\caption{Task-wise comparison of task-group-specific and elbow-based shared early-trajectory ranges for STEER-Emphasis on the Selective Audio Editing Benchmark. Succ. and SErr. denote success and selective error rates, respectively.}
	\label{table17}
	
	\vspace{-2mm}
	
	\begingroup
	\fontsize{6.2pt}{7.5pt}\selectfont
	\setlength{\tabcolsep}{3.2pt}
	\renewcommand{\arraystretch}{1.20}
	
	\resizebox{\textwidth}{!}{%
		\begin{tabular}{ll*{12}{c}}
			\toprule
			
			\multirow[c]{2}{*}{\textbf{Dataset}}
			& \multirow[c]{2}{*}{\textbf{Strategy}}
			& \textbf{Generate}
			& \textbf{Add}
			& \multicolumn{2}{c}{\textbf{Replace}}
			& \multicolumn{2}{c}{\textbf{Remove}}
			& \multicolumn{2}{c}{\textbf{Pitch}}
			& \multicolumn{2}{c}{\textbf{Volume}}
			& \multicolumn{2}{c}{\textbf{Overall}} \\
			
			\cmidrule(lr){3-3}
			\cmidrule(lr){4-4}
			\cmidrule(lr){5-6}
			\cmidrule(lr){7-8}
			\cmidrule(lr){9-10}
			\cmidrule(lr){11-12}
			\cmidrule(lr){13-14}
			
			& & Succ.$\uparrow$
			& Succ.$\uparrow$
			& Succ.$\uparrow$ & SErr.$\downarrow$
			& Succ.$\uparrow$ & SErr.$\downarrow$
			& Succ.$\uparrow$ & SErr.$\downarrow$
			& Succ.$\uparrow$ & SErr.$\downarrow$
			& Succ.$\uparrow$ & SErr.$\downarrow$ \\
			
			\midrule
			
			
			& Emph. (elbow)
			& 93.80
			& 94.10
			& \textbf{79.80} & \textbf{13.80}
			& 87.80 & 8.40
			& \textbf{89.33} & \textbf{10.13}
			& 60.40 & \textbf{6.50}
			& 83.98 & \textbf{9.68} \\
			
			\rowcolor{LightPurple}
			\cellcolor{white}\multirow{-2}{*}{\textbf{Selective}}
			& \textbf{STEER-Emphasis}
			& \textbf{94.00}
			& \textbf{94.60}
			& 78.90 & 14.00
			& \textbf{88.10} & \textbf{7.90}
			& 87.90 & 11.70
			& \textbf{61.60} & 6.80
			& \textbf{84.02} & 9.99 \\
			
			\bottomrule
		\end{tabular}%
	}
	
	\endgroup
	\vspace{-2mm}
\end{table*}
We conduct this ablation on our base editor in both dataset settings. The alternative variant, STEER-Emphasis$_{\mathrm{elbow}}$, replaces the task-group-specific ranges defined by the cumulative-effect threshold $\rho=0.8$ with a shared early-trajectory range. Specifically, we apply an elbow-based rule to the cumulative corrective-effect curve averaged across the four selective editing tasks, obtaining a shared range covering the first 15 of 50 denoising steps, as shown in Figure~\ref{figure10}. Accordingly, STEER-Emphasis$_{\mathrm{elbow}}$ samples $\sigma$ uniformly from $[\sigma_{15},1]$ for removal, replacement, pitch modification, and volume modification, while generation and addition retain logit-normal sampling. Both variants start from the same base-editor checkpoint and follow the same dataset-specific STEER-Emphasis training configurations as described in Section~\ref{AE2} and \ref{AE3}. Detailed task-wise results corresponding to Table~\ref{table6} are provided in Table~\ref{table16} and \ref{table17}.

\section{Details on Extension to Fine-Grained Selective Control}
\label{AH}
To complement the experiments in Section~\ref{5.6}, this appendix provides additional experimental details and extended results on two fine-grained selective-control tasks: transcript-conditioned target speaker removal (TSR) with our base editor and target speaker separation (TSS) with SAM Audio. Unlike category-level audio editing, both tasks require identifying the target speaker based on spoken content. Appendix~\ref{AH1} and Appendix~\ref{AH2} present the respective task settings, model adaptations, training configurations, evaluation metrics, and experimental results.

\subsection{Details on Target Speaker Removal with Our Base Editor}
\label{AH1}
In this section, we first introduce the TSR task setting, followed by the model adaptation and training configuration. Finally, we analyze the corresponding training results.

\paragraph{Task Setting.}
Existing audio editing models and tasks often specify the editing target at the category level, e.g., ``Remove the sound of [class label]'' from a mixture containing sources from different categories. This is also the setting adopted by our base editor, where the target source is specified by a class label. In many practical scenarios, however, the target must be identified from substantially finer-grained conditions. To investigate whether early-trajectory emphasis also benefits fine-grained selective control, we extend our base editor to transcript-conditioned \textit{Target Speaker Removal (TSR)}, where the model identifies and removes a target speaker based on spoken content while preserving the remaining speech. Given a speech mixture

\begingroup
\setlength{\abovedisplayskip}{5pt}
\setlength{\belowdisplayskip}{5pt}
\begin{equation}
	x=s_{\mathrm{target}}+s_{\mathrm{residual}},
\end{equation}
\endgroup
where $s_{\mathrm{target}}$ denotes the speech signal of the target speaker to be removed and $s_{\mathrm{residual}}$ denotes the remaining speech that should be preserved, the model learns
\begingroup
\setlength{\abovedisplayskip}{5pt}
\setlength{\belowdisplayskip}{5pt}
\begin{equation}
	\hat{s}_{\mathrm{residual}}=f_\theta(x,c),
\end{equation}
\endgroup
rather than generating $s_{\mathrm{target}}$. The condition $c$ is constructed from the transcription of the target speaker. During training, the instruction takes the form
\begingroup
\setlength{\abovedisplayskip}{5pt}
\setlength{\belowdisplayskip}{5pt}
\begin{equation}
	\texttt{\{Mix audio\} Remove the sound of \{transcription\}},
\end{equation}
\endgroup
where \texttt{\{Mix audio\}} denotes the \textit{input mixed audio}, while \texttt{\{transcription\}} denotes the \textit{text transcription} of the target speaker's speech to be removed. For example, if the target speaker says ``Please close the door,'' the corresponding instruction is ``\{Mix audio\} Remove the sound of Please close the door.'' The model must therefore use the textual content to locate the corresponding speaker within the mixture, remove the sound of that speaker, and preserve the remaining speech.

\paragraph{Model Adaptation, Training Strategy, and Trainable Parameters.}\mbox{}\par
\begin{wrapfigure}{r}{0.36\textwidth}
	\vspace{-20pt}
	\centering
	\includegraphics[width=\linewidth]{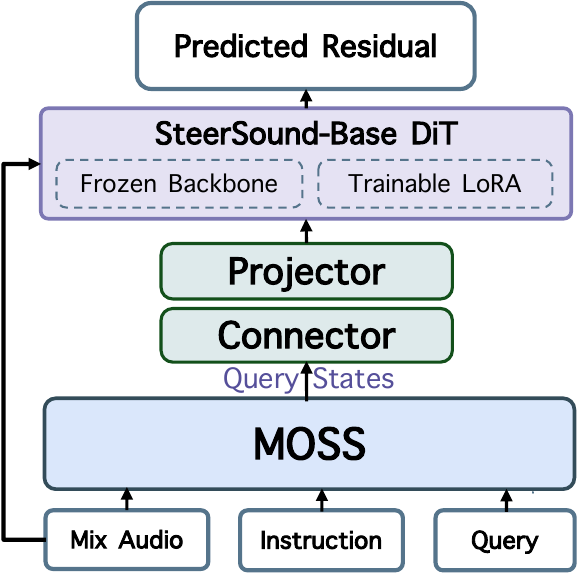}
	\caption{Architecture of our base editor adapted for the TSR task.}
	\label{figure37}
	\vspace{-0.8em}
\end{wrapfigure}
Since this task requires the model to understand the linguistic content of the input audio, we replace the original T5 text encoder in our base editor with a multimodal large language model (MLLM) and connect the MLLM to the DiT backbone of our base editor through a Meta Query~\cite{pan2025transfer} mechanism. Specifically, we use MOSS-Audio-4B-Thinking~\cite{yang2026moss} as the audio-content understanding module. MOSS jointly receives the mixed audio, the text removal instruction, and 50 learnable queries inserted after the \texttt{<SEG>} token. We extract the corresponding query hidden states before the output head of MOSS and pass them through a one-layer Transformer Connector followed by a LayerNorm--MLP Projector. The resulting representations are mapped into conditioning features and replace the original T5-based textual conditioning of our base editor.

This adaptation introduces newly initialized components, including the learnable queries, Transformer Connector, and Projector. Moreover, we observe that standard Flow Matching training barely converges on this fine-grained task and fails to learn effective target selection. Therefore, rather than first training these newly introduced components with standard Flow Matching and subsequently applying STEER-Emphasis as a post-training stage, we directly compare standard training with training that incorporates the early-trajectory emphasis principle from the beginning. This setting allows us to more clearly assess the effect of emphasizing the early trajectory throughout training. The architecture of our base editor adapted for the TSR task is shown in Figure~\ref{figure37}.

Since fine-grained speaker editing differs substantially from the original audio editing tasks, we adapt the DiT backbone of our base editor with LoRA to avoid disrupting its previously learned editing capability. Specifically, LoRA modules are inserted into the \texttt{qkv} and \texttt{linear} layers of the attention blocks, with rank 64, alpha 128, and dropout 0.05. We jointly train the DiT LoRA parameters, learnable queries, Transformer Connector, and Projector, while keeping all other components frozen. The numbers of trainable parameters for different modules are summarized in Table~\ref{table18}.

\paragraph{Training Configuration.}
\begin{wraptable}[9]{r}{0.27\textwidth}
	\vspace{-16pt}
	\centering
	
	\setlength{\abovecaptionskip}{0pt}
	\setlength{\belowcaptionskip}{4pt}
	
	\caption{Trainable parameters.}
	\label{table18}
	
	\footnotesize
	\setlength{\tabcolsep}{4.5pt}
	\renewcommand{\arraystretch}{1.10}
	
	\begin{tabular}{@{}lr@{}}
		\toprule
		\textbf{Module} & \textbf{Params.} \\
		\midrule
		DiT LoRA     & 15.73M \\
		Queries      & 0.13M \\
		Connector    & 52.46M \\
		Projector    & 8.53M \\
		\midrule
		\textbf{Total} & \textbf{76.84M} \\
		\bottomrule
	\end{tabular}
	
	\vspace{-2pt}
\end{wraptable}
We initialize the model from our base editor and train it for 20 epochs using either Standard FM or the early-emphasis training strategy. Both settings use AdamW with a weight decay of $0.01$, a constant learning rate of $1\times10^{-4}$, no warmup or learning-rate decay, a batch size of 64, and gradient clipping with a maximum norm of $1.0$.

Since several trainable components, including the DiT LoRA, Connector, and Projector, are newly initialized, this setting does not fully satisfy the original assumption of STEER-Emphasis that the model already possesses an initial editing capability. We therefore adopt a mixed sampling strategy: while most training states follow the STEER-Emphasis $\sigma$ sampling, a small proportion are sampled using the standard logit-normal distribution to retain coverage of the full generation trajectory. Specifically, Standard FM uses the standard logit-normal sampling strategy $\sigma\sim\operatorname{sigmoid}\bigl(\mathcal{N}(0,1)\bigr).$ For the early-emphasis training, we use
$$
\sigma\sim
\begin{cases}
	\mathcal{U}(0.97,1), & \text{with probability } 0.8,\\
	\operatorname{sigmoid}\bigl(\mathcal{N}(0,1)\bigr), & \text{with probability } 0.2.
\end{cases}
$$
Thus, $80\%$ of the training states are concentrated in the high-noise early-trajectory region, while the remaining $20\%$ retain standard logit-normal sampling to preserve coverage of the full generation trajectory. The training objective remains the standard Flow Matching objective in both settings.
\paragraph{Training Dataset.}
We use the PromptTSE~\cite{seki2025language} dataset for training and evaluation. In PromptTSE, each mixed-audio sample contains speech from two distinct speakers and is accompanied by the target-speaker audio to be extracted, the residual speech after removing the target speaker, and the transcription of the target speaker. In our TSR setting, the provided residual audio is used as the generation target, while the target transcription is used to construct the fine-grained removal instruction.

\paragraph{Evaluation Metric.}
We directly use CLAP Audio--Audio similarity for TSR ($\operatorname{CLAP\text{-}AA}_{\mathrm{TSR}}$) to measure the similarity between the generated residual audio and the ground-truth residual audio. Let $E_a(\cdot)$ denote the CLAP~\cite{elizalde2023clap} audio encoder. For sample $i$,
$$
\operatorname{CLAP\text{-}AA}_{\mathrm{TSR},i}
=
\frac{
	E_a(\hat{s}_{\mathrm{residual},i})^\top
	E_a(s_{\mathrm{residual},i})
}{
	\left\|E_a(\hat{s}_{\mathrm{residual},i})\right\|_2
	\left\|E_a(s_{\mathrm{residual},i})\right\|_2
}.
$$
The final score is averaged over all $N$ evaluation samples:
$$
\operatorname{CLAP\text{-}AA}_{\mathrm{TSR}}
=
\frac{1}{N}
\sum_{i=1}^{N}
\operatorname{CLAP\text{-}AA}_{\mathrm{TSR},i}.
$$
$\operatorname{CLAP\text{-}AA}_{\mathrm{TSR}}$ therefore directly measures the audio similarity between the generated residual audio and the ground-truth residual audio.
\paragraph{Experimental Results.} 
\begin{wrapfigure}{r}{0.48\textwidth}
	\vspace{-1.5em}
	\centering
	
	\includegraphics[width=\linewidth]{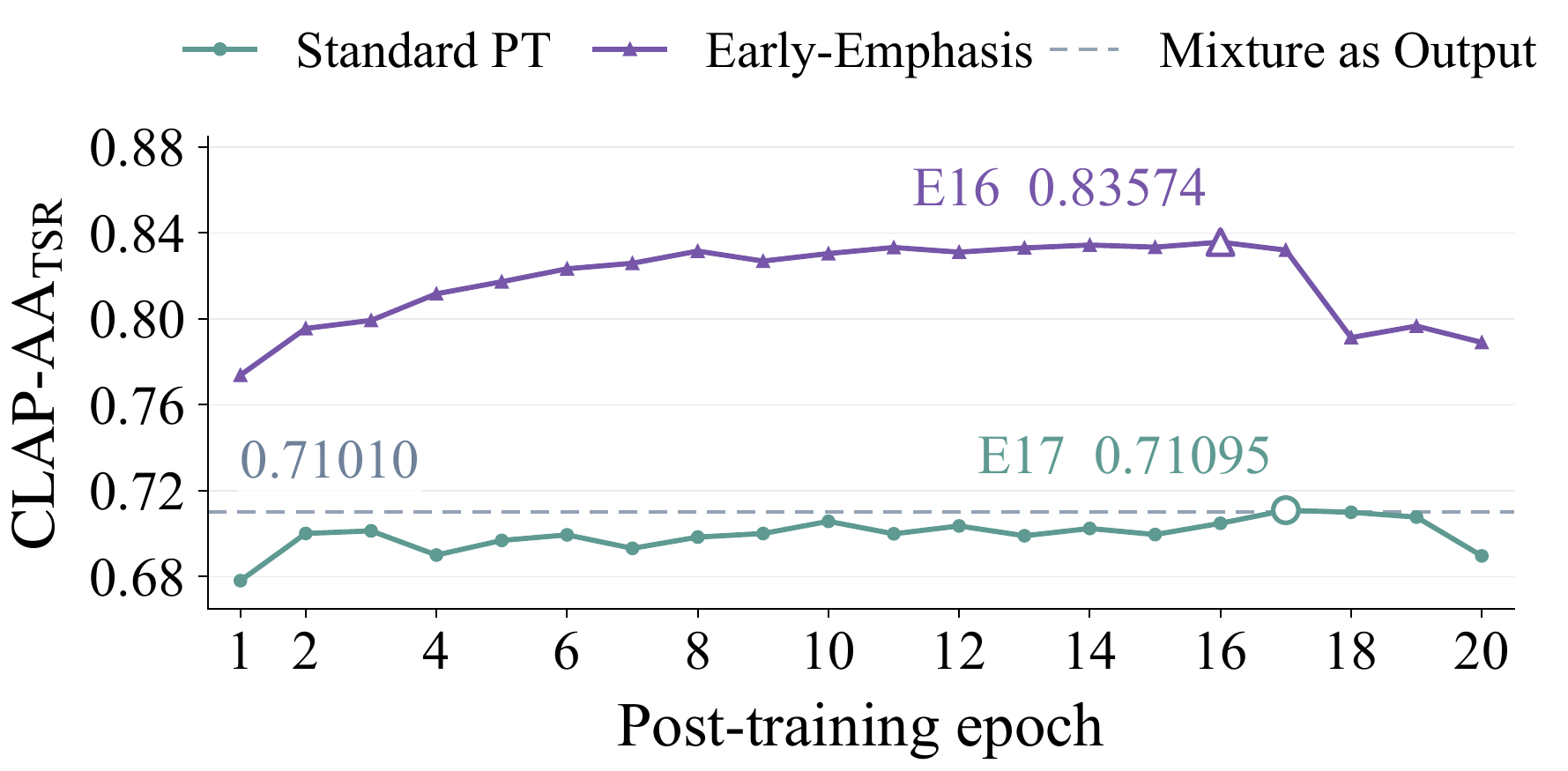}
	
	\vspace{-2mm}
	\caption{
		Validation $\operatorname{CLAP\text{-}AA}_{\mathrm{TSR}}$ during training
		on the TSR task with our adapted base editor. The horizontal line denotes
		the $\operatorname{CLAP\text{-}AA}_{\mathrm{TSR}}$ of the mixture audio.
	}
	\label{figure38}
	
	\vspace{-1em}
\end{wrapfigure}
Figure~\ref{figure38} shows the validation $\operatorname{CLAP\text{-}AA}_{\mathrm{TSR}}$ of Standard FM and early trajectory emphasis training over 20 training epochs. Compared with Standard FM, early trajectory emphasis training learns substantially more effectively. This result further demonstrates the importance of emphasizing the training on critical early-trajectory states for fine-grained selective control task.

The horizontal line denotes the $\operatorname{CLAP\text{-}AA}_{\mathrm{TSR}}$ obtained by directly using
the mixed audio as the model output and comparing it with the ground-truth residual audio. Notably, without emphasizing the training on critical early trajectory, Standard FM performs close to the direct-mixture baseline, whereas emphasizing the training on critical early trajectory substantially improves performance on this fine-grained selective control task.

\subsection{Details on Target Speaker Separation with SAM Audio}
\label{AH2}
In this section, we first introduce the TSS task setting, followed by the model adaptation and training configuration. Finally, we analyze the corresponding training results.

\paragraph{Task Setting.}

Following the experiment with our base editor, we further investigate whether early-trajectory emphasis benefits fine-grained selective control on SAM Audio. We consider transcript-conditioned \textit{Target Speaker Separation (TSS)}, in which the model identifies the target speaker based on spoken content and separates the mixture into target and residual speech. Given a speech mixture
\begingroup
\setlength{\abovedisplayskip}{5pt}
\setlength{\belowdisplayskip}{5pt}
\begin{equation}
	x=s_{\mathrm{target}}+s_{\mathrm{residual}},
\end{equation}
\endgroup
where $s_{\mathrm{target}}$ denotes the speech signal of the target speaker to be separated and $s_{\mathrm{residual}}$ denotes the remaining speech, the model learns
\begingroup
\setlength{\abovedisplayskip}{5pt}
\setlength{\belowdisplayskip}{5pt}
\begin{equation}
	[\hat{s}_{\mathrm{target}},\hat{s}_{\mathrm{residual}}]
	=
	f_\theta(x,c),
\end{equation}
\endgroup
with supervision applied to both $\hat{s}_{\mathrm{target}}$ and $\hat{s}_{\mathrm{residual}}$. The condition $c$ is constructed from the transcription of the target speaker. During training, the instruction takes the form
\begingroup
\setlength{\abovedisplayskip}{5pt}
\setlength{\belowdisplayskip}{5pt}
\begin{equation}
	\texttt{\{Mix audio\} Separate the sound of \{transcription\}}.
\end{equation}
\endgroup
The meanings of the individual components in the instruction are the same as those described in Section~\ref{AH1}.

\paragraph{Model Adaptation, Training Strategy, and Trainable Parameters.}\mbox{}\par
\begin{wraptable}{r}{0.31\textwidth}
	\vspace{-30pt}
	\centering
	
	\caption{Trainable parameters.}
	\label{tab:sam_tse_trainable_params}
	
	\vspace{0.3em}
	\footnotesize
	\setlength{\tabcolsep}{5pt}
	\renewcommand{\arraystretch}{1.04}
	
	\begin{tabular}{lr}
		\toprule
		\textbf{Module} & \textbf{Params.} \\
		\midrule
		SAM Audio LoRA & 127.40M \\
		Queries        & 0.13M \\
		Connector      & 65.54M \\
		Projector      & 18.09M \\
		\midrule
		\textbf{Total} & \textbf{211.17M} \\
		\bottomrule
	\end{tabular}
	
	\vspace{-0.5em}
\end{wraptable}
Since this task requires the model to understand the linguistic content of the input audio, we follow the adaptation used for our base editor and replace the original T5 text encoder in SAM Audio with MOSS-Audio-4B-Thinking~\cite{yang2026moss}. We connect MOSS to the SAM Audio DiT through the same Meta Query~\cite{pan2025transfer} mechanism. The overall architecture is identical to that shown in Figure~\ref{figure37}, except that the DiT backbone of our base editor is replaced by the SAM Audio DiT.

We similarly adapt the SAM Audio DiT with LoRA, using rank 256, alpha 512, and dropout 0.05. We also directly compare standard Flow Matching training with training that incorporates the early-trajectory emphasis principle from the beginning. During training, we use the same set of trainable components as described in Section~\ref{AH1}. The numbers of trainable parameters for different modules are summarized in Table~\ref{tab:sam_tse_trainable_params}.

\paragraph{Training Configuration.}
We initialize the model from the publicly released SAM Audio Small~\cite{shi2025sam} checkpoint and train it on this task for 20 epochs using either Standard FM or the early-emphasis training strategy. Both settings use AdamW with a weight decay of $0.01$, a peak learning rate of $5\times10^{-5}$, 1,000 optimizer-step warmup followed by cosine decay, a batch size of 56, and gradient clipping with a maximum norm of $1.0$.

Since several trainable components, including the DiT LoRA, Connector, and Projector, are newly initialized, this setting does not fully satisfy the original assumption of STEER-Emphasis that the model already possesses an initial task capability. We therefore adopt the same mixed sampling strategy as in the experiment with our base editor: while most training states follow the STEER-Emphasis $t$ sampling, a small proportion are sampled using the standard logit-normal distribution to retain coverage of the full generation trajectory. Specifically, Standard FM uses the standard logit-normal timestep sampling strategy $t\sim\operatorname{sigmoid}\bigl(\mathcal{N}(0,1)\bigr).$ For early-emphasis training, we use
$$
t\sim
\begin{cases}
	\mathcal{U}(0,0.03), & \text{with probability } 0.8,\\
	\operatorname{sigmoid}\bigl(\mathcal{N}(0,1)\bigr), & \text{with probability } 0.2.
\end{cases}
$$
As described in Appendix~\ref{AE1}, SAM Audio adopts the opposite time parameterization from our base editor, where $t=0$ corresponds to noise and $t=1$ corresponds to clean audio. Accordingly, $80\%$ of the training states are concentrated in the high-noise early-trajectory region, while the remaining $20\%$ retain standard logit-normal sampling to preserve coverage of the full generation trajectory. The training objective remains the standard Flow Matching objective in both settings.

\paragraph{Training Dataset.}
We use the same PromptTSE~\cite{seki2025language} dataset as in the experiment with our base editor. In our TSS setting, the target-speaker audio and residual speech are used as the supervision targets for the two output streams.

\paragraph{Evaluation Metric.}
We use CLAP Audio--Audio similarity for TSS
($\operatorname{CLAP\text{-}AA}_{\mathrm{TSS}}$) to evaluate the separation results of SAM Audio. Unlike $\operatorname{CLAP\text{-}AA}_{\mathrm{TSR}}$, which only measures the similarity between the generated residual audio and the ground-truth residual audio, $\operatorname{CLAP\text{-}AA}_{\mathrm{TSS}}$ jointly evaluates both separated outputs while penalizing similarity to the incorrect source. Let $E_a(\cdot)$ denote the CLAP~\cite{elizalde2023clap} audio encoder, and define the cosine similarity between two audio signals as
$$
C(a,b)
=
\frac{
	E_a(a)^\top E_a(b)
}{
	\left\|E_a(a)\right\|_2
	\left\|E_a(b)\right\|_2
}.
$$
For sample $i$, the model produces the predicted target speech
$\hat{s}_{\mathrm{target},i}$ and predicted residual speech
$\hat{s}_{\mathrm{residual},i}$, with corresponding ground-truth signals
$s_{\mathrm{target},i}$ and $s_{\mathrm{residual},i}$. We define
$$
\begin{aligned}
	\operatorname{CLAP\text{-}AA}_{\mathrm{TSS},i}
	=
	\frac{1}{2}\Big[
	& C(\hat{s}_{\mathrm{target},i},s_{\mathrm{target},i})
	- C(\hat{s}_{\mathrm{target},i},s_{\mathrm{residual},i}) \\
	& + C(\hat{s}_{\mathrm{residual},i},s_{\mathrm{residual},i})
	- C(\hat{s}_{\mathrm{residual},i},s_{\mathrm{target},i})
	\Big].
\end{aligned}
$$
The final score is averaged over all $N$ evaluation samples:
$$
\operatorname{CLAP\text{-}AA}_{\mathrm{TSS}}
=
\frac{1}{N}
\sum_{i=1}^{N}
\operatorname{CLAP\text{-}AA}_{\mathrm{TSS},i}.
$$
$\operatorname{CLAP\text{-}AA}_{\mathrm{TSS}}$ therefore encourages the predicted target speech to be more similar to the target ground truth than to the residual speech, while simultaneously encouraging the predicted residual speech to be more similar to the residual ground truth than to the target speech. A higher score indicates better separation of the target and non-target speakers.

\Needspace{0.32\textheight}

\paragraph{Experimental Results.}\mbox{}\par
\vspace{-0.4em}

\begin{wrapfigure}{r}{0.48\textwidth}
	\vspace{-3em}
	\centering
	
	\includegraphics[width=\linewidth]{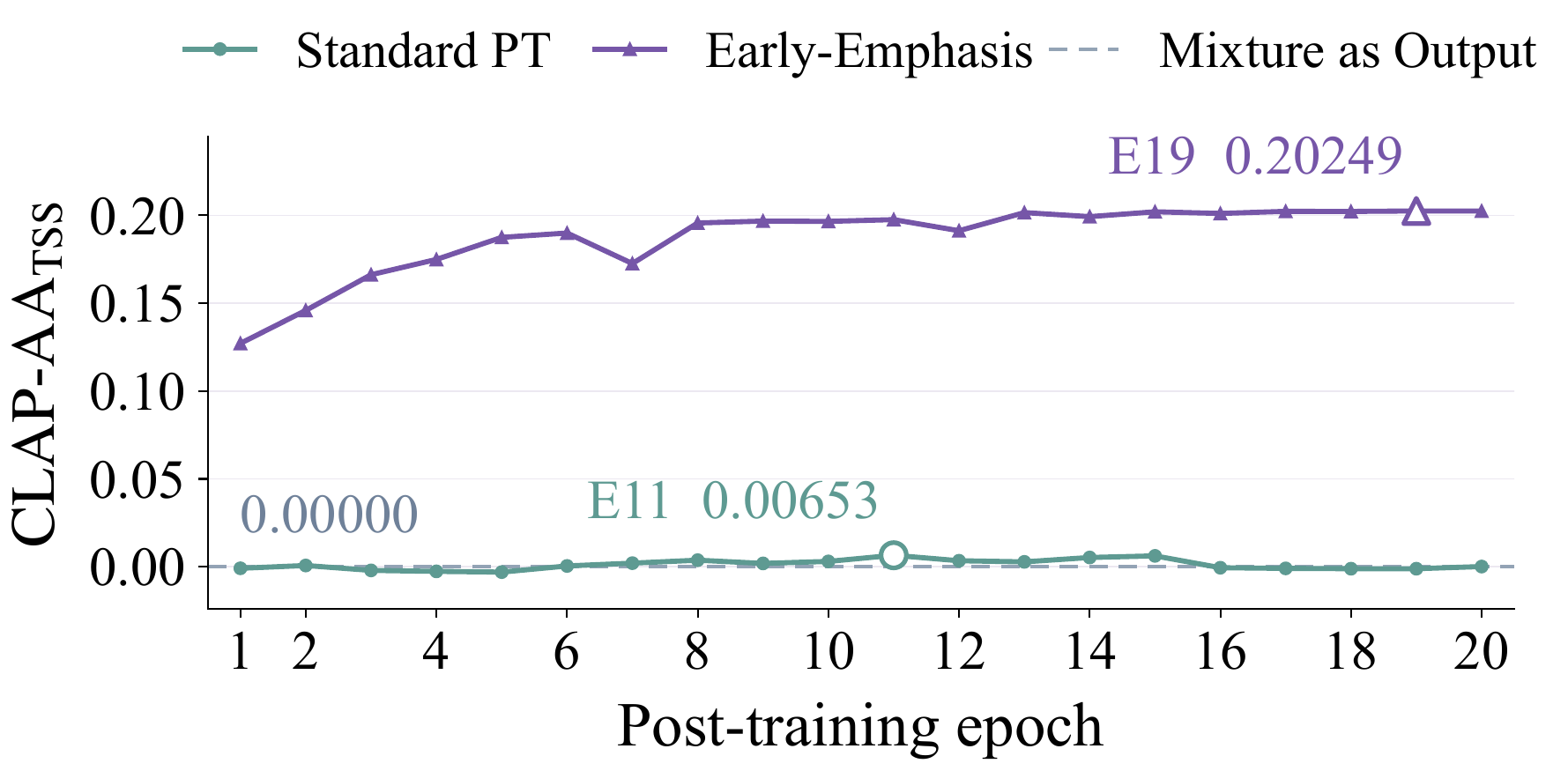}
	
	\vspace{-1.5mm}
	\caption{
		Validation $\operatorname{CLAP\text{-}AA}_{\mathrm{TSS}}$ during training
		on the TSS task with SAM Audio. The horizontal line denotes
		the $\operatorname{CLAP\text{-}AA}_{\mathrm{TSS}}$ of the mixture audio.}
	\label{figure39}
	
	\vspace{-0.8em}
\end{wrapfigure}

Figure~\ref{figure39} shows the validation
$\operatorname{CLAP\text{-}AA}_{\mathrm{TSS}}$ of Standard FM and early trajectory emphasis training over 20 training epochs. Compared with Standard FM, early trajectory emphasis training learns this fine-grained speaker separation task substantially more effectively. This result further demonstrates the importance of emphasizing the training on critical early-trajectory states for fine-grained selective control. Together with the results for our base editor, it further shows that this early-trajectory emphasis training principle remains effective across different model architectures and task formulations.

The horizontal line denotes the $\operatorname{CLAP\text{-}AA}_{\mathrm{TSS}}$ obtained by directly using the unseparated mixed audio as the model output. Standard FM achieves a CLAP-AA score near zero, whereas early-trajectory emphasis substantially improves performance on this fine-grained selective-control task.

\section{Human Evaluation and Website Design}

\subsection{Human Annotation Procedure and Criteria}
As reported in Sections~\ref{5.4}, and Appendices~\ref{AF2}, we conduct human evaluations of our base editor and its post-trained variants on the Complex Audio Editing Test Set and of task-adapted versions of our base editor and SAM Audio on the TSR and TSS tasks.

For the Complex Audio Editing Data, we largely follow the error-type definitions in Appendix~\ref{AC2}. The only difference is that the input audio may contain multiple sources. Therefore, a sample is categorized as Unintended Non-Target Edit whenever the target source undergoes the intended edit while at least one non-target source is also modified. For the TSR and TSS tasks, annotators are asked to determine whether the output audio satisfies the task requirement and to label each result as either successful or unsuccessful. For both annotation tasks, annotators are allowed to abstain when a sample is too complex to judge reliably. Such samples are excluded from the final statistics. During annotation, annotators are not informed which model produced each audio sample.

\subsection{Website Design}

\begin{figure*}[!b]
	\centering
	\includegraphics[width=0.85\textwidth]{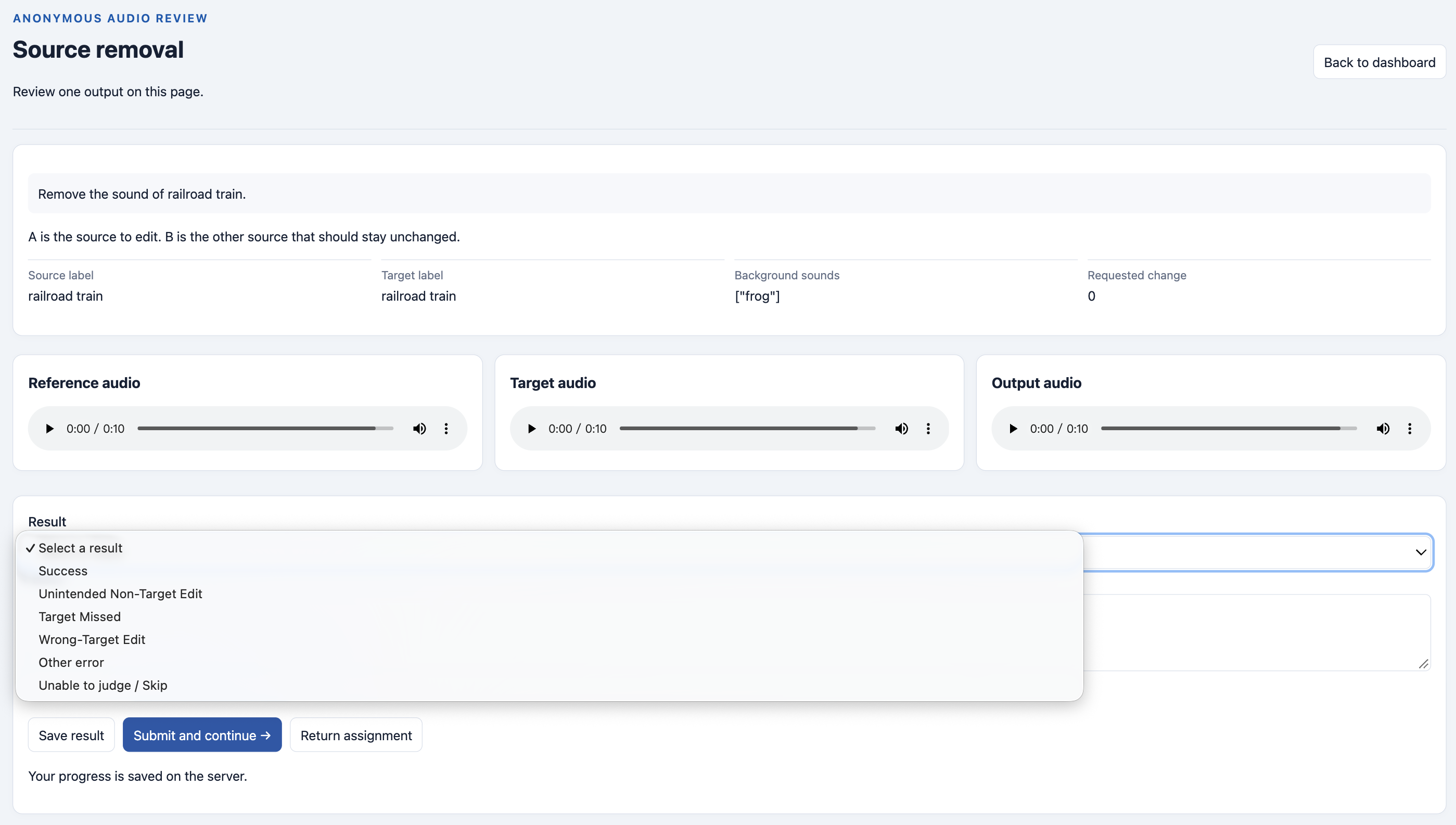}
	\vspace{-2mm}
	\caption{
		Human evaluation interface for the Complex Audio Editing Test Set.
	}
	\label{figure40}
	\vspace{-3mm}
\end{figure*}

\begin{figure*}[!b]
	\centering
	\includegraphics[width=0.85\textwidth]{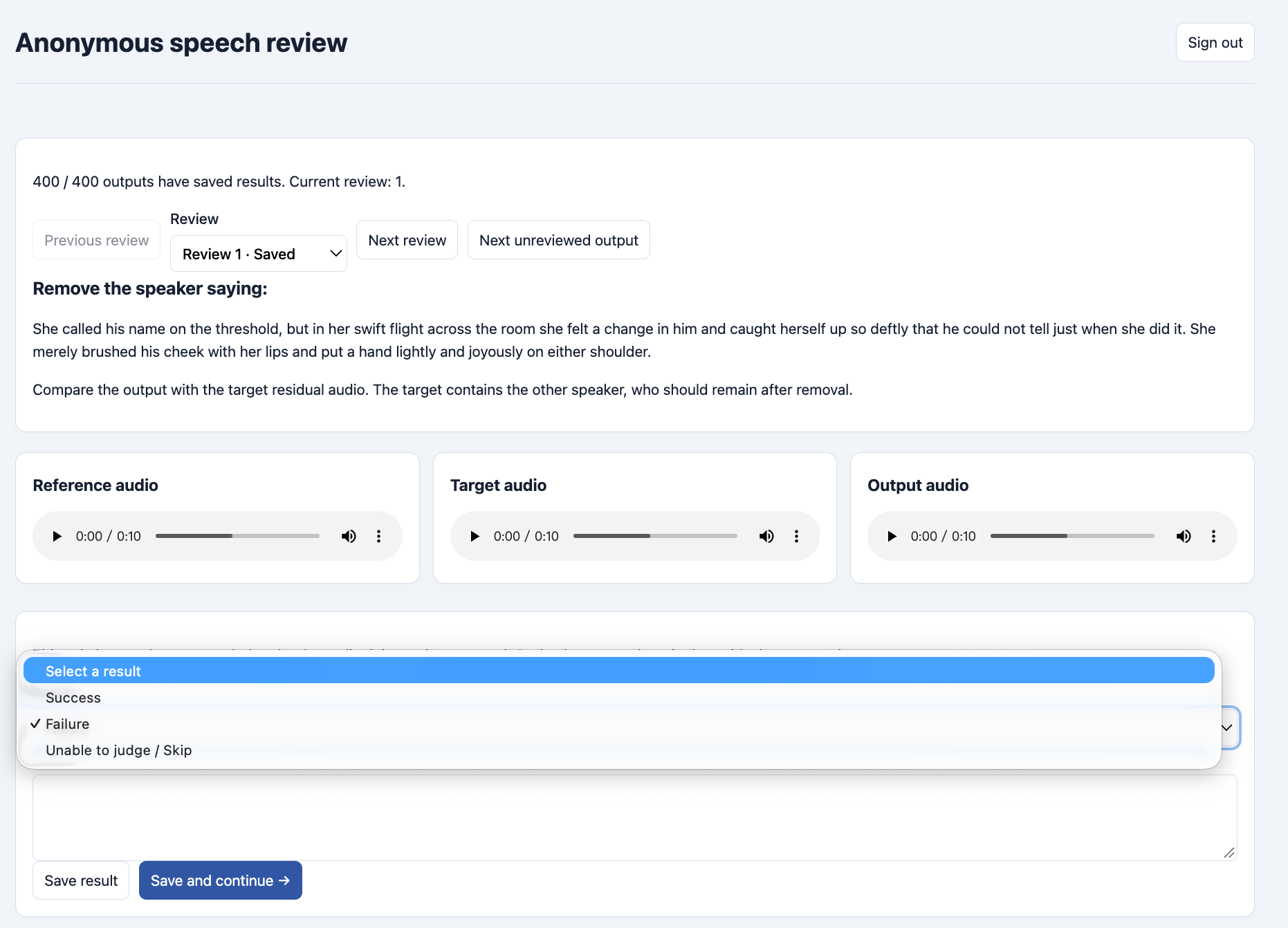}
	\vspace{-2mm}
	\caption{
		Human evaluation interface for the TSR and TSS tasks.
	}
	\label{figure41}
	\vspace{-3mm}
\end{figure*}

We develop two web-based annotation interfaces for human evaluation. For the Complex Audio Editing Test Set, annotators are shown the editing instruction and the reference, target, and output audio, and classify each result according to the success and selective-error categories (Figure~\ref{figure40}). For the TSR and TSS tasks, we use a simplified interface in which annotators judge each output as {Success}, {Failure}, or {Unable to Judge} (Figure~\ref{figure41}). Model identities are hidden in both interfaces.

\end{document}